\documentclass{article}

\usepackage{arxiv}

\usepackage[utf8]{inputenc} 
\usepackage[T1]{fontenc}    
\usepackage{amsmath,amssymb,amsfonts}
\usepackage{hyperref}       
\usepackage{cleveref}       
\usepackage{url}            
\usepackage{booktabs}       
\usepackage{nicefrac}       
\usepackage{microtype}      
\usepackage{lipsum}         
\usepackage{graphicx}
\usepackage{doi}
\usepackage{epsfig}
\usepackage[usenames, dvipsnames]{color}
\usepackage{mathtools}
\usepackage{bm}
\usepackage{stmaryrd}
\usepackage{tikz}
\usepackage{marvosym}
\usepackage{pgfplotstable}
\usepackage{pgfplots}
\usepackage{courier}
\usepackage{geometry}
\usepackage{filecontents}
\usepackage{booktabs}
\usepackage{tabularx}
\usepackage{multirow}
\usepackage{grffile}
\usepackage{subcaption}
\usepackage{algpseudocode}
\usepackage{algorithm}
\usepackage{xspace}
\usepackage[wby]{callouts}
\usepackage{siunitx}

\usetikzlibrary{spy}
\usetikzlibrary{patterns}
\usetikzlibrary{calc,shadings}
\usetikzlibrary{snakes,arrows}
\usepgfplotslibrary{patchplots}
\usetikzlibrary{plotmarks}
\usetikzlibrary{calc,positioning}
\usetikzlibrary{arrows.meta}

\newcommand{\eref}[1]{Eq.~\ref{#1}}
\newcommand{\fref}[1]{Fig.~\ref{#1}}
\newcommand{\frefs}[2]{Figs.~\ref{#1} and \ref{#2}}

\newcommand{\lj}{\mbox{$[\kern-0.1478125em[$}}
\newcommand{\rj}{\mbox{$]\kern-0.1478125em]$}}
\newcommand{\la}{\mbox{$\langle\kern-0.2325em\langle$}}
\newcommand{\ra}{\mbox{$\rangle\kern-0.2325em\rangle$}}

\title{Phase-field modeling of multicomponent vesicles in viscoelastic fluid}

\date{December 1, 2025}

\newif\ifuniqueAffiliation
\uniqueAffiliationtrue

\ifuniqueAffiliation 
\author{{Zuowei Wen} \\
	Institute of Photonics\\
	Leibniz University Hannover\\
	Welfengarten 1A, 30167 Hannover, Germany \\
	\texttt{zuoweiwen@iop.uni-hannover.de} \\
	\And
	{Navid Valizadeh} \\
	Institute of Photonics\\
	Leibniz University Hannover\\
	Welfengarten 1A, 30167 Hannover, Germany \\
	\texttt{valizadeh@iop.uni-hannover.de} \\
	\And
	{Timon Rabczuk} \\
	Institute of Structural Mechanics\\
	Bauhaus University Weimar\\
	Marienstrasse 15, 99423 Weimar \\
	\texttt{timon.rabczuk@uni-weimar.de} \\
	\And
	{Xiaoying Zhuang} \\
	Institute of Photonics\\
	Leibniz University Hannover\\
	Welfengarten 1A, 30167 Hannover, Germany \\
	\texttt{zhuang@iop.uni-hannover.de} \\
}
\else
\usepackage{authblk}

\newbox{\orcid}\sbox{\orcid}{\includegraphics[scale=0.06]{orcid.pdf}} 
\fi

\begin{document}
\maketitle

\begin{abstract}
	Multicomponent vesicles suspended in viscoelastic fluids are crucial for understanding a variety of physiological processes. In this work, we develop a continuum surface force (CSF) phase-field model to investigate the hydrodynamics of inextensible multicomponent vesicles in viscoelastic fluid flows with inertial forces. Our model couples a fluid field comprising both Newtonian and Oldroyd-B fluids, a surface concentration field representing the multicomponent distribution on the vesicle membrane, and a phase-field variable governing the membrane evolution. The viscoelasticity effect of extra stress is well incorporated into the full Navier-Stokes equations in the fluid field. The surface concentration field is determined by Cahn-Hilliard equations, while the membrane evolution is governed by a nonlinear advection-diffusion equation. The membrane is coupled to the surrounding fluid through the continuum surface force (CSF) framework. To ensure stable numerical solutions of the highly nonlinear multi-field model, we employ a residual-based variational multiscale (RBVMS) method for the Navier–Stokes equations, a Streamline-Upwind Petrov–Galerkin (SUPG) method for the Oldroyd-B equations, and a standard Galerkin finite element framework for the remaining equations. The system of PDEs is solved using an implicit, monolithic scheme based on the generalized-$\alpha$ time integration method. To enhance spatial accuracy, we employ isogeometric analysis (IGA). We present a series of two-dimensional numerical examples in shear and Poiseuille flows to elucidate the influence of membrane composition and fluid viscoelasticity on the hydrodynamics of multicomponent vesicles.
\end{abstract}

\keywords{ Isogeometric analysis \and Hydrodynamics \and Multicomponent vesicle \and Viscoelastic fluid \and Phase-field modeling \and Continuum surface force}

\section{Introduction}
In biological systems, multicomponent vesicles are fluid-filled compartments bounded by lipid bilayer membranes composed of a complex mixture of lipids, proteins, cholesterol, and additional molecular constituents. They are involved in key physiological processes, including cellular homeostasis, infection propagation, cancer development, and cardiovascular diseases \cite{Herrmann2021}. In addition to their biological significance, multicomponent vesicles are widely utilized in biomedical applications, including drug delivery, biomolecule production, and the design of artificial cell-like systems \cite{Elani2014}. Furthermore, due to their structural and dynamic similarities to biological cells, they are often employed as simplified models of red blood cells (RBCs) under various physiological and pathological conditions. RBCs resist shear deformation due to the presence of a membrane cytoskeleton and also exhibit resistance to bending and area dilatation owing to the nature of their lipid bilayer \cite{Abkarian2008,VLAHOVSKA2013451,Wan2011}. In contrast, the lipid bilayer membranes of vesicles are fluid-like and exhibit no shear resistance but similarly resist bending and remain nearly inextensible \cite{Lipowsky1991475,Seifert199713,ALAND201432}.  The presence of diverse components on the membrane alters its surface physical properties, spontaneous curvature, and interfacial energy, leading to a wide variety of topological and morphological shape transitions under flow conditions with varying geometrical and physical characteristics. All these effects are evident in phenomena such as tank-treading, swinging, and tumbling motions observed in shear flow \cite{Keller_Skalak_1982, Noguchi2007, Gera_2022, WEN2024117390}. Furthermore, RBCs are immersed in viscoelastic fluids that exhibit polymeric stresses beyond those of Newtonian fluids, resulting in distinct hydrodynamic behavior and significantly increasing the complexity of their analysis. The intricate fluid–structure interactions in vesicle hydrodynamics, coupled with their broad range of applications, continue to motivate extensive research in both experimental and computational domains.\par

While most previous studies have focused on the hydrodynamics of homogeneous vesicles in Newtonian fluid flows \cite{ALAND201432, VALIZADEH2022114191, Gannon2021}, or on particles suspended in viscoelastic fluids \cite{YUE_FENG_LIU_SHEN_2005,ESCOTT2024105262, Naseer_Izbassarov_Ahmed_Muradoglu_2024}, these studies represent simplified models of the complex bio-membrane hydrodynamics. In recent years, the hydrodynamics of multicomponent vesicles in Newtonian fluid flows has garnered significant attention and has been extensively studied using various methodologies \cite{GERA2018362, Bachini_Krause_Nitschke_Voigt_2023, venkatesh2024shapedynamicsnearlyspherical, WEN2024117390}. These studies have shown that the highly complex membrane heterogeneity has a significant impact on the flow-induced morphological and compositional dynamics of vesicles and can even alter their motion patterns. In this study, we investigate the hydrodynamics of inextensible multicomponent vesicles immersed in viscoelastic fluid flows, which provide a more realistic representation of in-vivo environments such as blood plasma, mucus, and other polymer-rich biological media. In contrast to Newtonian fluids, viscoelastic fluids possess both viscous and elastic characteristics, arising from the presence of macromolecules, and exhibit complex rheological phenomena such as polymeric stresses, normal stress differences, and stress relaxation. These properties can significantly influence vesicle deformation, orientation dynamics, and interfacial stability, thereby playing a crucial role in accurately capturing the mechanics of cellular and subcellular transport in physiological conditions. \par

A representative example of a multicomponent vesicle consists of three components: two distinct lipid species and cholesterol. In such systems, cholesterol preferentially associates with one of the lipid species (e.g., dipalmitoylphosphatidylcholine(DPPC)) \cite{PhysRevLett.94.148101,Davis2009,Pradeep2010}, leading to the formation of an ordered (stiff) phase, while the remaining components form a disordered (soft) phase \cite{VEATCH2005172}. Such a two-phase multicomponent vesicle is filled with a Newtonian fluid and suspended in a viscoelastic medium, which can be modeled using the Oldroyd-B constitutive framework. The Oldroyd-B model assumes that the viscoelastic fluid consists of an elastic polymeric component dispersed in a viscous Newtonian solvent. It captures both the solvent viscosity and the polymeric stress contribution through a linear elastic spring-dashpot system. This model accounts for stress relaxation and elastic recoil, key features in many biological fluids, while neglecting shear-thinning and finite extensibility effects, making it suitable for moderate deformation regimes relevant to our study \cite{Bird1987}.\par

In the continuum modeling of multicomponent vesicle hydrodynamics in viscoelastic fluids, the lipid bilayer membrane is described by an extension of the classical Canham–Helfrich model \cite{Canham1970, Helfrich1973693}, which captures both fluid-like and elastic solid-like behaviors. The membrane deforms to minimize its energy, while strictly preserving surface inextensibility and conserving the enclosed volume. Meanwhile, phase separation and mild migration of membrane components occur on the vesicle surface, governed by the Cahn–Hilliard equation \cite{Cahn1958}. This process captures the dynamic redistribution of different components in response to membrane deformation and external fluid flow. The complex nonlinear coupling among vesicle deformation, fluid flow, and the distribution of different components on the membrane has recently been the primary focus of several computational studies. Salac et al. \cite{SALAC20118192, salac_miksis_2012} developed a continuum surface force (CSF) model that couples the Navier–Stokes equations with a level-set formulation governing membrane evolution. Gera et al. \cite{GERA2018362} augmented the model by introducing an additional Cahn–Hilliard equation to investigate the hydrodynamics of multicomponent vesicles. In the realm of phase field modeling, Du et al. \cite{Du2007,DU2009923} developed a thermodynamically consistent framework that couples the Navier--Stokes equations with an evolution equation for a phase-field variable representing the vesicle membrane. Aland et al. \cite{ALAND201432} enhanced the model with a local inextensibility constraint within the diffuse interface formulation. Wen et al. \cite{WEN2024117390} further extended the model to investigate the hydrodynamics of multicomponent vesicles by employing an energy variation approach.\par

In most vesicle studies, the vesicles are suspended in Stokes flow, which significantly simplifies the governing nonlinear equations by neglecting inertial effects and reducing them to linear form \cite{C6SM02452A, Gera_2022,SOHN2010119}. However, under moderate Reynolds numbers, inertial effects can play a critical role in altering vesicle dynamics \cite{ALAND201432, salac_miksis_2012, Laadhari2012}. Given the broad range of Reynolds number conditions encountered in vesicle-related applications, it is essential to employ the full Navier–Stokes equations with inertial effects to accurately capture their hydrodynamic behavior. In addition to inertia, the viscoelastic response of polymeric components in the surrounding fluid significantly alters vesicle hydrodynamics, leading to distinctive non-Newtonian behaviors. For example, Seol et al. \cite{SEOL20191009} showed that a tumbling homogeneous vesicle can transition to tank-treading motion with a negative inclination angle under viscoelastic effects; with increasing inertial effects, this inclination angle becomes positive. However, the influence of viscoelastic effects on the hydrodynamics of multicomponent vesicles remains largely unexplored and warrants thorough investigation. To the best of our knowledge, this is the first unified computational framework to investigate the hydrodynamics of inextensible multicomponent vesicles immersed in viscoelastic fluid flows. This work captures the coupling among membrane heterogeneity, viscoelastic stress fields, and flow-induced morphological and compositional dynamics, revealing complex phenomena and providing mechanics-driven insight into behaviors that are absent in extensively studied simplified models, such as single-component or Newtonian-flow formulations.\par

The numerical solution of moving-interface problems poses a significant challenge, primarily because the location of the evolving interface is not known a priori. The numerical methods for handling vesicles can be broadly categorized into several approaches. The boundary element method reformulates the governing equations into boundary integral equations instead of solving partial differential equations in the domain \cite{Gannon2021, Veerapaneni20115610}. The immersed boundary method embeds the Lagrangian vesicle membrane into an Eulerian fluid flow framework \cite{ CASQUERO2017646, CASQUERO2021109872, SEOL20191009}. The level-set method captures the moving surface implicitly by evolving a signed distance function \cite{GERA2018362, LAADHARI2017271}. The phase-field method introduces a continuous order parameter to distinguish between the membrane and the surrounding fluid \cite{VALIZADEH2022114191, ALAND201432, DU2009923, WEN2024117390}. This work adopts the phase-field method to implicitly define the vesicle membrane, leveraging its capability to use simple meshes while accurately capturing complex geometries and accommodating extreme topological deformations with desirable smoothness. The phase-field approach has found extensive applications across various fields, including crack propagation \cite{ GONG2025111039}, tumor growth \cite{XU2020112648}, topology optimization \cite{LOPEZ2022114564},  vesicle dynamics \cite{ALAND201432,VALIZADEH2022114191,WEN2024117390}, and fluid-structure interaction \cite{VALIZADEH2025117618,RATH2024117348}. \par

In this paper, we develop a continuum surface force (CSF) phase-field model to investigate the hydrodynamics of inextensible multicomponent vesicles immersed in viscoelastic fluid flows with inertial effects. Building upon previous models for vesicle hydrodynamics in fully Newtonian fluids \cite{VALIZADEH2022114191,WEN2024117390}, we introduce a more general CSF-based coupling approach and extend the framework to incorporate the additional complexities introduced by fluid viscoelasticity. The resulting model is inherently nonlinear and couples several interacting fields: the fluid velocity and pressure fields, the phase-field identifying the vesicle interface, the membrane surface tension, the viscoelastic stress, and the surface concentration field describing the distribution of membrane components. The fluid field is governed by the full Navier--Stokes equations augmented with the Oldroyd-B model to capture the viscoelastic effects arising from the polymeric stresses in the outer fluid, while the inner fluid remains Newtonian. The surface concentration field of the two membrane phases is described by a concentration variable governed by the Cahn--Hilliard equation. The phase-field is evolved using a nonlinear advection–diffusion equation, where advection is driven by the fluid velocity field. To prevent the accumulation of numerical errors in global area and volume conservation, two regularization terms are introduced on the right-hand side of the phase-field evolution equation. Consequently, our simulations effectively capture the coupled hydrodynamics between membrane deformation and Newtonian–Oldroyd-B fluid flow, along with the phases evolution on the membrane surface.  The system of PDEs is solved using an implicit, monolithic scheme based on the generalized-$\alpha$ time integration method, enabling faster convergence and permitting larger time step sizes compared to explicit or semi-implicit staggered schemes. To enhance spatial accuracy, we employ isogeometric analysis (IGA). For stable and efficient numerical solutions of the highly nonlinear multi-field model, we adopt a residual-based variational multiscale (RBVMS) method for the Navier–Stokes equations, a Streamline-Upwind Petrov–Galerkin (SUPG) method for the Oldroyd-B equations, and a standard Galerkin finite element framework for the remaining equations. A series of two-dimensional numerical examples in shear and Poiseuille flows are presented to elucidate the influence of membrane composition and fluid viscoelasticity on the hydrodynamics of multicomponent vesicles. To summarize, the key contributions of our work are as follows:\par

\begin{itemize}
	\item A CSF phase-field model is developed to investigate the viscoelastic effects on the hydrodynamics of multicomponent vesicles in various flow regimes in 2D. The model accounts for the effects of bending, surface tension, and line tension forces, with variations in surface properties influencing fluid flow, membrane deformation, and the distribution of surface phases.
	\item This study presents a fully monolithic and implicit algorithm based on the generalized-$\alpha$ time integration method.
	\item  It employs the robust RBVMS method to stabilize the Navier--Stokes equations for incompressible flows and the SUPG method to handle the Oldroyd-B equations. This stabilization eliminates the need for the inf–sup stability condition in the $\boldsymbol{\sigma}$–$\mathbf{u}$–$p$ system, thereby allowing the use of equal-order interpolations.
	\item Isogeometric analysis is adopted to enhance the accuracy of numerical results per degree of freedom.
	\item This is the first study to examine the influence of viscoelastic effects on the hydrodynamics of multicomponent vesicles.
	\item In shear flow, where the swinging and tumbling motions of multicomponent vesicles are primarily driven by bending rigidity variations, viscoelastic fluids damp these motions, prolong swinging periods, and trigger tank-treading.  This viscoelastic damping also consistently increases the phase treading period of the components on the multicomponent vesicle membrane. In Poiseuille flow, viscoelasticity has only minor effects, primarily stretching the vesicle along the flow direction.
\end{itemize}
The remainder of this paper is organized as follows. In Sec.~\ref{Sec:MathematicalModel}, we present a CSF phase-field model for multicomponent vesicles with viscoelastic effect. In Sec.~\ref{Sec:NumericalFormulation}, we focus on the numerical formulation within the framework of isogeometric analysis and present the spatial and temporal discretization of the system. In Sec.~\ref{NumericalExamples}, numerical results demonstrate the effectiveness of the model in simulating multicomponent vesicles across a variety of flow scenarios in 2D, both with and without the viscoelastic effect. Finally, conclusions are drawn in Sec.~\ref{Conclusions}.

\section{Mathematical model} \label{Sec:MathematicalModel}
\begin{figure}[H]
	\centering
	\includegraphics [scale=0.32]{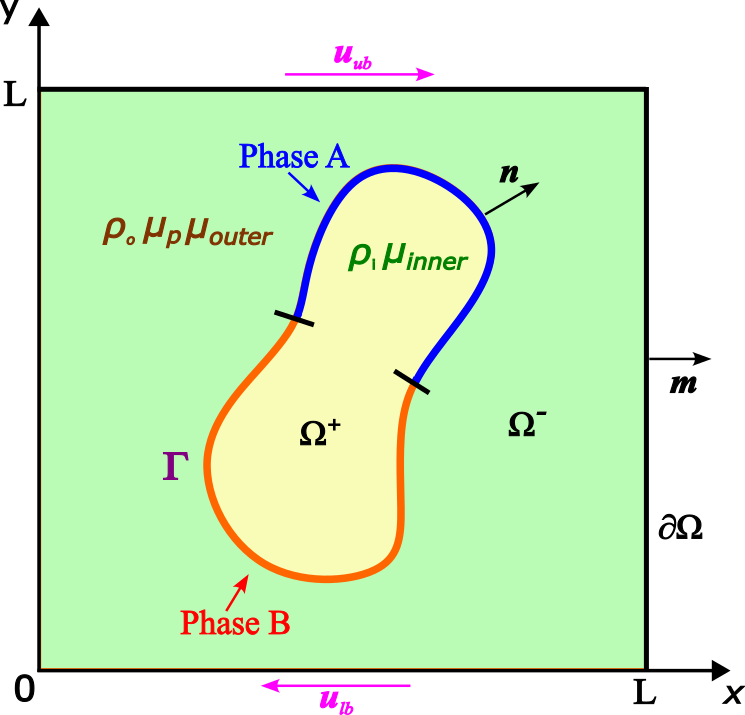}      
	\caption{The multicomponent vesicle, filled with Newtonian fluid, is suspended in an Oldroyd-B fluid within the computational domain.}
	\label{fig:shearFlow}
\end{figure}

This work focuses on giant inextensible multicomponent vesicles, filled with Newtonian fluid (depicted in yellow in Fig.~\ref{fig:shearFlow}) and suspended in an Oldroyd-B fluid (shown in green) flow that incorporates both inertial and viscoelastic effects. The characteristic length scale of such vesicles is on the order of $10^{-6} \: m$. The mathematical model comprises multiple interacting fields, with three primary fields coupled through the continuum surface force (CSF) framework: the fluid field, governed by the full Navier–Stokes equations augmented with the Oldroyd-B model; the surface concentration field of the two membrane phases, described by a concentration variable governed by the advective Cahn–Hilliard equation; and the phase-field, which evolves according to a nonlinear advection–diffusion equation driven by the fluid velocity. Lagrange multipliers are employed as regularization terms in the phase-field equation to prevent the accumulation of numerical errors in global area and volume conservation \cite{ALAND201432,VALIZADEH2022114191}.\par

Our diffuse-interface phase-field model builds upon the sharp-interface level-set framework of Laadhari et al.~\cite{LAADHARI2014328} and Gera et al.~\cite{GERA2018362}, and is further extended to incorporate viscoelastic effects, following the approaches of Yue et al. \cite{YUE_FENG_LIU_SHEN_2005}, who investigated viscoelastic effects on droplets, and Seol et al. \cite{SEOL20191009}, who studied the impact of viscoelasticity on homogeneous vesicles.\par

Let $\Omega\subset\mathbb{R}^\text{d}$, $\text{d}=2,3$, denote the computational domain which contains the multicomponent vesicle and fluid domain (see~\fref{fig:shearFlow}). $\partial\Omega$ denotes the boundary of computational domain, and
$\bm{m}$ the outward unit normal vector to $\partial\Omega$. The external Oldroyd-B fluid domain surrounding the vesicle is represented by $\Omega^{-}$ and internal Newtonian fluid domain within the vesicle is denoted by $\Omega^{+}$. The membrane of the vesicle is denoted by $\Gamma$ and the unit outward normal vector to $\Gamma$ is denoted by $\bm{n}$. The membrane consists of phase A and phase B. Let $\bm{u}$ denote the fluid velocity, $p$ the fluid pressure, $\rho$ the fluid density, $\mu_{inner}$ the viscosity of the inner Newtonian fluid, $\mu_p$ and $\mu_{outer}$ the viscosities of the polymer and Newtonian solvent components of the outer Oldroyd-B fluid, respectively. $c \in [0,1]$ is the concentration of phase B while concentration of phase A is indicated as $1-c$. We define the time as $t \in \left(0,T\right)$, with $T>0$. 

\subsection{Fluid field}
We begin with a sharp interface model for vesicle dynamics \cite{LAADHARI2014328,GERA2018362,SEOL20191009} and extend it into a continuum surface force (CSF) phase-field model. Assuming the fluid is incompressible, the full Navier–Stokes equations are given by:
{ \footnotesize
	\begin{align}
		\rho\left(\partial_t \bm{u}+\bm{u}\cdot\nabla\bm{u}\right) -\nabla\cdot \left[2\left(\left(1-\mathcal{H}\right)\mu_{inner}+\mathcal{H}\mu_{outer}\right) \bm{D}(\bm{u})+\mathcal{H}\bm{\sigma}\right]+\nabla p &=\bm{0}, &&\text{in} \; \Omega\times\left(0,T\right),\label{LBLME-SI}\\
		\nabla\cdot\bm{u} &=0, &&\text{in} \;\Omega\times\left(0,T\right), \label{LBME-SI}
\end{align}}
where $\mathcal{H}$ is the Heaviside function defined by:
\[
\mathcal{H}(\bm{x},t)= 
\begin{cases}
	1,\quad& \text{if } \bm{x} \;\text{is in Oldroyd-B fluid},\\
	0,& \text{if } \bm{x} \;\text{is in Newtonian fluid}.
\end{cases}
\]
Here $\bm{D}(\bm{u})=\frac{1}{2}\left(\nabla\bm{u}+\nabla^T\bm{u}\right)$ is the strain rate tensor, $\bm{I}$ is the identity tensor, $\rho$ is the fluid density which is assumed to be the same inside and outside of the vesicle, $\bm{\sigma}$ is the dumbbell elastic stress tensor and is only active in the Oldroyd-B fluid region where $\mathcal{H} = 1$. \eref{LBLME-SI} corresponds to the momentum equation, and \eref{LBME-SI} enforces the incompressibility condition by ensuring mass conservation. The boundary and initial conditions are specified as follows:
\begin{align}
	\bm{u} &= \bm{u}_b, &&\text{on} \quad\partial\Omega_{D}\times\left(0,T\right), \label{Dirichlet-u-SI}\\
	\left(2\left(\left(1-\mathcal{H}\right)\mu_{inner}+\mathcal{H}\mu_{outer}\right) \bm{D}(\bm{u})+\mathcal{H}\bm{\sigma}-p \, \bm{I}\right) \cdot \bm{m} &=\bm{0}, &&\text{on} \quad\partial\Omega_{N}\times\left(0,T\right), \label{traction-SI}\\
	\bm{u}(0) &= \bm{u}_0,&&\text{in} \quad\Omega , \label{Initial-u-SI}
\end{align}
where $\partial\Omega_N$ and $\partial\Omega_D$ are the natural and essential parts of $\partial\Omega$ for which we have $\partial \Omega=\partial\Omega_D \cup \partial\Omega_N$. \par
The dumbbell elastic stress tensor $\bm{\sigma}$ is symmetric and obeys the Maxwell equation \cite{YUE_FENG_LIU_SHEN_2005}:
\begin{align}
	&\xi_\sigma\overset{\nabla}{\bm{\sigma}}+\bm{\sigma}= \mu_p \bm{D}(\bm{u}),&&\text{in} \quad \Omega\times\left(0,T\right),  \label{VS-SI}\\
	&\overset{\nabla}{\bm{\sigma}}=\partial_t  \bm{\sigma}+\bm{u}\cdot\nabla\bm{\sigma}-\left(\nabla\bm{u} \, \bm{\sigma}+\bm{\sigma}\,\nabla^T\bm{u}\right). \label{UC-VS-SI}
\end{align}
$\overset{\nabla}{\bm{\sigma}}$ is the upper convected time derivative of $\bm{\sigma}$ defined in \eref{UC-VS-SI}. $\xi_\sigma$ is the polymer relaxation time. The total viscosity of the Oldroyd-B fluid is determined by the combined contributions of the Newtonian solvent viscosity $\mu_{outer}$ and the polymer viscosity $\mu_p$.\par
\subsubsection{Membrane interface}
We employ the continuum surface force (CSF) method to couple the interactions between the vesicle membrane interface and the surrounding fluid into the Navier--Stokes momentum equation (\eref{LBLME-SI}). Originally introduced by Brackbill et al. \cite{BRACKBILL1992335}, the CSF method was subsequently employed by Salac et al. \cite{GERA2018362, SALAC20118192} to model vesicle dynamics within the level-set framework. We begin by presenting a sharp interface formulation and subsequently extend it into our diffuse interface phase-field framework.\par
We focus on the impermeable multicomponent vesicle, whose lipid bilayer membrane exhibits in-plane resistance to stretching \cite{Morris2001} and out-of-plane resistance to bending. We assume that the vesicle conserves both its total surface area and enclosed volume throughout its evolution. In addition, the membrane satisfies a local inextensibility constraint, which is mathematically expressed in the sharp interface formulation as:
\begin{equation}
	\nabla_s\cdot\bm{u} =0, \quad\text{in}\quad \Gamma\times\left(0,T\right) , \label{IC-SI}
\end{equation} 
where $\nabla_s$ is the surface gradient operator.\par
At the membrane interface between the vesicle and the surrounding fluid, various forces contribute to membrane deformation, which in turn influences the fluid flow in the surroundings. These forces must balance the jump in hydrodynamic stress across the membrane interface \cite{ALAND201432,VALIZADEH2022114191,SALAC20118192}:
{\footnotesize
	\begin{align}
		&\lj \bm{u} \rj =0, &&\text{in}\quad \Gamma\times\left(0,T\right), \label{Jump-u-SI}\\
		&\lj 2\left(\left(1-\mathcal{H}\right)\mu_{inner}+\mathcal{H}\mu_{outer}\right) \bm{D}(\bm{u})+\mathcal{H}\bm{\sigma}-p \, \bm{I} \rj \cdot \bm{n} =\lj \bm{q_s} \rj\cdot\bm{n} =\bm{F_s}, && \text{in}\quad \Gamma \times\left(0,T\right) , \label{Jump-stress-SI}
	\end{align}
}
where $\lj \cdot \rj$ denotes the jump in a quantity across the interface. The most crucial force at the membrane interface is the bending force, typically derived from the variational derivative of the Canham--Helfrich bending energy. Another force is the surface tension, which enforces the local inextensibility constraint according to \eref{IC-SI}. Together with contributions from Gaussian bending energy and line tension energy, the total sharp interface force is expressed as:
\begin{align}
	\bm{F_s} &=-\frac{\partial E}{\partial \Gamma}, \label{F-SI-I}\\
	E &= E_b + E_g + E_{\lambda} + E_c, \nonumber\\
	E_b &= \int_{\Gamma}  \frac{b_c(c)}{2} (H-H_0(c))^2 ~\text{d}a, \nonumber\\
	E_g &= \int_{\Gamma}  b_g(c) K ~\text{d}a,\nonumber\\ 
	E_{\lambda} &= \int_{\Gamma} \lambda ~\text{d}a,\nonumber\\
	E_c &= \int_{\Gamma} \left(h_c(c)+\frac{k_f}{2}|\nabla_s c|^2 \right)~\text{d}a. \nonumber
\end{align}
Here, $E$ is the free energy functional of the membrane, and $E_b$ represents the total bending energy. $H$ is the total curvature of the membrane surface, $H_0(c)$ is the spontaneous curvature that depends on $c$. $b_c(c)$ represents the bending rigidity of the membrane, which depends on the distribution of phase A and phase B, denoted as $b_c$ in the remainder of this work. $E_g$ is the Gaussian bending energy, $b_g$ the Gaussian bending rigidity, and $K$ is the Gaussian curvature. In this work, the Gaussian bending rigidity of the two lipid species are assumed to be of the same order of magnitude \cite{liu2011advances}, and we further assume $b_g$ to be constant. According to the Gauss-Bonnet theorem, for a membrane surface of fixed topology, $E_g$ becomes a constant and thus can be disregarded. $E_{\gamma}$ is the energy contribution due to surface tension and $\lambda$ is the surface tension variable. $E_c$ is the line energy, $k_f=\epsilon_{c}^2$, and $\epsilon_{c}$ is the regularization parameter that controls the thickness of the interface between the two phases,  $h_c(c)$ is the double-well potential that represents the mixing energy. We denote $H_0(c)$ and $h_c(c)$ as $H_0$ and $h_c$, respectively, in the remainder of this work. In this model, we neglect the spontaneous curvature $H_0$ and express \eref{F-SI-I} in its explicit form:
\begin{align}
	\bm{F_s} &=-\left(\frac{\partial E_b}{\partial \Gamma}+\frac{\partial E_\lambda}{\partial \Gamma}+\frac{\partial E_c}{\partial \Gamma}\right) \nonumber\\
	&=b_c \left(\frac{1}{2}H^3-2HK+\Delta_sH\right)\bm{n}+\frac{1}{2}H^2\nabla_sb_c+\bm{n}H\Delta_sb_c \nonumber\\
	&\quad+\nabla_s\lambda-\lambda H\bm{n} \nonumber\\
	&\quad+k_f\left(\nabla_sc\cdot \bm{L}\nabla_sc\right)\bm{n}-\frac{k_f}{2}|\nabla_sc|^2H\bm{n}-k_f(\nabla_sc)\Delta_sc , \label{F-SI-EX}
\end{align}
where $\Delta_s=\nabla_s\cdot\nabla_s$ is the Laplace-Beltrami operator. $\bm{L} = \nabla_s \bm{n}$ is the surface curvature tensor, and the Gaussian curvature is given by $K = \frac{1}{2} \left( \left( \mathrm{tr} \left( \bm{L} \right) \right)^2 - \mathrm{tr} \left( \bm{L}^2 \right) \right)$ where $\mathsf{tr}(\cdot)$ is the trace operator. Full details of the derivation for the above expressions can be found in Gera et al. \cite{GERA2018362}, where redundant terms similar to surface tension are neglected \cite{C8SM01087K}.\par
We derive the CSF model by considering smooth test functions $\bm{w}$ defined on the domain $\Omega$, with $\bm{w} = 0$ on the domain boundary $\partial\Omega$, for the velocity field $\bm{u}$ \cite{Gomez2017}. From \eref{Jump-u-SI}, we have $\lj \bm{w} \rj = 0$ on $\Gamma$. We now proof that both \eref{LBLME-SI} and \eref{Jump-stress-SI} are equivalent to the requirement:
\begin{align}
	\rho \frac{d}{dt}\int_{\Omega}\bm{w}\cdot\bm{u} \;d\Omega+\rho\int_{\Omega} \bm{w}\cdot\left(\bm{u}\cdot\nabla\bm{u}\right)d\Omega=-\int_{\Omega}\nabla\bm{w}:\bm{q_s}d\Omega \label{CSF-P}
\end{align}
where the left-hand side of the equation, upon applying Reynolds’ theorem, can be expressed as:
\begin{align*}
	& \rho \frac{d}{dt}\int_{\Omega^-\cup\,\Omega^+}\bm{w}\cdot\bm{u} \;d\Omega +\rho\int_{\Omega^-\cup\,\Omega^+} \bm{w}\cdot\left(\bm{u}\cdot\nabla\bm{u}\right)d\Omega=\\
	&\rho\int_{\Omega^-\cup\,\Omega^+}\bm{w}\cdot\partial_t\bm{u}\;d\Omega +\rho\int_{\Omega^-\cup\,\Omega^+} \bm{w}\cdot\left(\bm{u}\cdot\nabla\bm{u}\right)d\Omega-\int_{\Gamma} \bm{w}\cdot \lj \bm{u} \rj \bm{u_n}\;d\Gamma
\end{align*}
Substituting \eref{Jump-u-SI}, we obtain:
\begin{align}
	\rho \frac{d}{dt}\int_{\Omega^-\cup\,\Omega^+}\bm{w}\cdot\bm{u} \;d\Omega +&\rho\int_{\Omega^-\cup\,\Omega^+} \bm{w}\cdot\left(\bm{u}\cdot\nabla\bm{u}\right)d\Omega= \nonumber\\
	&\rho\int_{\Omega^-\cup\,\Omega^+ }\bm{w}\cdot\partial_t\bm{u}\;d\Omega +\rho\int_{\Omega^-\cup\,\Omega^+} \bm{w}\cdot\left(\bm{u}\cdot\nabla\bm{u}\right)d\Omega \label{CSF-L-P}
\end{align}
\noindent The right hand side of \eref{CSF-P} equals:
\begin{align}
	-\int_{\Omega^-\cup\,\Omega^+}\nabla\bm{w}:\bm{q_s}d\Omega&=\int_{\Omega^-\cup\,\Omega^+}\bm{w}\cdot\left(\nabla\cdot \bm{q_s}\right)+\int_{\Gamma}\bm{w}\cdot\left(\lj \bm{q_s} \rj\cdot\bm{n}\right)d\Gamma \nonumber\\
	&=\int_{\Omega^-\cup\,\Omega^+}\bm{w}\cdot\left(\nabla\cdot \bm{q_s}\right)+\int_{\Gamma}\bm{w}\cdot\bm{F_s}d\Gamma \label{CSF-R-P}
\end{align}
By combining \eref{CSF-L-P} and \eref{CSF-R-P}, we obtain the equivalent forms of \eref{LBLME-SI} and \eref{Jump-stress-SI}:
\begin{align}
	\rho\int_{\Omega}\bm{w}\cdot\left(\partial_t\bm{u}+\bm{u}\cdot\nabla\bm{u}\right)\;d\Omega  = \int_{\Omega}\bm{w}\cdot\left(\nabla\cdot \bm{q_s}\right)+\int_{\Gamma}\bm{w}\cdot\bm{F_s}d\Gamma \label{CSF-S}
\end{align}\par
To construct the phase-field model, we introduce a phase-field variable $\phi(\bm{x},t) := \tanh\left(\frac{-d(\bm{x},t)}{\sqrt{2}\epsilon}\right)$, where $d(\bm{x},t)$ is the signed distance function to the membrane interface. The parameter $\epsilon$ controls the thickness of the diffuse interface. By definition, $d(\bm{x},t)$ is negative inside the vesicle, positive outside, and zero at the membrane. Consequently, the phase-field variable satisfies $\phi \approx 1$ inside the vesicle, $\phi \approx -1$ outside, and $\phi \approx 0$ at the membrane interface $\Gamma$. The unit outward normal vector to the membrane can be defined as $\bm{n}=-\frac{\nabla\phi}{|\nabla\phi|}$. By utilizing the phase-field variable and Dirac delta functions, the hydrodynamic equations of binary fluids with different properties in \eref{CSF-P} can be rewritten into a single {\it ‘diffusified’} formulation. The strong form of the diffuse interface version of \eref{CSF-S} is stated as:
{ \footnotesize
	\begin{align*}
		\rho\left(\partial_t \bm{u}+\bm{u}\cdot\nabla\bm{u}\right) -\nabla\cdot \left[2\left(\frac{1+\phi}{2}\mu_{inner}+\frac{1-\phi}{2}\mu_{outer}\right) \bm{D}(\bm{u})+\frac{1-\phi}{2}\bm{\sigma}\right]+\nabla p =\bm{F_D}, \;\text{in} \; \Omega\times\left(0,T\right),
\end{align*}}
where $\bm{F_D}=\delta_{\phi} \bm{F_s}$ and $\delta_\phi = \frac{3}{4\sqrt{2}\epsilon}(1-\phi^2)^2$ is the approximate delta function casting the surface integral as an equivalent volume integral in diffuse interface form \cite{Dolow2013}. The sharp-interface body force $\bm{F_s}$, derived according to the principle of thermodynamic consistency, is recovered in the limit $\epsilon \rightarrow 0$ by the diffuse-interface body force $\bm{F_D}$. Additionally, within the CSF phase-field model, we have:
\begin{align*}
	\nabla_s \bm{u} &= \bm{P} \;\nabla \bm{u},\\
	\nabla_s \cdot \bm{u} &= \bm{P} : \nabla \bm{u},\\
	\bm{P}&=\bm{I}-\bm{n}\otimes\bm{n},\\
	H=-\frac{\sqrt{2}}{\left(1-\phi^2\right)}f&=-\frac{\sqrt{2}}{\left(1-\phi^2\right)}\left(\epsilon\Delta\phi-\frac{1}{\epsilon}\left(\phi^2-1\right)\phi\right) ,
\end{align*}
where $\bm{P}$ denotes the surface projection operator, and the total curvature $H$ is smoothly defined in the diffuse interface formulation \cite{wang2005phase} with $f=\epsilon\Delta\phi-\frac{1}{\epsilon}\left(\phi^2-1\right)\phi$.\par
The diffuse interface representation of the Navier–Stokes equations and inextensibility constraint is given by:
\begin{align}
	&\rho\left(\partial_t \bm{u}+\bm{u}\cdot\nabla\bm{u}\right) -\nabla\cdot \bm{\mathcal{D}}+\nabla p =\bm{F_D}, &&\text{in} \; \Omega\times\left(0,T\right), \label{CSF-DI}\\
	&\nabla\cdot\bm{u} =0, &&\text{in} \;\Omega\times\left(0,T\right), \label{LBME-DI}\\
	&\xi_\sigma\overset{\nabla}{\bm{\sigma}}+\bm{\sigma}= \mu_p \bm{D}(\bm{u}),&&\text{in} \; \Omega\times\left(0,T\right),  \label{VS-DI}\\
	&\xi\,\epsilon^2\nabla \cdot\left(\phi^2\nabla\lambda\right)+|\nabla\phi|\,\bm{P}:\nabla\bm{u} = 0, && \text{in}\;\Omega\times\left(0,T\right),\label{eq:lambda}
\end{align}
where
\begin{align}
	&\overset{\nabla}{\bm{\sigma}}=\partial_t  \bm{\sigma}+\bm{u}\cdot\nabla\bm{\sigma}-\left(\nabla\bm{u} \, \bm{\sigma}+\bm{\sigma}\, \nabla^T\bm{u}\right),\nonumber\\
	&\bm{\mathcal{D}} =2\left(\frac{1+\phi}{2}\mu_{inner}+\frac{1-\phi}{2}\mu_{outer}\right) \bm{D}(\bm{u})+\frac{1-\phi}{2}\bm{\sigma},  \nonumber\\
	&\bm{F_D} =\delta_{\phi}\left(b_c \left(\frac{1}{2}H^3-2HK+\Delta_sH\right)\bm{n}+\frac{1}{2}H^2\nabla_sb_c+\bm{n}H\Delta_sb_c\right) \nonumber\\
	&\quad\quad\;+\delta_{\phi}\left(\nabla_s\lambda-\lambda H\bm{n}\right) \nonumber\\
	&\quad\quad\;+\delta_{\phi}\left(k_f\left(\nabla_sc\cdot \bm{L}\nabla_sc\right)\bm{n}-\frac{k_f}{2}|\nabla_sc|^2H\bm{n}-k_f(\nabla_sc)\Delta_sc\right). \label{F-DI-EX}
\end{align}
Here, \eref{eq:lambda} represents an equation for $\lambda$ that enforce inextensibility constraint  $\nabla_s\cdot\bm{u}=\bm{P}:\nabla\bm{u}=0$ \cite{ALAND201432}. Away from the interface, $\phi^2\approx1$ and $|\nabla\phi|\approx0$, so the equation becomes $\nabla^2\lambda=0$. Near the interface, $\phi^2\approx0$ and $|\nabla\phi|$ is nonzero, therefore $\bm{P}:\nabla\bm{u}=0$ is locally enforced. $\xi>0$ is a constant parameter. We assume inner fluid and outer fluid have the same density $\rho=\rho_{inner}=\rho_{outer}$. For cases involving fluids with differing densities, maintaining thermodynamic consistency in the diffuse interface model requires modifications to the current approach, as described in the literature (e.g., see \cite{Abels2012}).

\subsection{Phase-field evolution}
To capture the evolution of the membrane interface in the CSF phase-field model, we introduce an evolution equation for the phase-field variable $\phi$. This equation must effectively couple the fluid field with membrane forces, while also incorporating constraints to ensure global area and volume conservation. Most phase-field models for vesicle hydrodynamics are derived based on an energetic variational approach, originally proposed in \cite{Du2007,DU2009923}, which considered only global volume and surface area constraints. This framework was later extended by \cite{ALAND201432}, who incorporated a local inextensibility constraint into the phase-field formulation (referred to as Model B in \cite{ALAND201432}). Building on this model, \cite{VALIZADEH2022114191} developed a fully implicit, monolithic isogeometric analysis approach to study the hydrodynamics of homogeneous vesicles. Subsequently, \cite{WEN2024117390} extended the model to multicomponent vesicles using the energetic variational framework. Following this line of development, we consider a nonlinear advection--diffusion equation for the phase field, where the advection is driven by the fluid velocity field. To ensure consistent energy dissipation and to prevent the accumulation of numerical errors in global area and volume conservation, two regularization terms are introduced on the right-hand side of the equation. The phase-field equation without area and volume conservation is given by:
\begin{align} 
	&\partial_t \phi+\bm{u}\cdot\nabla\phi=-\gamma_c \tilde{g}, \quad &&\text{in}\quad \Omega\times\left(0,T\right),\label{eq:phi-N-AV}
\end{align}
where $\gamma_c$ is a small constant mobility parameter, $\tilde{g}$ is the regularization term to ensure consistent energy dissipation and is derived by variation of the diffuse form of the total membrane energy $\mathcal{E}$ with respect to $\phi$:
\begin{align}
	\mathcal{E}&=\int_{\Omega}\frac{1}{2}\frac{3}{4\sqrt{2}\epsilon}b_c\left(\epsilon\Delta\phi-\mathcal{Q}\right)^2 \text{d} v +\int_{\Omega}\mathcal{M}\delta_\phi \text{d} v, \label{E-DI}\\
	\mathcal{Q}&=\frac{1}{\epsilon}(\phi^2-1)\phi, \nonumber\\
	\mathcal{M} &= h_c+\frac{k_f}{2}|\nabla_s c|^2, \nonumber\\
	\tilde{g}&=\frac{\partial \mathcal{E}}{\partial \phi} \nonumber\\
	&= \Delta(\frac{3}{4\sqrt{2}}b_cf)-\frac{3}{4\sqrt{2}\epsilon}b_c f\frac{d \mathcal{Q}}{d \phi}+\frac{\partial \mathcal{M}}{\partial \phi}\delta_\phi+\mathcal{M}\delta_\phi'. \label{g-tide}
\end{align}
Note that in \eref{E-DI}, the surface tension energy is neglected, as local inextensibility of the fluid is imposed through an auxiliary equation in which the surface tension serves as a Lagrange multiplier. The diffuse interface form of energy $\mathcal{E}$ is equivalent to the sharp representation of energy $E$ as $\epsilon \rightarrow 0$ \cite{Dolow2013,ALAND201432,LOWENGRUB2016112,VALIZADEH2022114191,WEN2024117390}. The boundary and initial conditions are given as:
\begin{align}
	& \phi = \phi_b, \quad&&\text{on}\quad \partial\Omega\times\left(0,T\right),\nonumber\\
	& \lambda = 0, &&\text{on}\quad \partial\Omega\times\left(0,T\right),\nonumber\\
	& \phi(0) = \phi_0. &&\text{in}\quad \Omega,  \nonumber\\
	& \mathcal{A}= \mathcal{A}_0, &&\text{for}\quad \left(0,T\right), \label{A-LM}\\
	& \mathcal{V}= \mathcal{V}_0, &&\text{for}\quad \left(0,T\right), \label{V-LM}
\end{align}
where $\mathcal{A}$ and $\mathcal{V}$ are the area and volume of the multicomponent vesicle. We introduce Lagrange multipliers $\lambda_{\text{global}} = \lambda_{\text{global}}(t) \in \mathbb{R}$, which is a spatially constant multiplier enforcing total surface area conservation (\eref{A-LM}), and $\lambda_{\text{volume}} = \lambda_{\text{volume}}(t) \in \mathbb{R}$, which is a spatially constant multiplier enforcing total volume conservation (\eref{V-LM}). They serve as correction terms in \eref{CSF-DI} and \eref{eq:phi-N-AV}, and we obtain:
\begin{align}
	&\rho\left(\partial_t \bm{u}+\bm{u}\cdot\nabla\bm{u}\right) -\nabla\cdot \bm{\mathcal{D}}+\nabla p =\bm{F}, &&\text{in} \quad \Omega\times\left(0,T\right), \label{CSF-W-AV}\\
	&\partial_t \phi+\bm{u}\cdot\nabla\phi=-\gamma_c g,  &&\text{in}\quad \Omega\times\left(0,T\right),\label{eq:phi-W-AV}\\
	&\bm{F}=\bm{F_D} -\left(\lambda_{\text{global}}f+\lambda_{\text{volume}}\right)\nabla\phi, \nonumber\\
	&g=\tilde{g}-\lambda_{\text{global}}f-\lambda_{\text{volume}}. \nonumber
\end{align}
To obtain $\lambda_{\text{volume}}$ and $\lambda_{\text{global}}$, we impose the following constraints:
\begin{align}
	& \frac{d}{dt}\mathcal{V}[\phi] = \frac{d}{dt} \int_{\Omega}\,\frac{1}{2}\left(\phi+1\right)  ~\text{d} v = 0 \quad \text{(Volume Constraint)}\\
	& \frac{d}{dt}\mathcal{A}[\phi] = \frac{d}{dt} \int_{\Omega}\,\frac{3}{2\sqrt{2}}\left(\frac{\epsilon}{2}|\nabla\phi|^2+\frac{1}{4\epsilon}(\phi^2-1)^2\right)  ~\text{d} v = 0 \quad \text{(Area Constraint)}
\end{align}
Taking the time derivatives and using \eref{eq:phi-W-AV}, we obtain the following equations which should be solved for $\lambda_{\text{volume}}$ and $\lambda_{\text{global}}$
\begin{align}
	&\lambda_{\text{volume}} \int_{\Omega}~\text{d} v + \lambda_{\text{global}} \int_{\Omega}\, f ~\text{d} v= \int_{\Omega}\, \left(\frac{1}{\gamma}\,\bm{u}\cdot\nabla\phi+g\right) ~\text{d} v +\frac{1}{2\Delta t}\left(\mathcal{V}_0-\mathcal{V}\right) \\
	& \lambda_{\text{volume}} \int_{\Omega} f ~\text{d} v + \lambda_{\text{global}} \int_{\Omega}\, f^2 ~\text{d} v= \int_{\Omega}\, \left(\frac{1}{\gamma}\,\bm{u}\cdot\nabla\phi+g\right) f ~\text{d} v -\frac{1}{2\Delta t}\left(\mathcal{A}_0-\mathcal{A}\right)
\end{align}
where we have further added additional penalty terms to the right-hand side of these equations in order to improve accuracy and avoid error accumulations \cite{ALAND201432,VALIZADEH2022114191,DU2006757}. $\Delta t$ denotes the time step size.
\subsection{Surface concentration field}
The evolution of phase A and B on the membrane surface is modeled by an advective Cahn–Hilliard equation with variable $c \in [0,1]$, which is the surface concentration of phase B, while the concentration of phase A is defined as $1 - c$. This equation incorporates an advection term driven by fluid flow and a diffusion term determined by surface energy. The Eulerian formulation of the Cahn-Hilliard equation is stated as  \cite{Dolow2013,GREER2006216}:
\begin{align*}
	&\partial_t c+\bm{u}\cdot\nabla_s c =\nabla_s\cdot(M_c\nabla_s\tilde{\beta}), &&\text{in}\quad \Gamma\times\left(0,T\right),\\
	&\tilde{\beta} =\frac{\partial E}{\partial c}=\frac{1}{2}b_c'H^2+h_c'-k_f\Delta_s c, &&\text{in}\quad \Gamma\times\left(0,T\right),\\
	&c(0) = c_0, &&\text{in}\quad \Gamma,
\end{align*} 
where  $M_c$ is the mobility, $\tilde{\beta}$ is the chemical potential which is computed by the variational derivative of the free energy functional with respect to the surface concentration. However, while the Eulerian representation of surface PDEs is versatile, it leads to a severely degenerate diffusion equation within the embedding space due to the absence of diffusion perpendicular to the surface \cite{WEN2024117390}. To overcome this limitation, the diffuse interface Eulerian form of the Cahn--Hilliard equation, pioneered by \cite{Voigt05} and subsequently refined in various studies \cite{LOWENGRUB2016112,WEN2024117390}, has been developed. Based on these diffuse interface strategies, the Cahn–Hilliard equation is written as:
\begin{align}
	&\delta_t\partial_t c+\nabla \cdot\left(\bm{u}c\delta_c\right) =\nabla\cdot(M_c\delta_c\nabla\beta)+\alpha_P\nabla\cdot(\delta_c (\nabla c \cdot \bm{n})\bm{n}), &&\text{in}\quad \Omega\times\left(0,T\right),\label{eq:c-DI}\\
	&\beta =\frac{\partial E}{\partial c}\delta_{\phi}=\left(\frac{1}{2}b_c'H^2+h_c'-k_f\Delta_s c\right)\delta_{\phi}, &&\text{in}\quad \Omega\times\left(0,T\right), \nonumber\\
	&c(0) = c_0, &&\text{in}\quad \Omega, \nonumber
\end{align}
where the last term in \eref{eq:c-DI} is an augmented penalty term adopted in \cite{LOWENGRUB2016112,KLOPPE2024117090,VALIZADEH2019599,WEN2024117390}. This term is equivalent to imposing the constraint $\nabla c \cdot \bm{n} = 0$ within the neighborhood of the membrane interface $\Gamma$. As a result, the surface concentration $c$ (the concentration at the zero level set of the phase field) remains unaffected by concentrations at other level sets during diffusion, thereby avoiding data transfer between different level sets. The $\delta_c$ and $\delta_t$ are defined as:
\begin{align}
	&\delta_c  = \left(1-\phi^2\right)^2, \nonumber\\
	&\delta_t  = \max\{\delta_c,\epsilon_{\text{tol}}\}.\label{Delta-t}
\end{align}
Note in \eref{Delta-t}, we truncate $\delta_t$ with a tolerance $\epsilon_{\text{tol}}$ to ensure the problem is well-posed and to improve convergence. Note that we consider a penalty term in the vicinity of the interface, 
$ \alpha_P \nabla \cdot \big( \delta_c (\nabla c \cdot \bm{n}) \bm{n} \big)$, 
to enforce $\nabla c \cdot \bm{n} = 0$. 
This ensures that $c$ is extended off the interface $\Gamma$ as a constant along the normal direction.\par
The $b_c$ and $h_c$ are defined as:
\begin{align}
	b_c&= 0.5\,(b_B-b_A)\left(1+\tanh\left[3\,(c-0.5)\right]\right)+b_A, \label{eq:BR}\\
	h_c&=\frac{1}{2\Theta}\left(c \log c+(1-c)\log(1-c)\right)+c\,(1-c), \label{eq:LE}
\end{align}
where $b_A$ and $b_B$ are constant bending rigidity of phase A and phase B, respectively. \eref{eq:BR} defines an interpolation of surface bending rigidity in terms of bending rigidities of phase A and phase B. We define double well potential in \eref{eq:LE} to have a logarithmic form to ensure the concentration is stricted bounded $c \in [0,1]$. $\Theta$ is a constant and we assume $\Theta = 1.5$ to ensure the $h_c$ has the double-well potential property as studied in \cite{VALIZADEH2019599}.\par

\subsection{CSF phase-field model} \label{diffuse}
We non-dimensionalize the mathematical model following the approach in \cite{GERA2018362,WEN2024117390,SEOL20191009}. All fluid properties are normalized with respect to the outer fluid. $L_0$ is the characteristic length and $T_0$ is the characteristic time. $u_0=\frac{L_0}{T_0}$ is the characteristic velocity. $\beta_0$ is the characteristic chemical potential and $\tilde{M_c}$ is the characteristic mobility. $b_{max}$ is the bending rigidity of stiffer phase. $\alpha=\frac{b_{max}}{k_f}$ measures the strength of the bending forces to line tension. $\alpha_{\mu_n}=\frac{\mu_{inner}}{\mu_{outer}}$ is the viscosity contrast between the interior Newtonian fluid and the Newtonian solvent of outer Oldroyd-B fluid. $\alpha_{\mu_p}=\frac{\mu_p}{\mu_{outer}}$ is the viscosity contrast between the polymer and the Newtonian solvent in Oldroyd-B fluid.\par
We define Peclet number $Pe=\frac{L_0^2}{T_0\beta_0\tilde{M_c}}$ that measures the speed at which the surface phases evolve compared to the characteristic time. Cahn number $Cn^2=\frac{k_f}{\beta_0L_0^3}$ measures the intensity of line tension. Reynolds number $Re=\frac{\rho_0u_0L_0}{\mu_{outer}}$ measures the ratio of fluid inertial forces to the viscous forces. Bending capillary parameter $Ca=\frac{\mu_{outer}L_0^3}{b_{max}T_0}$ measures the viscous effects compared to the strength of the membrane bending. Weissenberg number  $\mathcal{W}i=\frac{\xi_\sigma}{T_0}$ measures the ratio of elastic force and viscous force in Oldroyd-B fluid. The final non-dimentionalized system of equations in the diffuse-interface framework is stated as:\\
Find $\bm{u}$, $p$, $\phi$, $f$, $\lambda$, $c$, and $\bm{\sigma}$ such that
\begin{align}
	&\partial_t \bm{u}+\bm{u}\cdot\nabla\bm{u} -\frac{1}{Re}\nabla\cdot \bm{\mathcal{D}}+\nabla p =\bm{F}, && \text{in} \; \Omega\times\left(0,T\right), \label{NS-MM-F}\\
	&\nabla\cdot\bm{u} =0, &&\text{in} \;\Omega\times\left(0,T\right), \label{NS-MS-F}\\
	&\partial_t \phi+\bm{u}\cdot\nabla\phi=-\gamma_c g,  &&\text{in}\; \Omega\times\left(0,T\right),\label{phi-F}\\
	&f=\epsilon \Delta \phi -\frac{1}{\epsilon}\left(\phi^2-1\right)\phi, && \text{in}\; \Omega\times\left(0,T\right), \label{f-F}\\
	&\xi\,\epsilon^2\nabla \cdot\left(\phi^2\nabla\lambda\right)+|\nabla\phi|\,\bm{P}:\nabla\bm{u} = 0,  &&\text{in}\;\Omega\times\left(0,T\right),\label{lambda-F}\\
	&\delta_t\partial_t c+\nabla \cdot\left(\bm{u}c\delta_c\right) =\frac{1}{Pe}\nabla\cdot(M_c\delta_c\nabla\beta)+\alpha_P\nabla\cdot(\delta_c (\nabla c \cdot \bm{n})\bm{n}), &&\text{in}\; \Omega\times\left(0,T\right),\label{c-F}\\
	&\mathcal{W}i\,\overset{\nabla}{\bm{\sigma}}+\bm{\sigma}= \alpha_{\mu_p} \bm{D}(\bm{u}),&&\text{in} \; \Omega\times\left(0,T\right),  \label{VS-F}
\end{align}
where we have:
\begin{align}
	&\overset{\nabla}{\bm{\sigma}}=\partial_t  \bm{\sigma}+\bm{u}\cdot\nabla\bm{\sigma}-\left(\nabla\bm{u}\,\bm{\sigma}+\bm{\sigma}\,\nabla^T\bm{u}\right),\nonumber\\
	&\bm{\mathcal{D}} =2\left(\frac{1+\phi}{2}\alpha_{\mu_n}+\frac{1-\phi}{2}\right) \bm{D}(\bm{u})+\frac{1-\phi}{2}\bm{\sigma},  \nonumber\\
	&\bm{F} =\frac{\delta_{\phi}}{ReCa}\left(b_c \left(\frac{1}{2}H^3-2HK+\Delta_sH\right)\bm{n}+\frac{1}{2}H^2\nabla_sb_c+\bm{n}H\Delta_sb_c\right) \nonumber\\
	&\quad\quad+\mathcal{P}_L\delta_{\phi}\left(\nabla_s\lambda-\lambda H\bm{n}\right) \nonumber\\
	&\quad\quad+\frac{\delta_{\phi}}{\alpha ReCa}\left(\left(\nabla_sc\cdot \bm{L}\nabla_sc\right)\bm{n}-\frac{1}{2}|\nabla_sc|^2H\bm{n}-(\nabla_sc)\Delta_sc\right)\nonumber\\
	&\quad\quad-\mathcal{P}_C\left(\lambda_{\text{global}}f+\lambda_{\text{volume}}\right)\nabla\phi,\label{F-F}\\
	&g=\frac{1}{ReCa}\left(\Delta(\frac{3}{4\sqrt{2}}b_cf)-\frac{3}{4\sqrt{2}\epsilon}b_c f\frac{\partial \mathcal{Q}}{\partial \phi}\right)+\frac{1}{2\alpha ReCa}\frac{\partial \mathcal{M}}{\partial \phi}\delta_\phi \nonumber\\
	&\quad+\frac{1}{\alpha ReCa}\left(\frac{1}{Cn^2}h_c+\frac{1}{2}|\nabla_s c|^2\right)\delta_\phi'-\lambda_{\text{global}}f-\lambda_{\text{volume}}, \nonumber\\
	&\beta =\left(\alpha \frac{Cn^2}{2}b_c'H^2+h_c'-Cn^2\Delta_s c\right)\delta_{\phi}, \nonumber\\
	&H=-\frac{\sqrt{2}}{\left(1-\phi^2\right)}f. \nonumber
\end{align}
Note that $\mathcal{P}_L$ and $\mathcal{P}_C$ are the coefficients controlling the strength of local inextensibility and the conservation of global area and volume, respectively, after non-dimensionalization. A large value of either coefficient enforces the corresponding constraint more strictly but can lead to numerical convergence difficulties. In practice, $\mathcal{P}_L = \mathcal{P}_C = 1$ is typically sufficient for moderately constrained cases, as considered in \cite{ALAND201432,VALIZADEH2022114191,WEN2024117390}. The boundary and initial condiitions are given:
\begin{align}
	& \bm{u} = \bm{u}_b, \quad&&\text{on}\quad \partial\Omega_{D}\times\left(0,T\right), \label{Dirichlet-u-f}\\
	& \phi = \phi_b, &&\text{on}\quad \partial\Omega_{D}\times\left(0,T\right), \label{Dirichlet-phi-f}\\
	&f=f_b, &&\text{on}\quad \partial\Omega_{D}\times\left(0,T\right), \label{Dirichlet-fb}\\
	& \lambda = 0, &&\text{on}\quad \partial\Omega_{D}\times\left(0,T\right), \label{Dirichlet-f}\\
	&\nabla \phi\cdot\bm{m}=0, &&\text{on}\quad \partial\Omega_{N}\times\left(0,T\right), \label{traction-phi}\\
	&\left(\mathcal{D}-p \, \bm{I}\right) \cdot \bm{m} =\bm{0}, &&\text{on}\quad \partial\Omega_{N}\times\left(0,T\right), \label{traction-f}\\
	&\nabla c\cdot\bm{m}=0, &&\text{on}\quad \partial\Omega_{N}\times\left(0,T\right), \label{traction-c}\\
	&\nabla \bm{\sigma}\cdot\bm{m}=0, &&\text{on}\quad \partial\Omega_{N}\times\left(0,T\right), \label{traction-beta}\\
	&\bm{u}(0) = \bm{u}_0, &&\text{in}\quad \Omega, \label{Initial-u0}\\
	& \phi(0) = \phi_0, &&\text{in}\quad \Omega, \label{Initial-phi-f}\\
	& c(0) = c_0, &&\text{in}\quad \Omega. \label{Initial-c}
\end{align}

\section{Numerical formulation} \label{Sec:NumericalFormulation}
In this section, we present the numerical formulation of our diffuse interface model. The spatial and temporal discretization methods are detailed in the following subsections.
\subsection{Weak form}
For the spatial discretization, we adopt isogeometric analysis. To avoid the need for satisfying the inf-sup condition arising from the coupling of multiple fields, we employ the stabilized residual-based variational multiscale (RBVMS) method \cite{bazilevs2013computational} for the Navier–Stokes equations augmented with viscoelastic effects. For the Oldroyd-B constitutive equation \eref{VS-F}, the Streamline-Upwind/Petrov-Galerkin (SUPG) stabilization method \cite{BROOKS1982199} is used to suppress numerical instabilities caused by advection-dominated behavior. The remaining equations are discretized using the standard Galerkin finite element method. The RBVMS method is a robust computational fluid dynamics (CFD) technique designed to handle a wide range of flow regimes governed by the Navier–Stokes equations. It is particularly effective for enabling equal-order interpolation of velocity and pressure fields, overcoming the limitations posed by the inf-sup condition. \par
As described in Section~\ref{diffuse}, our CSF phase-field model requires solving for the fields $\bm{u}, p, \phi, f, \lambda, c$, and $\bm{\sigma}$. In this subsection, we derive the RBVMS formulation for the Navier–Stokes equations and the SUPG formulation for the Oldroyd-B constitutive equation, respectively. Let us define $\mathcal{S}_u$, $\mathcal{S}_p$, and $\mathcal{S}_\sigma$ as the sets of trial functions for the velocity $\bm{u}$, pressure $p$, and dumbbell elastic stress $\bm{\sigma}$, respectively, as follows:
\begin{align*}
	&\mathcal{S}_u = \{\bm{u}|\bm{u}\in (H^{1}(\Omega))^d, \quad \bm{u} = \bm{u}_b \quad \text{on} \quad \partial\Omega_{D}\}, \\
	&\mathcal{S}_p = \{p|p \in L^{2}(\Omega), \quad \int_{\Omega} p ~\text{d}\Omega = 0 \quad \text{if} \quad \partial\Omega=\partial\Omega_{D}\},\\
	&\mathcal{S}_\sigma = \{\bm{\sigma}|\bm{\sigma}\in (H^{1}(\Omega))^{d\times d}\},
\end{align*}
where $L^{2}(\Omega)$ denotes the space of square-integrable scalar-valued functions over the domain $\Omega$, while $(H^{1}(\Omega))^d$ and $(H^{1}(\Omega))^{d\times d}$ represent the spaces of vector-valued and tensor-valued functions, respectively, whose components and their first derivatives are square-integrable on $\Omega$.\par
The sets of weighting functions $\bm{w}$, $q$, and $\bm{y}$, corresponding to $\bm{u}$, $p$, and $\bm{\sigma}$, are defined as:
\begin{align*}
	&\mathcal{V}_u = \{\bm{w}|\bm{w} \in (H^{1}(\Omega))^d, \quad \bm{w} = \bm{0} \quad \text{on} \quad \partial\Omega_{D}\}, \\
	& \mathcal{V}_p = \mathcal{S}_p,\\
	&\mathcal{V}_\sigma =  \mathcal{S}_\sigma.
\end{align*}  \par
The weak form of the Navier-Stokes equation is given by: 
\begin{align} \label{eq:NS-weak}
	\int_{\Omega} \bm{w} \cdot \left( \partial_t \bm{u}+ \bm{u}\cdot \nabla\bm{u} - \bm{F}\right) ~\text{d}v &+ \int_{\Omega} \bm{D}(\bm{w}):\bm{S}(\bm{u},p,\bm{\sigma}) ~\text{d}v \nonumber\\
	&- \int_{\partial\Omega_{N}} \bm{w} \cdot \bm{h}_N ~\text{d}a + \int_{\Omega} q \, \nabla \cdot \bm{u} ~\text{d}v = 0,
\end{align} 
where $\bm{h}_N$ is the prescribed traction vector on $\partial\Omega_{N}$. $\bm{F}$ is the additional force term in \eref{F-F}. $\bm{S}(\bm{u},p,\bm{\sigma})$ is given by:
\begin{align*}
	&\bm{S}(\bm{u},p,\bm{\sigma})=\frac{1}{\mathrm{Re}}\left[2\left(\frac{1+\phi}{2}\alpha_{\mu_n}+\frac{1-\phi}{2}\right) \bm{D}(\bm{u})+\frac{1-\phi}{2}\bm{\sigma}\right]-p\bm{I}.
\end{align*}\par
In the RBVMS and SUPG formulation, the weighting and trial function spaces are decomposed into sub-spaces containing coarse and fine scales via a multiscale direct-sum decomposition:
\begin{align*}
	&	\mathcal{S}_u = \mathcal{S}^{h}_{u} \oplus \mathcal{S}^{\prime}_{u}, \quad
	\mathcal{S}_p = \mathcal{S}^{h}_{p} \oplus \mathcal{S}^{\prime}_{p}, \quad
	\mathcal{S}_\sigma = \mathcal{S}^{h}_{\sigma} \oplus \mathcal{S}^{\prime}_{\sigma},\\
	&	\mathcal{V}_u = \mathcal{V}^{h}_{u} \oplus \mathcal{V}^{\prime}_{u}, \quad
	\mathcal{V}_p = \mathcal{V}^{h}_{p} \oplus \mathcal{V}^{\prime}_{p}, \quad
	\mathcal{V}_\sigma = \mathcal{V}^{h}_{\sigma} \oplus \mathcal{V}^{\prime}_{\sigma},
\end{align*}  
where the spaces with $h$ superscript denote coarse scale space while spaces with ``$\prime$'' superscript denote fine scale space. The members of $\mathcal{S}_u$, $\mathcal{S}_p$, $\mathcal{S}_\sigma$, $\mathcal{V}_u$, $\mathcal{V}_p $, and $\mathcal{V}_\sigma$ can be written as:
\begin{align*}
	&\bm{u} = \bm{u}^{h} + \bm{u}^{\prime}, \quad
	p = p^h + p^{\prime}, \quad  
	\bm{\sigma} = \bm{\sigma}^{h} + \bm{\sigma}^{\prime}, \\
	&\bm{w} = \bm{w}^{h} + \bm{w}^{\prime}, \quad
	q = q^h + q^{\prime}, \quad
	\bm{y} = \bm{y}^{h} + \bm{y}^{\prime}.
\end{align*}\par
Based on the methodology introduced by Bazilevs {\it et al.} \cite{Bazilevs2007173}, we opt for $\bm{w} = \bm{w}^h$, $q = q^h$, and $\bm{y} = \bm{y}^h$. The fine-scale of $\bm{u}$, $p$, and $\bm{\sigma}$ are defined as:
\begin{align*} 
	&\bm{u}^{\prime} := -\tau_M\,\bm{r}_M(\bm{u}^h,p^h,\bm{\sigma}^h), \quad
	p^{\prime} := -\tau_C \, r_\mathrm{C}(\bm{u}^h), \quad \bm{\sigma}^{\prime} := -\tau_\sigma\,\bm{r}_\sigma(\bm{u}^h,\bm{\sigma}^h) .
\end{align*} \par
The residuals $\bm{r}_M$, $r_C$, and $\bm{r}_{\sigma}$ correspond to the linear momentum and continuity equations of the Navier–Stokes system, and the Oldroyd-B constitutive equation, respectively. These residuals are defined as:
\begin{align*}
	& \bm{r}_{M}(\bm{u}^h,p^h,\bm{\sigma}) = \left(\partial_t \bm{u}^h + \bm{u}^h \cdot \nabla\bm{u}^h - \bm{F}^h\right) - \nabla\cdot\bm{S}(\bm{u}^h,p^h),\\
	& r_\mathrm{C}(\bm{u}^h)=\nabla\cdot\bm{u}^h,\\
	& \bm{r}_{\sigma}(\bm{u}^h,\bm{\sigma}^h) = \mathcal{W}i\left(\partial_t \bm{\sigma}^h + \bm{u}^h \cdot \nabla\bm{\sigma}^h-\left(\bm{\sigma}^h\cdot\nabla\bm{u}^h+\nabla^T\bm{u}^h\cdot\bm{\sigma}^h\right)\right)-\alpha_{\mu_p} \bm{D}(\bm{u}^h).
\end{align*}\par
The stabilization parameters $\tau_{M}$, $\tau_{C}$ and $\tau_{\sigma}$ are given by:
\begin{align*} 
	\tau_{M} &= \left(\frac{4}{\Delta t^2}+\bm{u}^h\cdot\mathbf{G}\bm{u}^h+C_I\,(\frac{\mu^*}{\mathrm{Re}})^2\,\mathbf{G}:\mathbf{G}\right)^{-1/2}, \\
	\tau_C &= \left(\mathsf{tr}(\mathbf{G})\text{ }\tau_M\right)^{-1} ,\\
	\tau_{\sigma} &= \left(\frac{4}{\Delta t^2}+\mathcal{W}i^2\,\bm{u}^h\cdot\mathbf{G}\bm{u}^h\right)^{-1/2} ,
\end{align*}
where  $\mu^*=\frac{1+\phi}{2}\alpha_{\mu_n}+\frac{1-\phi}{2}$ and $\mathbf{G}$ is the element metric tensor which is defined as:
\begin{align*}
	\mathbf{G}=\left[\mathrm{G}_{ij}\right]=\sum_{k=1}^{\text{d}}\frac{\partial\xi_k}{\partial x_i}\frac{\partial\xi_k}{\partial x_j} 
\end{align*}
Here, $\bm{\xi}$ and $\bm{x}$ denote the parametric and physical coordinates, respectively. The mapping function $\bm{x}(\bm{\xi})$ transforms the parametric domain to the physical domain. The constant $C_I$ is a non-dimensional positive parameter used in the stabilization terms. In our work, we choose $C_I = 6 \, \mathfrak{p}^4$, where $\mathfrak{p}$ is the order of the basis functions.\par
We employ standard Galerkin finite element methods for the remaining equations and define the trial function space as follows:
\begin{align*}
	&\mathcal{S}^h_\phi = \{\phi|\phi \in H^{2}(\Omega), \quad \phi = \phi_b \quad \text{on} \quad \partial\Omega_{\phi}\} \\
	&\mathcal{S}^h_f = \{f|f \in H^{2}(\Omega), \quad f = f_b \quad \text{on} \quad \partial\Omega_{f}\} \\
	&\mathcal{S}^h_\lambda = \{\lambda|\lambda \in H^{1}(\Omega), \quad \lambda = 0 \;\,\quad \text{on} \quad \partial\Omega_{\lambda}\}\\
	&\mathcal{S}^h_c = \{c|c \in H^{2}(\Omega)\}
\end{align*}
while the weighting functions are denoted as $\mathcal{V}^h_\phi$, $\mathcal{V}^h_f$, $\mathcal{V}^h_\lambda$, and $\mathcal{V}^h_c$ with  homogeneous essential boundary conditions.\par 
The final weak form of the problem is stated as:  find $\bm{u}^h \in \mathcal{S}^{h}_{u}$, $p^h \in \mathcal{S}^{h}_{p}$, $\phi^h\in \mathcal{S}^h_\phi$, $f^h\in \mathcal{S}^h_f$, $\lambda^h\in \mathcal{S}^h_\lambda$, $c^h\in \mathcal{S}^h_c$, and $\bm{\sigma}^h\in \mathcal{S}^h_\sigma$ such that $\forall$ $\bm{w}^h \in \mathcal{V}^{h}_{u}$, $q^h \in \mathcal{V}^{h}_{p}$, $r^h\in \mathcal{V}^h_\phi$, $l^h\in \mathcal{V}^h_f$, $s^h\in \mathcal{V}^h_\lambda$,  $z^h \in \mathcal{V}^{h}_{c}$, and  $\bm{y}^h \in \mathcal{V}^{h}_{\sigma}$:
\begin{align} \label{eq:RBVMS1}
	B(\{\bm{w}^h,q^h,r^h,l^h,s^h,z^h,\bm{y}^h\},\{\bm{u}^h,p^h,\phi^h,f^h,\lambda^h,c^h,\bm{\sigma}^h\})-L(\{\bm{w}^h,r^h\})=0,
\end{align} 
where the explicit formulation of the final weak form is provided in \ref{App:A}.\par
The conservation of surface phase mass is not preserved in \eref{eq:RBVMS1} due to numerical diffusion and the accumulation of discretization errors. Even when the equation is expressed in a conservative form, accurate mass conservation is not guaranteed over time, as numerical errors may still accumulate (see \cite{KLOPPE2024117090,aland2023phasefieldmodelactivecontractile}). To compensate for these errors, we introduce a correction method originally proposed by \cite{XU2006590} and later extended to the diffuse interface framework in \cite{WEN2024117390}, which corrects the surface concentration after each time step, thereby ensuring mass conservation of the membrane. The corrected concentration after each time step, $c_{n+1} $, is obtained as:
\[c_{n+1} = c_n\frac{\int_{\Omega}c_0 \delta_{\phi} \; d\Omega}{\int_{\Omega}c_h \delta_{\phi} \; d\Omega}\]
where $c_h$ is the concentration computed by solving the residual in \eref{eq:RBVMS1} at each time step $t_n$ and $c_0$ is the initial surface concentration at $t_0$.

%

\subsection{Space and time discretization}
In this work, we utilize Isogeometric Analysis (IGA) with quadratic NURBS basis functions ($\mathfrak{p}=2$) to discretize the function spaces, taking advantage of optimal regularity over the entire domain for all fields. This fulfills the requirement of $C^1$ continuity in space for the equations shown in \eref{eq:RBVMS3}, particularly for the Cahn–Hilliard equation, which is a $4^{th}$-order PDE, as well as for the terms involving $\bm{F}$ that also require $C^1$ continuity. For time integration, we adopt the generalized-$\alpha$ method \cite{Chung1993371,Jansen2000305}, implemented in an implicit manner. As shown in \cite{Jansen2000305}, this method achieves second-order accuracy when the spectral radius at infinity is set to $\rho_\infty = 0.5$, a value consistently used in all our simulations. The nonlinear system of equations is solved using a fully monolithic scheme based on the Newton–Raphson method, employing a two-phase predictor–multicorrector algorithm. Detailed information on this algorithm can be found in \cite{Hughes2009}.

\section{Results}  \label{NumericalExamples}
In this section, we first present numerical examples of two-dimensional homogeneous vesicle hydrodynamics under shear flow, with and without viscoelastic effects, to validate our model and examine the influence of Newtonian viscosity contrast and viscoelasticity. We then extensively study multicomponent vesicles in shear flow, with and without viscoelastic effects, to highlight the impact of membrane compositional heterogeneity and to investigate the role of viscoelasticity in multicomponent vesicle hydrodynamics. Finally, we explore the viscoelastic effect on multicomponent vesicles in two-dimensional Poiseuille flow. The inner fluid is always modeled as a Newtonian fluid, while the outer fluid can be either a Newtonian or an Oldroyd-B viscoelastic fluid. These two-phase fluid configurations are denoted as N/N and N/O, respectively. The effect of viscoelasticity on vesicle hydrodynamics is controlled by the Weissenberg number $\mathcal{W}i$, as described in Section~\ref{diffuse}. When $\mathcal{W}i = 0$, the viscoelastic effects vanish, and the fluid configuration effectively reduces from N/O to N/N. The model is implemented in PetIGA \cite{PetIGA}, an open-source, high-performance software framework for NURBS-based isogeometric analysis of partial differential equations. PetIGA is built on top of the PETSc scientific computing library \cite{petsc-web-page,petsc-user-ref}. Following our previous works \cite{WEN2024117390,VALIZADEH2022114191}, Dirichlet boundary conditions for the fluid velocity field is imposed weakly.\par
In all cases, we initialize the 2D vesicle as an ellipse defined by:
\[(\frac{x}{a_E})^2+(\frac{y}{b_E})^2=1\]
where $b_E$ is the semi-major axis in $y$ direction with $b_E  > a_E$. Phase field variable is initialized by:
\[\phi_0(\bm{x}) =\frac{ \tanh(-d(\bm{x}))}{\sqrt{2}\epsilon}\]
where $d$ is the signed distance to ellipse in 2D. The reduced area $R_v\in\left(0,1\right]$ is defined as the ratio between the vesicle area and the area of a circle with the same perimeter as the vesicle. It is written as 
\[R_v=\frac{\mathcal{V}_0}{\pi}\left(\frac{2\pi}{\mathcal{A}_0}\right)^{2}\]
where in this case $\mathcal{A}_0$ and $\mathcal{V}_0$ denote the perimeter and area of the vesicle, respectively. Accordingly, we fix the enclosed area $\mathcal{V}_0 = \pi$ and vary the reduced area $R_v$ to initialize different vesicle shapes. Unless otherwise stated, we initialize the concentrations of phases A and B on vesicles as pre-segregated, with the concentration of phase B explicitly prescribed as:
\[c_0=\frac{2+\tanh(4(c_{1}y-c_{2}))-\tanh(4(c_{1}y+c_{2}))}{2}\]
where the interface of phase A and phase B is located at $y=\frac{c_{2}}{c_{1}}$ and $y=-\frac{c_{2}}{c_{1}}$ along $y$-axis. Spatial discretization is performed in $\Omega\in [-3,3]^2$ using uniform, $\mathcal{C}^1$-continuous quadratic NURBS elements. We assume $\gamma = 0.1$, $\xi = 1$, and $\epsilon = 2h$, where $h$ denotes the element size. The non-dimensionalized parameters are set to $Re = 1 \times 10^{-3}$, $Pe = 1$, $Cn^2 = 0.01$, $\alpha = 10$, $\alpha_{\mu_p} = 1$, and the time step size is $\Delta t = 2.5 \times 10^{-4}$, unless otherwise specified. For simplicity, the densities of the inner and outer fluids are assumed to be equal, and the surface mobility is taken as a constant, $M_c = 1$. \par
\subsection{Homogeneous vesicle dynamics in shear flow} 
We begin with the most common case: a homogeneous vesicle in shear flow. The vesicle exhibits either tank-treading or tumbling motion, depending on the membrane geometry, its physical properties, and the surrounding fluid conditions. In tank-treading motion, the homogeneous vesicle reaches a steady shape with a fixed inclination angle, while the surface lipids tread relative to the fluid flow direction. In tumbling motion, the vesicle rotates periodically, and the surface lipids simultaneously tread along the membrane \cite{Laadhari2012,VALIZADEH2022114191}. These behaviors are examined in this section under both matched ($\alpha_{\mu_n} = 1$) and unmatched ($\alpha_{\mu_n} \neq 1$) Newtonian viscosity conditions, within both N/N and N/O fluid environments. \par
\begin{figure}[h]
	\centering
	\includegraphics [scale=0.45]{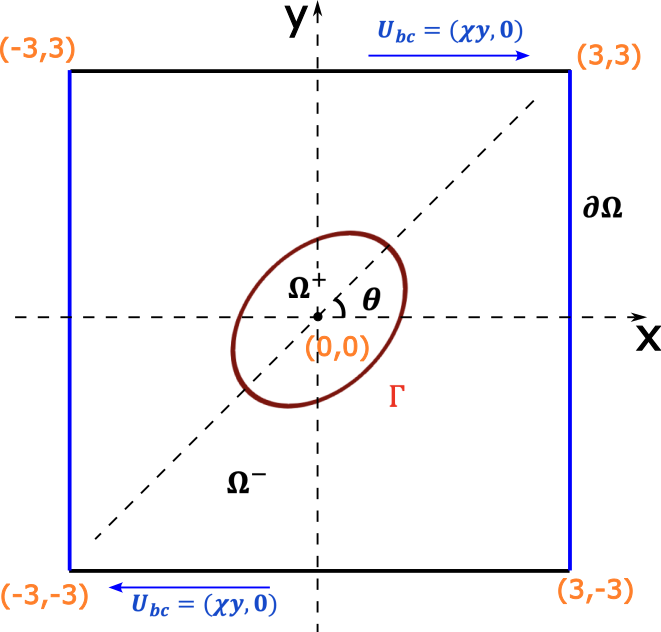}      
	\caption{Initialization of homogeneous vesicle in shear flow.}
	\label{fig:H2DshearflowInitial}	 		
\end{figure}
We set up a homogeneous vesicle in shear flow within a two-dimensional computational domain defined as $\Omega\in [-3,3]^2$, as shown in \fref{fig:H2DshearflowInitial}, which is discretized using $128 \times 128$ elements. The semi-major and semi-minor axes of the vesicle are $b_E = 1.2045$ and $a_E = 0.8302$, respectively, corresponding to a reduced volume $R_v = 0.95$. The vesicle is centered at $(0, 0)$. $\theta \in \left[-\frac{\pi}{2}, \frac{\pi}{2}\right]$ is the inclination angle, which denotes the angle between the major axis of the vesicle and the $x$-axis. The initial inclination angle of the vesicle is $\theta_0 = \frac{\pi}{4}$. As discussed in \cite{SEOL20191009}, $\theta_0$ is irrelevant to vesicle hydrodynamics once a quasi-steady state is reached. The externally applied shear flow is imposed by prescribing the velocity $\bm{U}_{bc} = (\chi y, 0)$ on the upper and lower boundaries (depicted in black in \fref{fig:H2DshearflowInitial}), where the shear rate is set to $\chi = 1$. Periodic boundary conditions are applied on the left and right boundaries (shown in blue). Additionally, on the upper and lower boundaries, we impose a Dirichlet boundary condition $\lambda = 0$, along with homogeneous Neumann boundary conditions for $\phi$, $f$, and $\bm{\sigma}$.\par
\subsubsection{Effect of Newtonian viscosity contrast between inner and outer fluids} \label{H2DNN}
\begin{figure}[h]
	\centering
	\begin{subfigure}{0.45\textwidth}
		\centering
		\includegraphics[width=\linewidth,trim={0cm 0cm 0cm 0cm},clip]{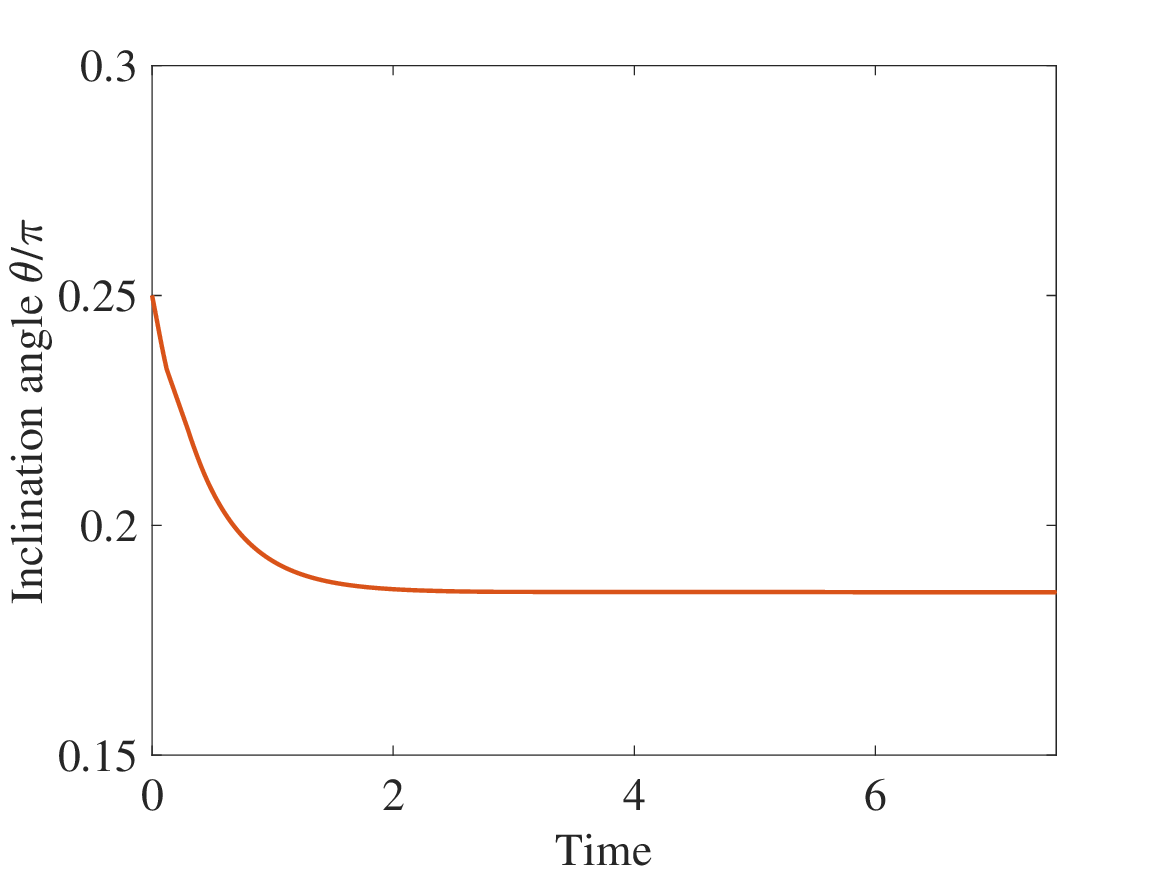}
		\caption{Tank-treading motion in N/N fluid with $\alpha_{\mu_n} = 1$}
	\end{subfigure}\quad
	\begin{subfigure}{0.45\textwidth}
		\centering
		\includegraphics[width=\linewidth,trim={0cm 0cm 0cm 0cm},clip]{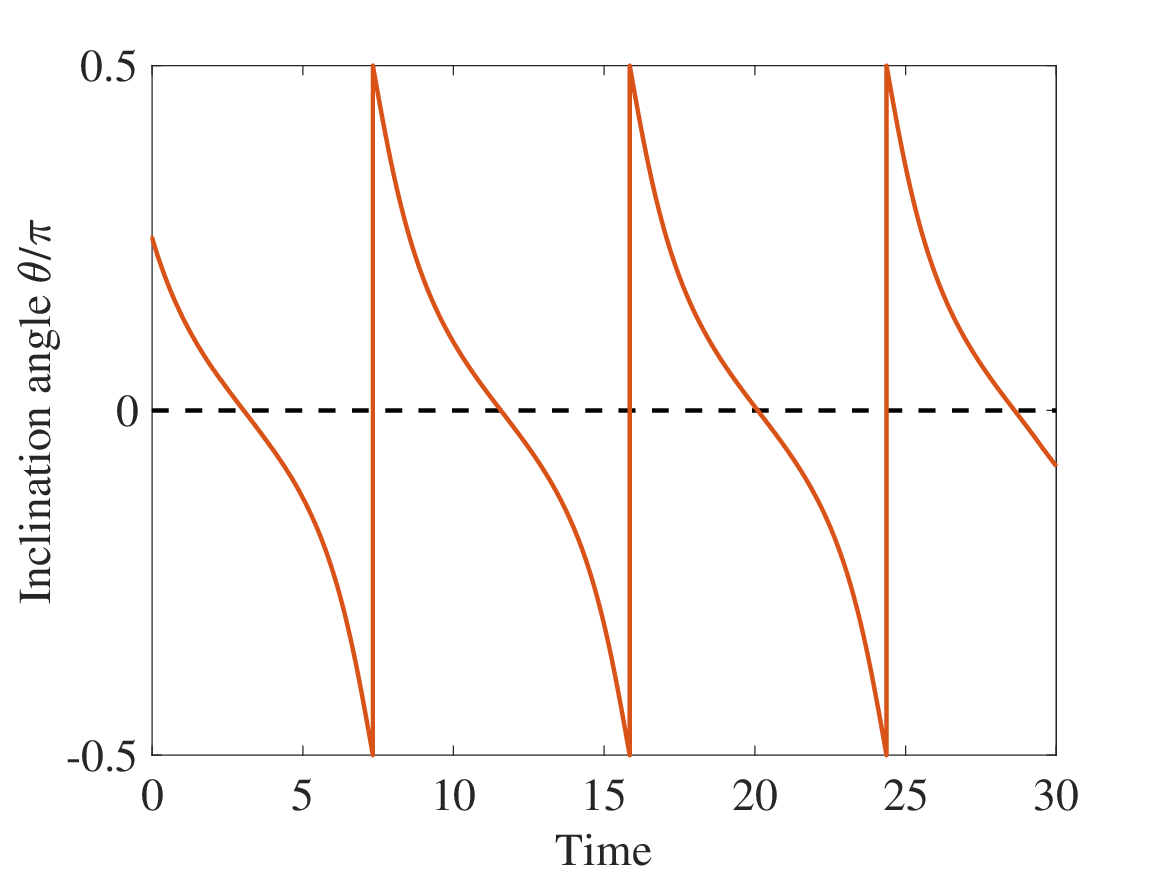}
		\caption{Tumbling motion in N/N fluid with $\alpha_{\mu_n} = 19$}
	\end{subfigure}
	\caption{Homogeneous vesicles in N/N shear flow. Evolution of the inclination angle over time for two cases: matched Newtonian viscosity ($\alpha_{\mu_n} = 1$) and high viscosity contrast ($\alpha_{\mu_n} = 19$).}	
	\label{fig:shearFlowIAH2D}	 		
\end{figure}

We first simulate two numerical examples in an N/N fluid domain with different values of $\alpha_{\mu_n}$ to illustrate the influence of Newtonian viscosity contrast. The unmatched viscosity case ($\alpha_{\mu_n}\neq1$) will later be compared with N/O scenarios to build an understanding of the viscoelastic effect on homogeneous vesicles, corresponding to the studies presented in \cite{SEOL20191009}. We set the Capillary number to $Ca = 1$ and the Weissenberg number to $\mathcal{W}i = 0$ for the two N/N cases: one with matched viscosity conditions ($\alpha_{\mu_n} = 1$), and the other with unmatched viscosity conditions ($\alpha_{\mu_n} = 19$), where the inner Newtonian viscosity is significantly higher than the outer Newtonian viscosity. The evolution of the inclination angle for the two cases is shown in \fref{fig:shearFlowIAH2D}, which serves as the most representative parameter for characterizing the hydrodynamics of a homogeneous vesicle.
 \par
We observe that a homogeneous vesicle with $R_v = 0.95$ and $\alpha_{\mu_n} = 1$ in an N/N shear flow exhibits tank-treading motion, as shown in \fref{fig:shearFlowIAH2D}.(a). The vesicle rotates rapidly and reaches a stationary inclination angle within time $t < 2$. Increasing the viscosity contrast to $\alpha_{\mu_n} = 19$ causes the vesicle to transition into a periodic tumbling motion, as illustrated in \fref{fig:shearFlowIAH2D}.(b). These results are qualitatively consistent with the numerical observations reported in Fig. 12 of \cite{SEOL20191009}.
\subsubsection{Viscoelastic effect on the hydrodynamics of homogeneous vesicles}\label{H2DNO}
In this subsection, we investigate the viscoelastic effect on the tumbling motion of homogeneous vesicles. The parameters are identical to those in the case presented in Section~\ref{H2DNN} and shown in \fref{fig:shearFlowIAH2D}.(b), except that the fluid domain is now N/O with $\mathcal{W}i = 1$. This case is particularly worth investigating, as Figs. 15 and 16 of \cite{SEOL20191009} report that a vesicle transitions from a tumbling motion to a tank-treading state with a negative inclination angle when viscoelastic effects are introduced. This transition occurs because the elastic stress counteracts the viscosity contrast that would otherwise induce tumbling. We aim to reproduce this case to validate the accuracy and robustness of our viscoelastic vesicle model. \par
\begin{figure}[h]
	\centering
	\includegraphics [scale=0.45]{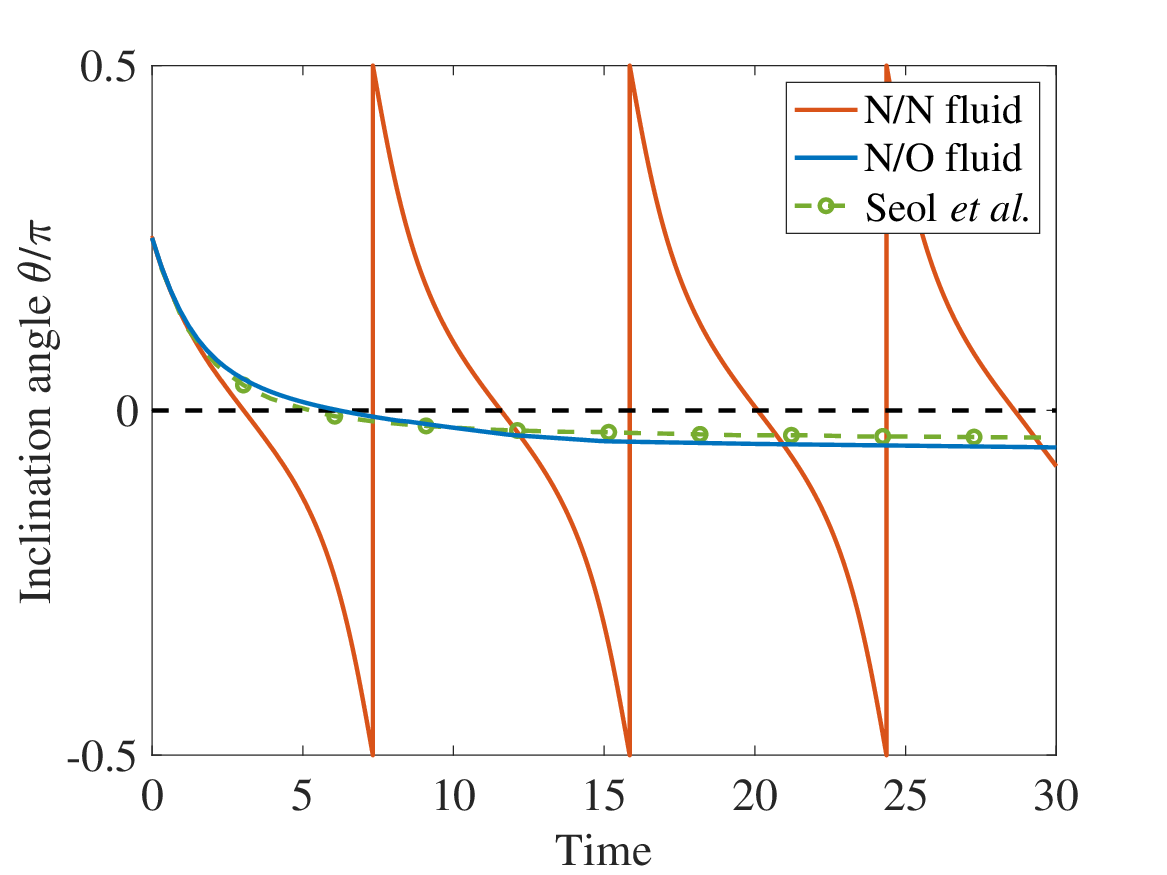}
	\caption{Unmatched homogeneous vesicle in shear flow. Evolution of the inclination angle over time for three cases: N/N fluid, N/O fluid, and the N/O case reported in \cite{SEOL20191009}, Fig. 16.}
	\label{fig:H2DshearflowValidP}	 		
\end{figure}
\begin{figure}[H]
	\centering
	\includegraphics [scale=0.08]{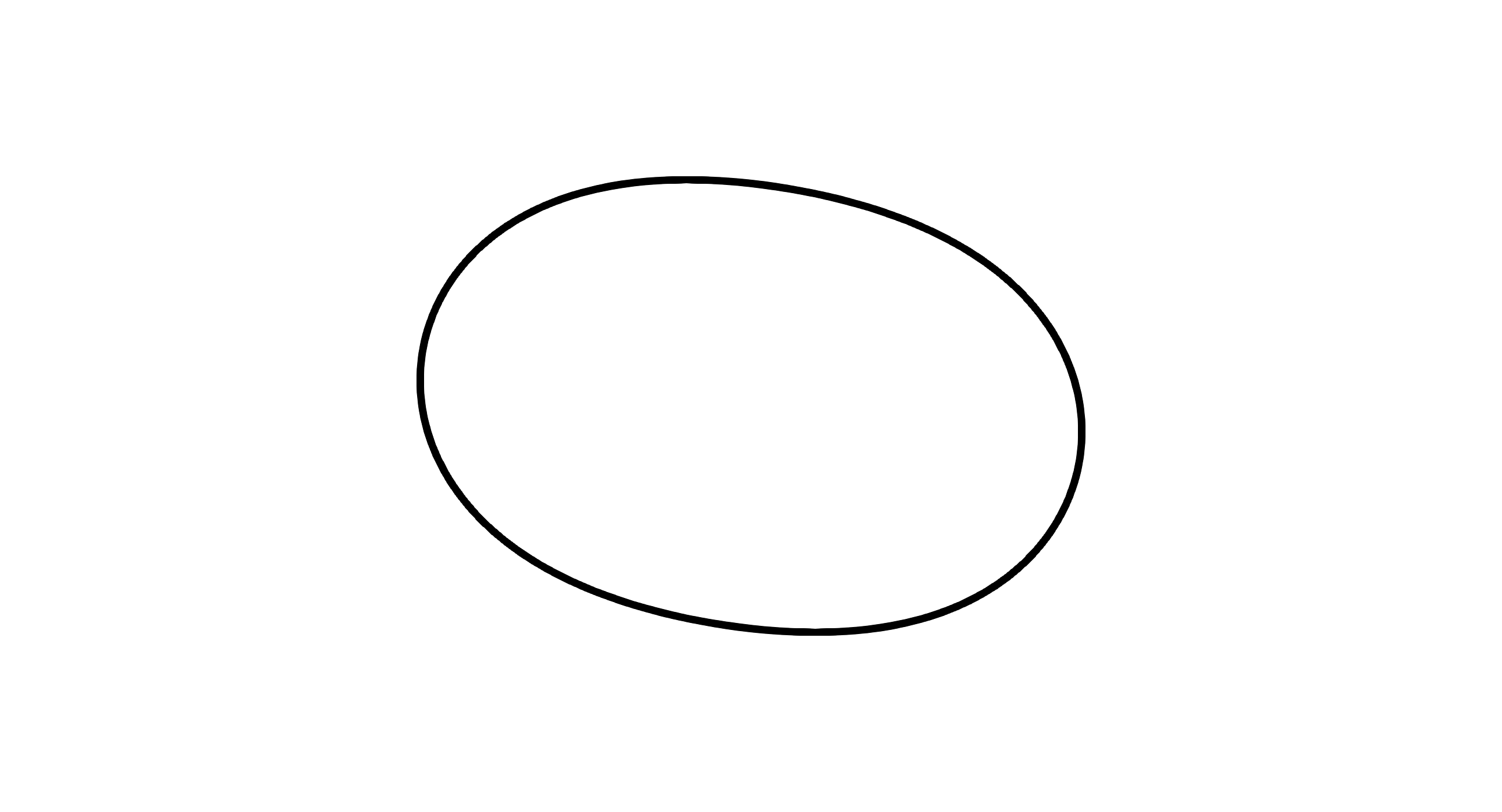}
	\caption{Negative inclination angle tank-treading motion at stationary state: $t = 15$.}
	\label{fig:H2DshearflowValid}	 		
\end{figure}

The evolution of the inclination angle for the unmatched viscosity homogeneous vesicle in both N/N and N/O fluid environments is presented in \fref{fig:H2DshearflowValidP}. For comparison, we also include the corresponding numerical result from Fig. 16 of \cite{SEOL20191009}, shown as a green dashed line. As expected, the tumbling motion observed in the N/N case (red solid line) transitions into a negative inclination angle tank-treading motion in the N/O case (blue solid line) due to the viscoelastic effects. Moreover, our result shows excellent agreement with the N/O case reported in \cite{SEOL20191009}. The stationary state of the homogeneous vesicle at $t = 15$, displayed in \fref{fig:H2DshearflowValid}, clearly exhibits negative inclination angle tank-treading behavior. These observations confirm the validity of our vesicle hydrodynamics model incorporating viscoelastic effects.
\subsubsection{Parametric study}
We study the sensitivity of the numerical solution by varying discretization parameters, specifically the element size $h$ and the time step size $\Delta t$. The parameters used for this parametric study correspond to the negative inclination angle tank-treading case discussed in Section~\ref{H2DNO}. We first examine the effect of spatial discretization by considering different mesh resolutions: $128 \times 128$, $144 \times 144$, $176 \times 176$, and $192 \times 192$, while keeping the time step size fixed at $\Delta t = 2.5 \times 10^{-4}$. Next, we assess the impact of time step size by using $\Delta t = 1 \times 10^{-4}$, $2.5\times 10^{-4}$, $4 \times 10^{-4}$, and $5 \times 10^{-4}$ with a fixed $128 \times 128$ mesh. For each case, we report the area error defined as:
\begin{align*}
	& A_{erorr}=\frac{|A-A_0|}{A_0}
\end{align*}
where $A_0$ is the prescribed area of the vesicle, and the inclination angle error is defined as:
\begin{align*}
	& \theta_{erorr}=\frac{|\theta-\theta_{ref}|}{|\theta_{ref}|}
\end{align*}
where $\theta_{ref}$ denotes the reference inclination angle obtained using the finest mesh resolution or the smallest time step size, respectively. The data are collected at $t = 1$. Errors in perimeter and mass conservation are negligible and thus not considered here. The results presented in Table~\ref{table:1} demonstrate that the relative error in area and inclination angle decreases as the discretization is refined, indicating good convergence behavior of the method.\\
\begin{table}[H]
	\centering
	\begin{tabular}{m{2 cm} m{1.5cm} m{1.5 cm} m{1.5 cm} m{2.5cm} m{1.5 cm} m{1.5cm}} 
		\hline
		Elements &h & $A_{error}$ & $\theta_{error}$ & $\Delta t \, (128\times128)$ & $A_{error}$ & $\theta_{error}$ \\ 
		\hline
		$128\times128$ &0.0469 & 0.323\% & 0.182\% & $5.0\times10^{-4}$ & 0.644\% & 0.194\% \\ 
		$144\times144$ &0.0417 & 0.285\% & 0.084\% & $4.0\times10^{-4}$  & 0.516\% & 0.138\% \\
		$176\times176$ &0.0341 & 0.231\% & 0.011\%  & $2.5\times10^{-4}$ & 0.323\% & 0.074\%\\ 
		$192\times192$ &0.0313 & 0.211\% & \quad -  & $1.0\times10^{-4}$ & 0.129\% & \quad -\\ 
		\hline
	\end{tabular}
	\caption{Convergence history of area and inclination angle error with respect to spatial and temporal resolution.}
	\label{table:1}
\end{table}
Overall, our model exhibits quantitative agreement with the numerical results presented in Seol et al. \cite{SEOL20191009}.

\subsection{Multicomponent vesicle dynamics in shear flow} \label{M2DSF}
In this section, we focus on the study of multicomponent vesicle hydrodynamics under various conditions, and ultimately investigate the influence of viscoelastic effects on different multicomponent vesicle behaviors. In numerical studies of multicomponent vesicle hydrodynamics \cite{Barrett2017, GERA2018362, Bachini_Krause_Nitschke_Voigt_2023, venkatesh2024shapedynamicsnearlyspherical,Gera_2022, WEN2024117390}, vesicles exhibit complex hydrodynamics and phases evolution on the membrane surface. In addition to the tank-treading and tumbling motions, a transitional regime known as swinging motion is also observed. In the swinging motion, the membrane phases undergo continuous phase treading, which induces periodic fluctuations in the inclination angle within the range $[0, \frac{\pi}{2}]$. These oscillations are driven by bending rigidity variations resulting from the migrating membrane phases \cite{Gera_2022, WEN2024117390}. As the influence of bending rigidity variation increases, the multicomponent vesicle transitions from swinging to tumbling motion. Notably, such behaviors—both swinging and tumbling—can occur even in the absence of viscosity contrast, as demonstrated in Section~\ref{H2DNN}, driven purely by the phase-induced bending rigidity variation. We will first demonstrate these observations with numerical examples. Subsequently, based on the observed hydrodynamic modes of multicomponent vesicles, we investigate the viscoelastic effects on their behavior—an aspect that, to the best of our knowledge, has not been previously studied.\par
\begin{figure}[h]
	\centering
	\includegraphics [scale=0.45]{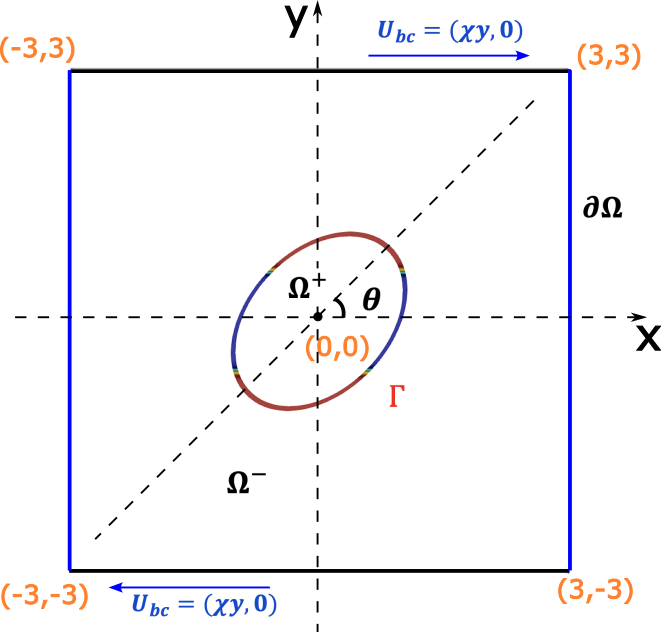}      
	\caption{Initialization of multicomponent vesicle in shear flow.}
	\label{fig:M2DshearflowInitial}	 		
\end{figure}

To better illustrate the effects of multicomponent composition on the membrane surface, we set $\alpha = 10$, such that bending rigidity and shear stress dominate the dynamics of phase treading and vesicle hydrodynamics, while the contribution from line tension is comparatively minor. The initial configuration is shown in \fref{fig:M2DshearflowInitial}, which matches that of the homogeneous vesicle cases, except with phase separation initialized at $y = \pm \frac{3}{5}$. Consequently, all multicomponent vesicle cases in shear flow exhibit two stiff domains (depicted in blue in \fref{fig:M2DshearflowInitial}) and two soft domains (shown in red), which we denote by $M_{\text{phases}} = 2$. In this configuration, the spatial distribution of material properties interacts strongly with the elongational deformation induced by the shear flow \cite{Gera_2022, WEN2024117390}.
\subsubsection{Effect of multicomponent composition on membrane surface dynamics} \label{M2DNN}
In this section, we investigate multicomponent vesicles in N/N fluid shear flow to validate the statement that swinging and tumbling motions are primarily driven by variations in bending rigidity. The membrane surface undergoes phase treading, which induces a periodic variation in bending rigidity at any fixed point on the surface. When the stiff phase (blue) reaches the high-curvature regions at the ends of the vesicle, it significantly increases the local bending energy and, consequently, the total membrane energy. As a result, the soft phase (red) energetically favoring the high-curvature regions to minimize the overall bending energy. Under the continuous shear stress applied in shear flow, the phases are forced to tread along the membrane surface. The periodic migration of the phases leads to periodic hydrodynamic patterns that manifest as swinging and tumbling motions. As indicated in \cite{Gera_2022}, the swinging and tumbling motions can be governed by the parameter $\mathcal{C} = \frac{|b_A - b_B|}{Ca}$, which depends on the bending rigidity amplitude and the Capillary number. To investigate the influence of multicomponent effects on the membrane surface, we fix $b_A = 1$ and vary both $b_B$ and $Ca$ to generate a series of numerical examples.\par
\begin{figure}[h]
	\begin{subfigure}{0.135\textwidth}
		\centering
		\includegraphics[width=\linewidth,trim={2.6cm 0cm 2.6cm 0cm},clip]{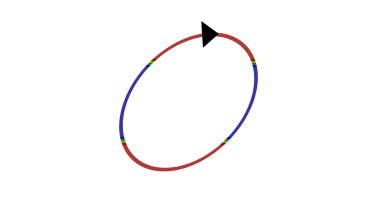}
		\caption*{$t=0$}
	\end{subfigure}
	\begin{subfigure}{0.135\textwidth}
		\centering
		\includegraphics[width=\linewidth,trim={2.6cm 0cm 2.6cm 0cm},clip]{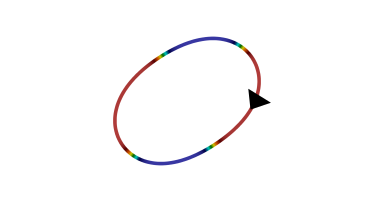}
		\caption*{$t=5$}
	\end{subfigure}
	\begin{subfigure}{0.135\textwidth}
		\centering
		\includegraphics[width=\linewidth,trim={2.6cm 0cm 2.6cm 0cm},clip]{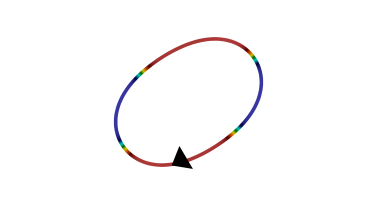}
		\caption*{$t=10$}
	\end{subfigure}
	\begin{subfigure}{0.135\textwidth}
		\centering
		\includegraphics[width=\linewidth,trim={2.6cm 0cm 2.6cm 0cm},clip]{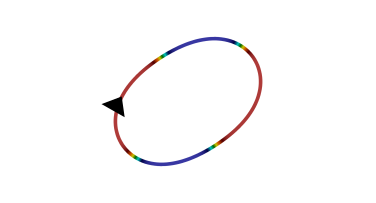}
		\caption*{$t=15$}
	\end{subfigure}
	\begin{subfigure}{0.135\textwidth}
		\centering
		\includegraphics[width=\linewidth,trim={2.6cm 0cm 2.6cm 0cm},clip]{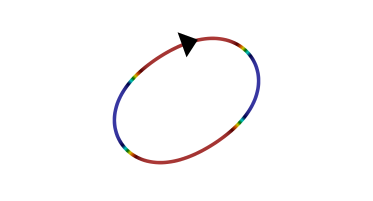}
		\caption*{$t=20$}
	\end{subfigure}
	\begin{subfigure}{0.135\textwidth}
		\centering
		\includegraphics[width=\linewidth,trim={2.6cm 0cm 2.6cm 0cm},clip]{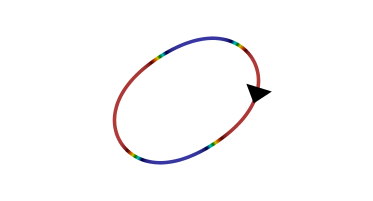}
		\caption*{$t=25$}
	\end{subfigure}
	\begin{subfigure}{0.135\textwidth}
		\centering
		\includegraphics[width=\linewidth,trim={2.6cm 0cm 2.6cm 0cm},clip]{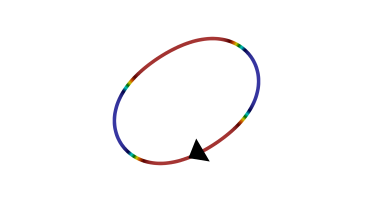}
		\caption*{$t=30$}
	\end{subfigure}
	\caption{Tank-treading motion of a multicomponent vesicle under shear flow for $Ca=4$ and $ b_B = 1$.}	
	\label{fig:M2Dbc1TT}	 		
\end{figure}

We use the same parameters as the tank-treading homogeneous vesicle in N/N fluid with matched viscosity condition ($\alpha_{\mu_n}=1$), as presented in Section~\ref{H2DNN}. For the first case, we set $Ca = 4$, $ b_B = 1$, and $\Delta t=5\times10^{-4}$, representing a multicomponent vesicle in N/N shear flow without bending rigidity variation. In this scenario, vesicle hydrodynamics are governed solely by line tension, surface tension, and stress from the surrounding fluid in \eref{NS-MM-F} and \eref{F-F}. Accordingly, \fref{fig:M2Dbc1TT} presents the resulting tank-treading motion of the multicomponent vesicle, during which the membrane exhibits constant-velocity phase treading along its surface.
 \par
\begin{figure}[h]
	\centering
	\includegraphics [scale=0.45]{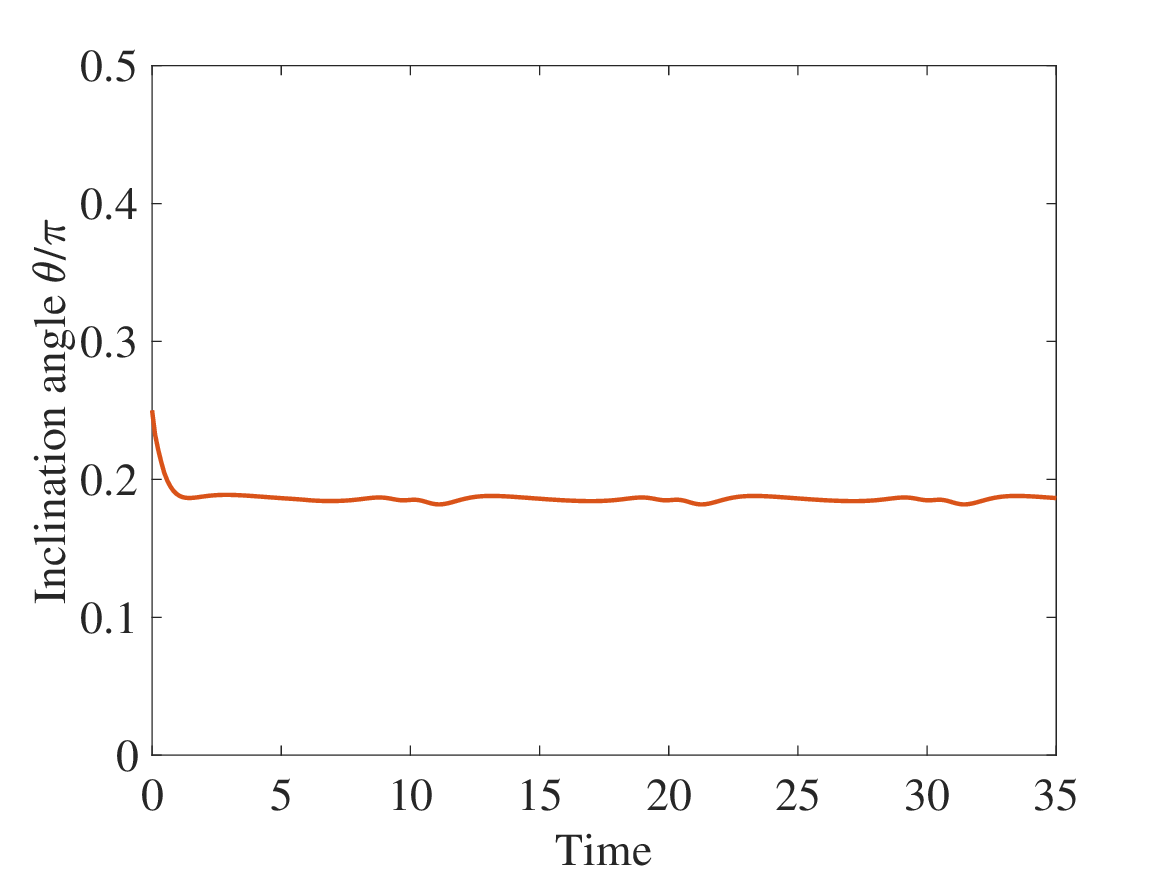}
	\caption{Tank-treading motion of a multicomponent vesicle under shear flow for $Ca=4$ and $ b_B = 1$. Evolution of the inclination angle over time.}
	\label{fig:M2Dbc1TTIA}	 		
\end{figure}

\fref{fig:M2Dbc1TTIA} shows the evolution of the inclination angle for this case. We observe that the multicomponent vesicle rapidly reaches a tank-treading motion, accompanied by minor periodic fluctuations. These fluctuations are attributed to the line energy concentrated at the interfaces between the two phases on the membrane surface; a feature absent in the homogeneous vesicle case shown in \fref{fig:shearFlowIAH2D}.(a), where no such interfaces exist. As $\mathcal{C} = 0$ in this case, implying the absence of bending rigidity variation, the vesicle undergoes steady tank-treading motion with uniform phase treading velocity.
\par
\begin{figure}[h]
	\centering
	\begin{subfigure}{0.45\textwidth}
		\centering
		\includegraphics[width=\linewidth,trim={0cm 0cm 0cm 0cm},clip]{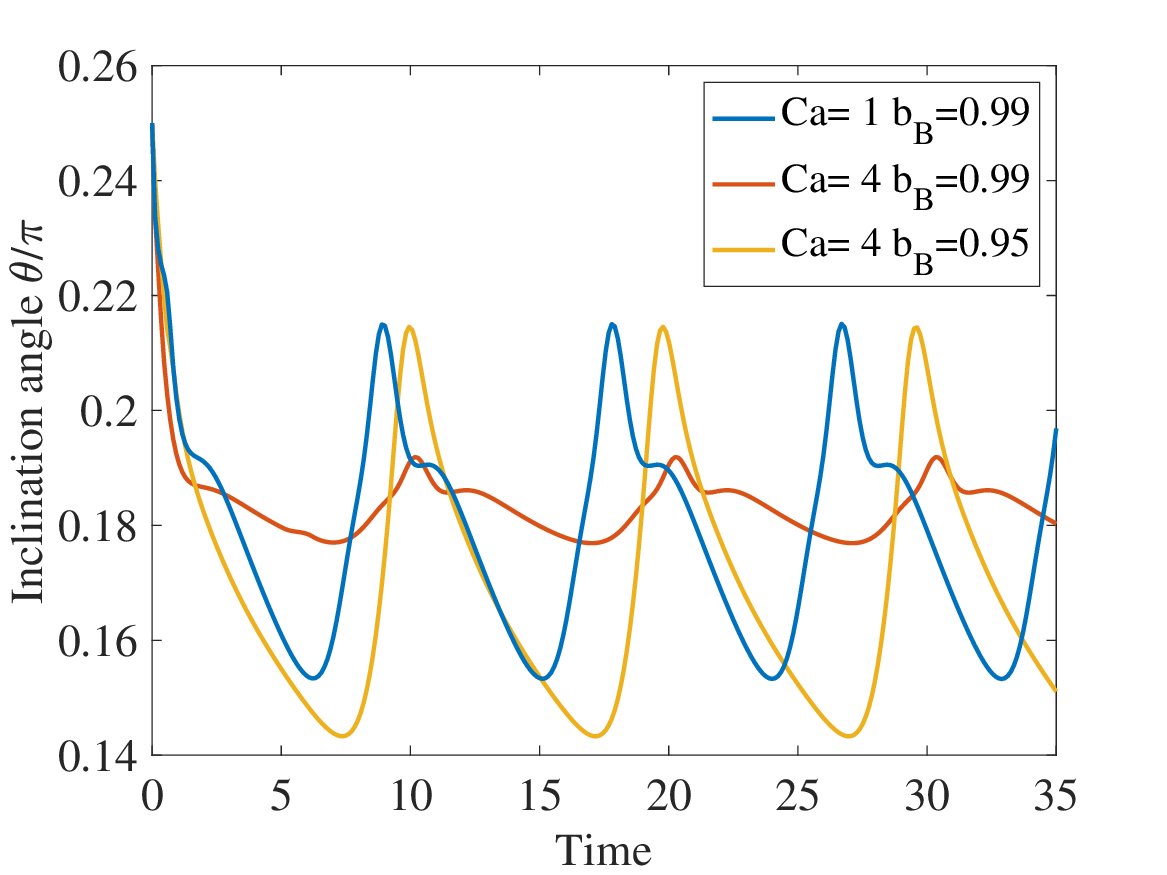}
		\caption{Swinging motions of multicomponent vesicles in N/N fluid}
	\end{subfigure}\quad
	\begin{subfigure}{0.45\textwidth}
		\centering
		\includegraphics[width=\linewidth,trim={0cm 0cm 0cm 0cm},clip]{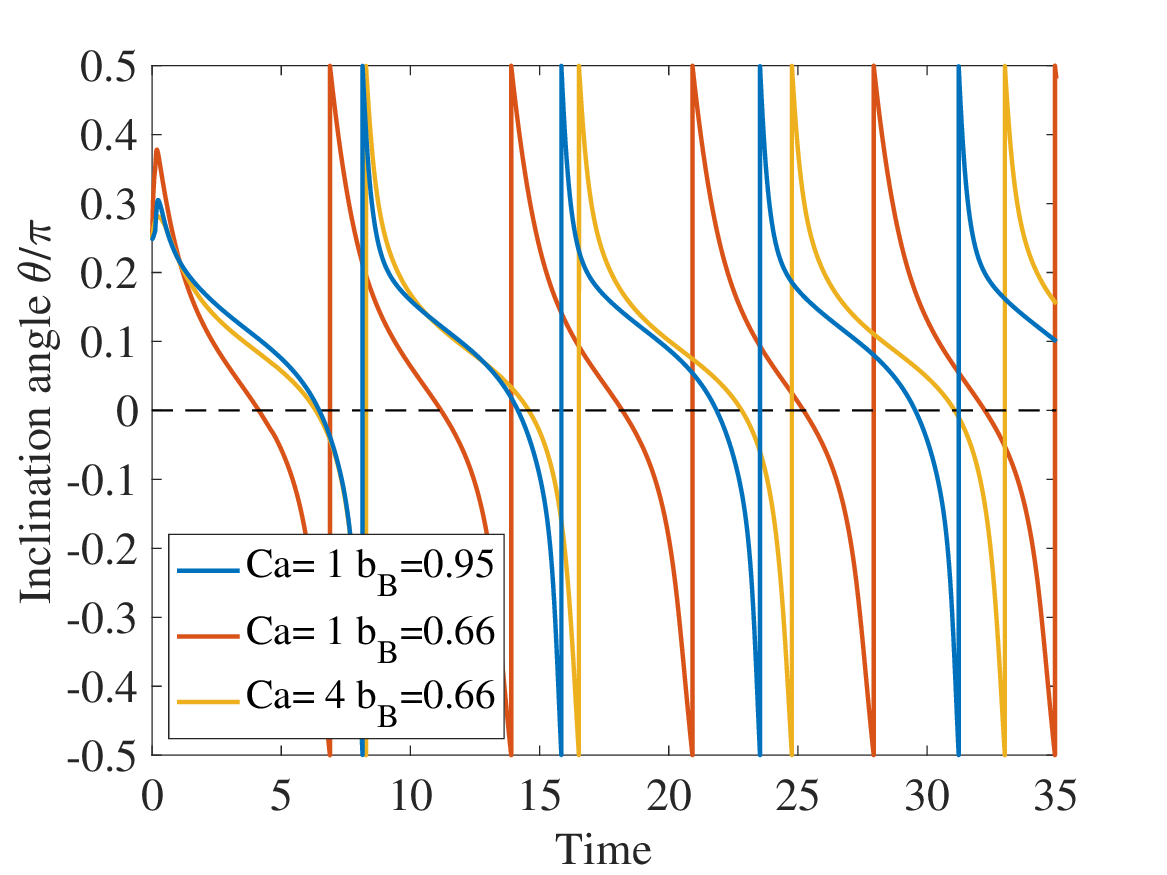}
		\caption{Tumbling motions of multicomponent vesicles in N/N fluid}
	\end{subfigure}
	\caption{Multicomponent vesicles in N/N shear flow. Evolution of the inclination angle over time for $\mathcal{C}>0$.}	
	\label{fig:shearFlowIAM2D}	 		
\end{figure}

To better understand the effect of bending rigidity variation on the swinging and tumbling motions of multicomponent vesicles, we increase the parameter $\mathcal{C}$ by conducting 6 numerical examples, setting $Ca = 1$ and $4$ for each of the following values of $b_B$: $0.99$, $0.95$, and $0.66$. The time step size is $\Delta t=1\times10^{-3}$. The evolution of the inclination angle over time for swinging and tumbling motions is shown in \fref{fig:shearFlowIAM2D}. In \fref{fig:shearFlowIAM2D}.(a), we observe that the following three cases result in swinging motion: $\mathcal{C}_1 = 0.0025$ ($Ca = 4$, $b_B = 0.99$), $\mathcal{C}_2 = 0.01$ ($Ca = 1$, $b_B = 0.99$), and $\mathcal{C}_3 = 0.0125$ ($Ca = 4$, $b_B = 0.95$). Furthermore, as the effect of bending rigidity variation increases (i.e., as $\mathcal{C}$ increases from 0.0025 to 0.0125), the corresponding swinging amplitude $\Delta \theta$ also increases: $2.7^\circ$, $11.1^\circ$, and $12.8^\circ$, respectively. If we further increase $\mathcal{C}$ to $\mathcal{C}_4 = 0.05$ ($Ca = 1$, $b_B = 0.95$), $\mathcal{C}_5 = 0.0833$ ($Ca = 4$, $b_B = 0.66$), and $\mathcal{C}_6 = 0.3333$ ($Ca = 1$, $b_B = 0.66$), the amplified effect of bending rigidity variation leads to tumbling motion of the multicomponent vesicle shown in \fref{fig:shearFlowIAM2D}.(b). The results of the six examples, along with the tank-treading case shown in \fref{fig:M2Dbc1TT}, support the conclusion that bending rigidity variation triggers the swinging motion of multicomponent vesicles. As this effect becomes more pronounced, the swinging motion intensifies with increasing amplitude and eventually transitions into tumbling motion. It is also worth noting that when the Capillary number is small (i.e., $Ca = 1$), the multicomponent vesicle exhibits greater sensitivity to variations in $b_B$. This behavior is evident in \fref{fig:shearFlowIAM2D} and can be interpreted directly from the definition of $\mathcal{C} = \frac{|b_A - b_B|}{Ca}$.
\par
\begin{figure}[h]
	\begin{subfigure}{0.135\textwidth}
		\centering
		\includegraphics[width=\linewidth,trim={2.6cm 0cm 2.6cm 0cm},clip]{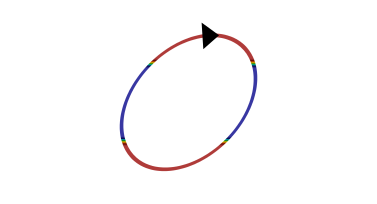}
		\caption*{$t=0$}
	\end{subfigure}
	\begin{subfigure}{0.135\textwidth}
		\centering
		\includegraphics[width=\linewidth,trim={2.6cm 0cm 2.6cm 0cm},clip]{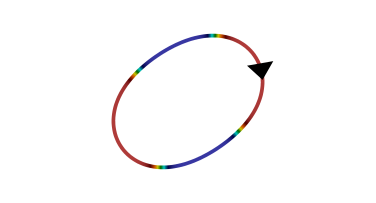}
		\caption*{$t=2$}
	\end{subfigure}
	\begin{subfigure}{0.135\textwidth}
		\centering
		\includegraphics[width=\linewidth,trim={2.6cm 0cm 2.6cm 0cm},clip]{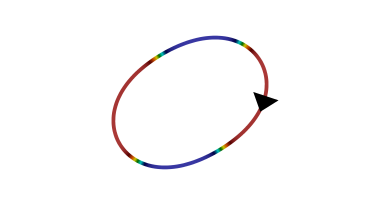}
		\caption*{$t=4$}
	\end{subfigure}
	\begin{subfigure}{0.135\textwidth}
		\centering
		\includegraphics[width=\linewidth,trim={2.6cm 0cm 2.6cm 0cm},clip]{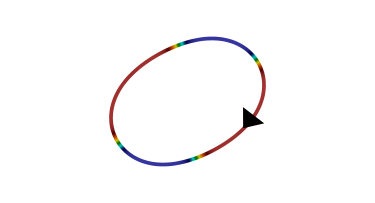}
		\caption*{$t=6$}
	\end{subfigure}
	\begin{subfigure}{0.135\textwidth}
		\centering
		\includegraphics[width=\linewidth,trim={2.6cm 0cm 2.6cm 0cm},clip]{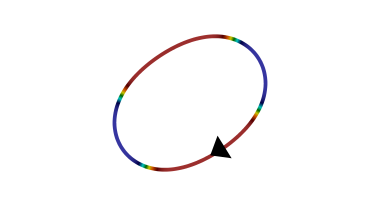}
		\caption*{$t=8$}
	\end{subfigure}
	\begin{subfigure}{0.135\textwidth}
		\centering
		\includegraphics[width=\linewidth,trim={2.6cm 0cm 2.6cm 0cm},clip]{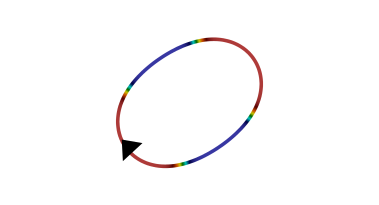}
		\caption*{$t=10$}
	\end{subfigure}
	\begin{subfigure}{0.135\textwidth}
		\centering
		\includegraphics[width=\linewidth,trim={2.6cm 0cm 2.6cm 0cm},clip]{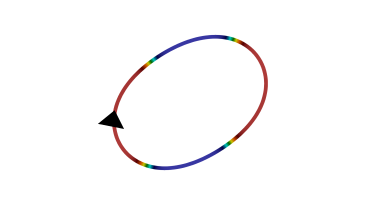}
		\caption*{$t=12$}
	\end{subfigure}
	\caption{Swinging motion of a multicomponent vesicle under shear flow for $\mathcal{C}_2$.}	
	\label{fig:M2Dbc1SW}	 		
\end{figure}
\begin{figure}[H]
	\begin{subfigure}{0.135\textwidth}
		\centering
		\includegraphics[width=\linewidth,trim={2.6cm 0cm 2.6cm 0cm},clip]{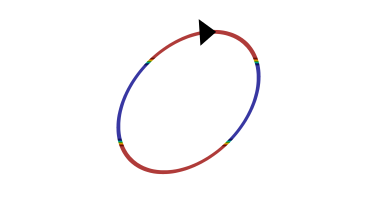}
		\caption*{$t=0$}
	\end{subfigure}
	\begin{subfigure}{0.135\textwidth}
		\centering
		\includegraphics[width=\linewidth,trim={2.6cm 0cm 2.6cm 0cm},clip]{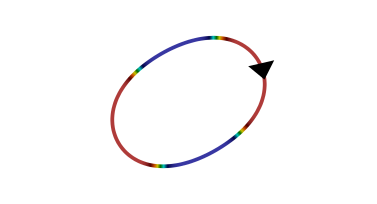}
		\caption*{$t=2$}
	\end{subfigure}
	\begin{subfigure}{0.135\textwidth}
		\centering
		\includegraphics[width=\linewidth,trim={2.6cm 0cm 2.6cm 0cm},clip]{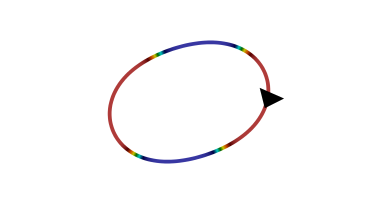}
		\caption*{$t=4$}
	\end{subfigure}
	\begin{subfigure}{0.135\textwidth}
		\centering
		\includegraphics[width=\linewidth,trim={2.6cm 0cm 2.6cm 0cm},clip]{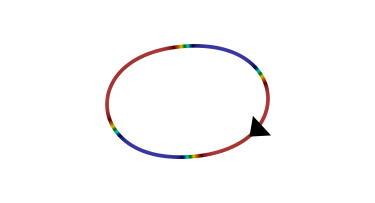}
		\caption*{$t=6$}
	\end{subfigure}
	\begin{subfigure}{0.135\textwidth}
		\centering
		\includegraphics[width=\linewidth,trim={2.6cm 0cm 2.6cm 0cm},clip]{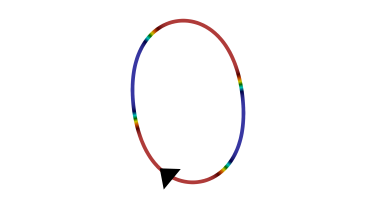}
		\caption*{$t=8$}
	\end{subfigure}
	\begin{subfigure}{0.135\textwidth}
		\centering
		\includegraphics[width=\linewidth,trim={2.6cm 0cm 2.6cm 0cm},clip]{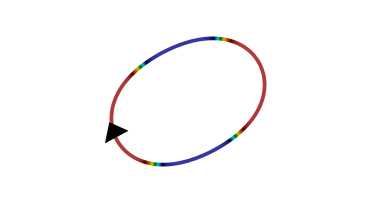}
		\caption*{$t=10$}
	\end{subfigure}
	\begin{subfigure}{0.135\textwidth}
		\centering
		\includegraphics[width=\linewidth,trim={2.6cm 0cm 2.6cm 0cm},clip]{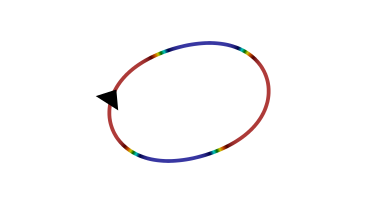}
		\caption*{$t=12$}
	\end{subfigure}
	\caption{Tumbling motion of a multicomponent vesicle under shear flow for $\mathcal{C}_4$.}	
	\label{fig:M2Dbc1TM}	 		
\end{figure}

We present the swinging motion at $\mathcal{C}_2 = 0.01$ ($Ca = 1$, $b_B = 0.99$) in \fref{fig:M2Dbc1SW}, and the tumbling motion at $\mathcal{C}_4 = 0.05$ ($Ca = 1$, $b_B = 0.95$) in \fref{fig:M2Dbc1TM}, to visualize the hydrodynamics of multicomponent vesicles coupled with phase evolution on the membrane. Note that the phase treading velocity is no longer constant; instead, the soft phase (red) tends to remain at the two ends of the vesicle, where the curvature is highest, for a longer duration within each period. This behavior, attributed to the effect of bending rigidity variation on phase evolution along the membrane surface, has also been reported in \cite{GERA2018362, Gera_2022, WEN2024117390}. These two representative cases will be used to investigate the viscoelastic effects on multicomponent vesicles in shear flow later. \par
 \par
We also plot the evolution of perimeter error, area error, and mass conservation error for the two cases shown in \frefs{fig:M2Dbc1SW}{fig:M2Dbc1TM}, where the mass conservation error is defined as:
\[M_{error} = \frac{\int_{\Omega}c_h \delta_{\phi} \; d\Omega - \int_{\Omega}c_0 \delta_{\phi} \; d\Omega}{\int_{\Omega}c_0 \delta_{\phi} \; d\Omega}\]\par
\begin{figure}[H]
	\centering
	\begin{subfigure}{0.4\textwidth}
		\centering
		\includegraphics[width=\linewidth,trim={0cm 0cm 0cm 0cm},clip]{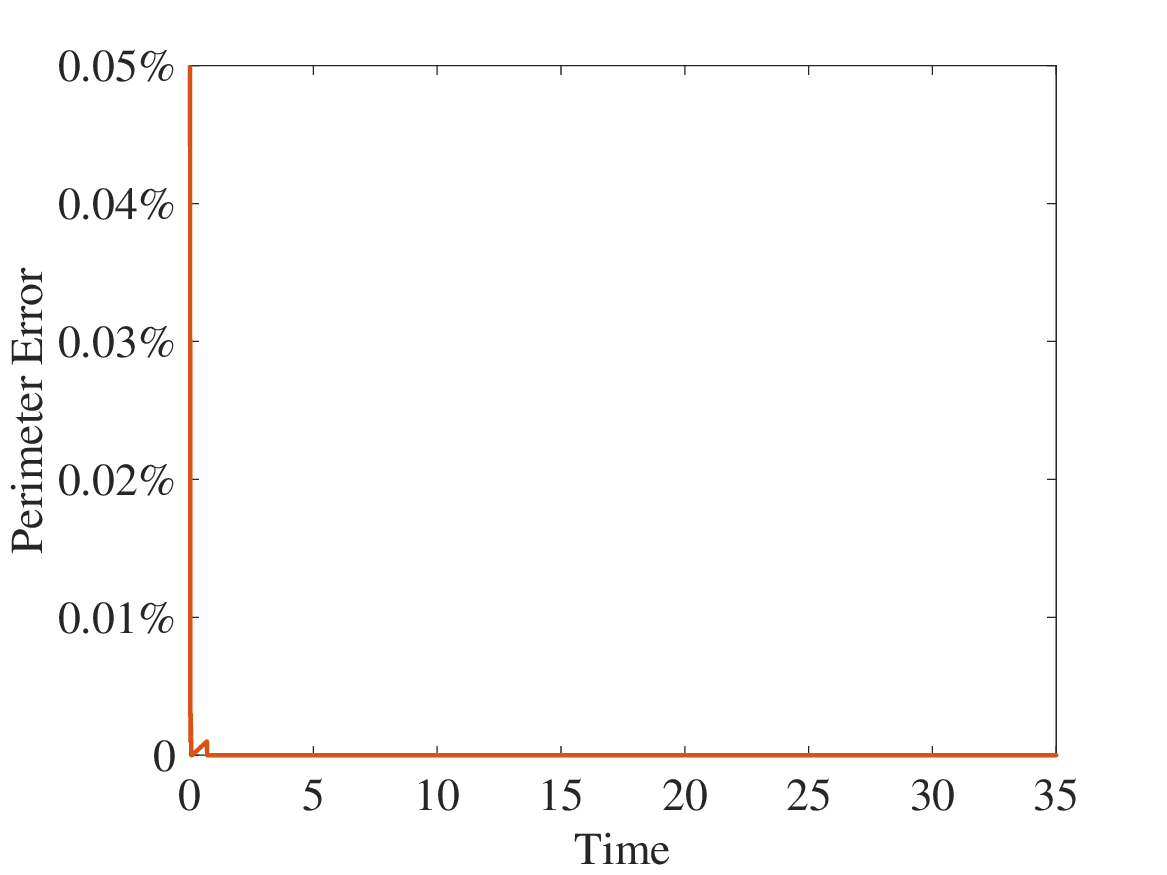}
		\caption{Swinging motion with $\mathcal{C}_2$}
	\end{subfigure}\quad
	\begin{subfigure}{0.4\textwidth}
		\centering
		\includegraphics[width=\linewidth,trim={0cm 0cm 0cm 0cm},clip]{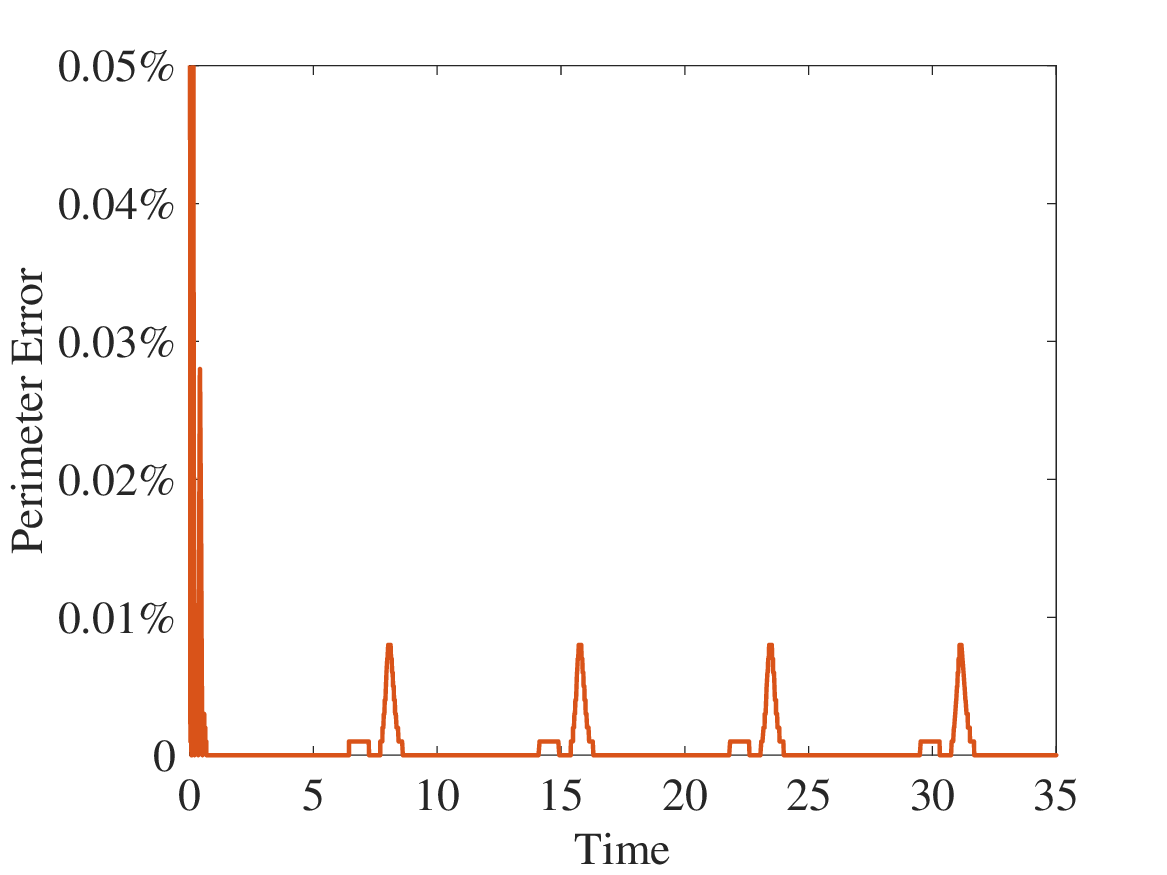}
		\caption{Tumbling motion with $\mathcal{C}_4$}
	\end{subfigure}
	\caption{Swinging and tumbling vesicles in shear flow. Evolution of the perimeter error for $\mathcal{C}_2$ and $\mathcal{C}_4$.}	
	\label{fig:shearFlowP}	 		
\end{figure}

From \frefs{fig:shearFlowP}{fig:shearFlowM}, we observe that the perimeter, area, and mass conservation are accurately maintained throughout the simulation, with only minor fluctuations occurring during the initial and rapid deformation time period. These initial fluctuations are attributed to the vesicle initialization process and the influence of the penalty functions used in our model. The errors quickly decay to near-zero values after a few time steps. During the stable state of the swinging vesicle, the maximum perimeter, area, and mass conservation errors are $0.000\%$, $1.233\%$, and $0.002\%$, respectively. For the tumbling vesicle in its relatively stable phase, the corresponding errors are $0.008\%$, $1.441\%$, and $0.041\%$. These results demonstrate that our model effectively enforces global perimeter and area conservation in 2D, as well as mass conservation for inextensible multicomponent vesicles even with large time step size.

\begin{figure}[h]
	\centering
	\begin{subfigure}{0.4\textwidth}
		\centering
		\includegraphics[width=\linewidth,trim={0cm 0cm 0cm 0cm},clip]{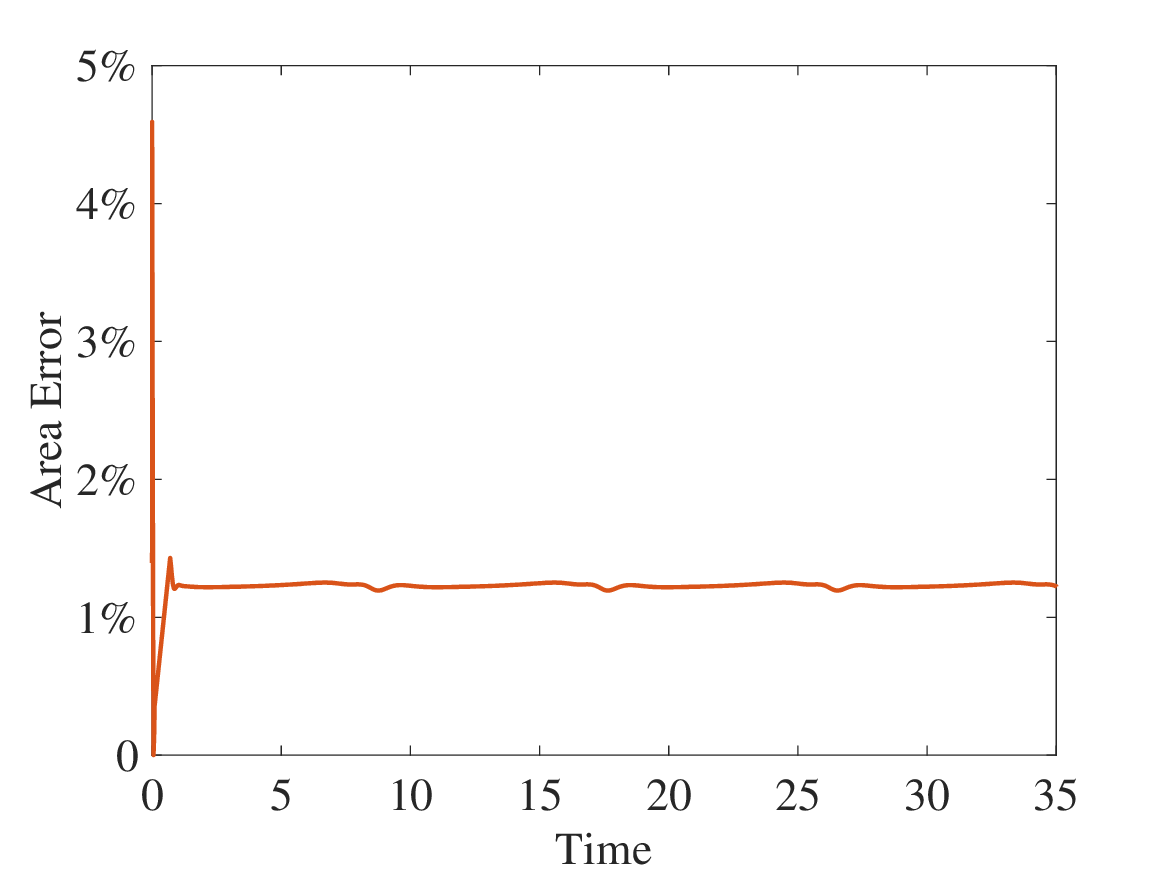}
		\caption{Swinging motion with $\mathcal{C}_2$}
	\end{subfigure}\quad
	\begin{subfigure}{0.4\textwidth}
		\centering
		\includegraphics[width=\linewidth,trim={0cm 0cm 0cm 0cm},clip]{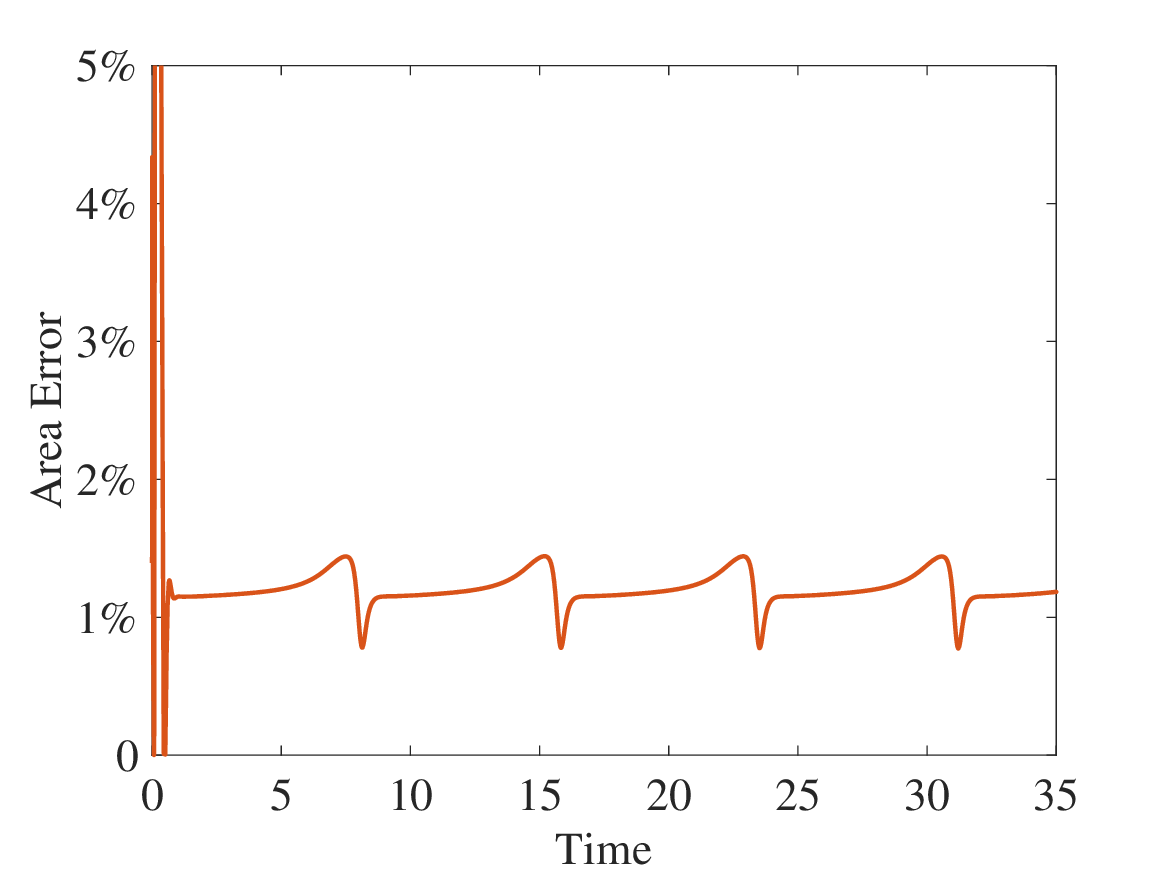}
		\caption{Tumbling motion with $\mathcal{C}_4$}
	\end{subfigure}
	\begin{subfigure}{0.4\textwidth}
		\centering
		\includegraphics[width=\linewidth,trim={0cm 0cm 0cm 0cm},clip]{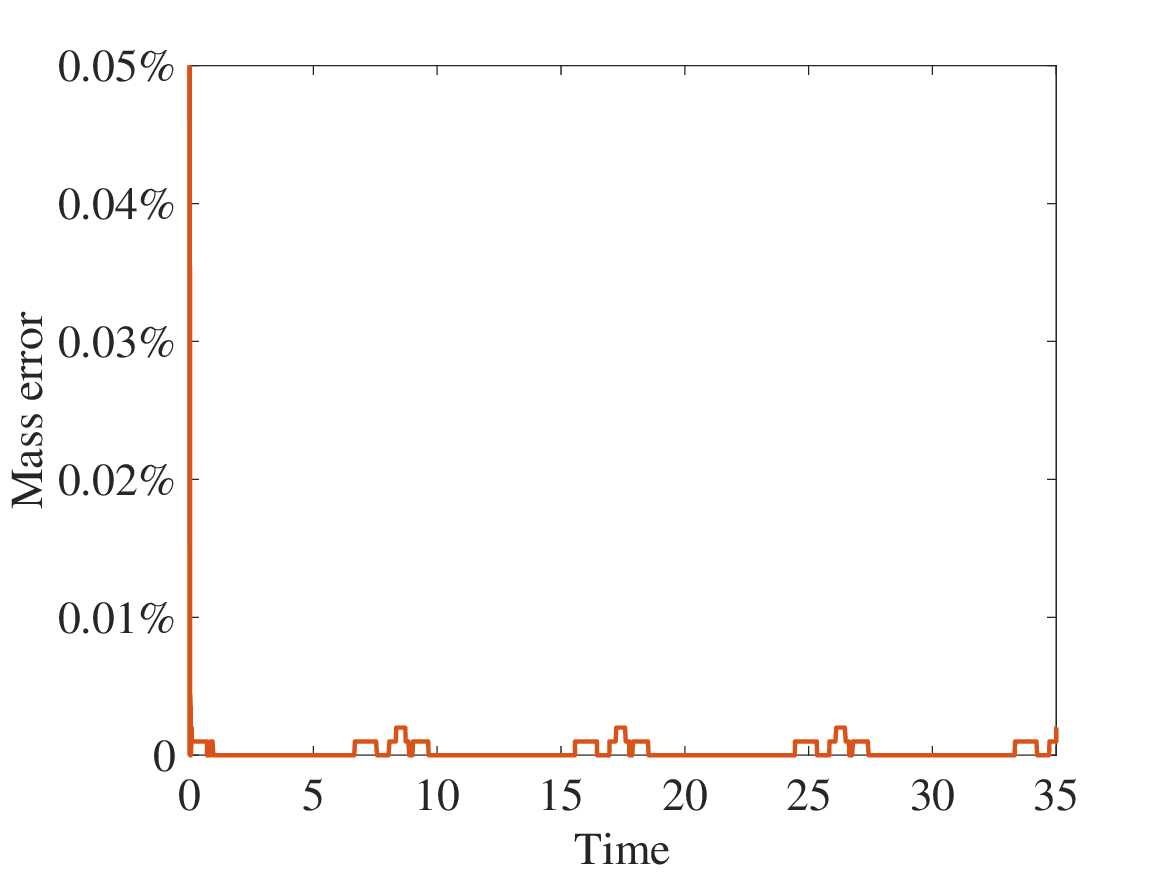}
		\caption{Swinging motion with $\mathcal{C}_2$}
	\end{subfigure}\quad
	\begin{subfigure}{0.4\textwidth}
		\centering
		\includegraphics[width=\linewidth,trim={0cm 0cm 0cm 0cm},clip]{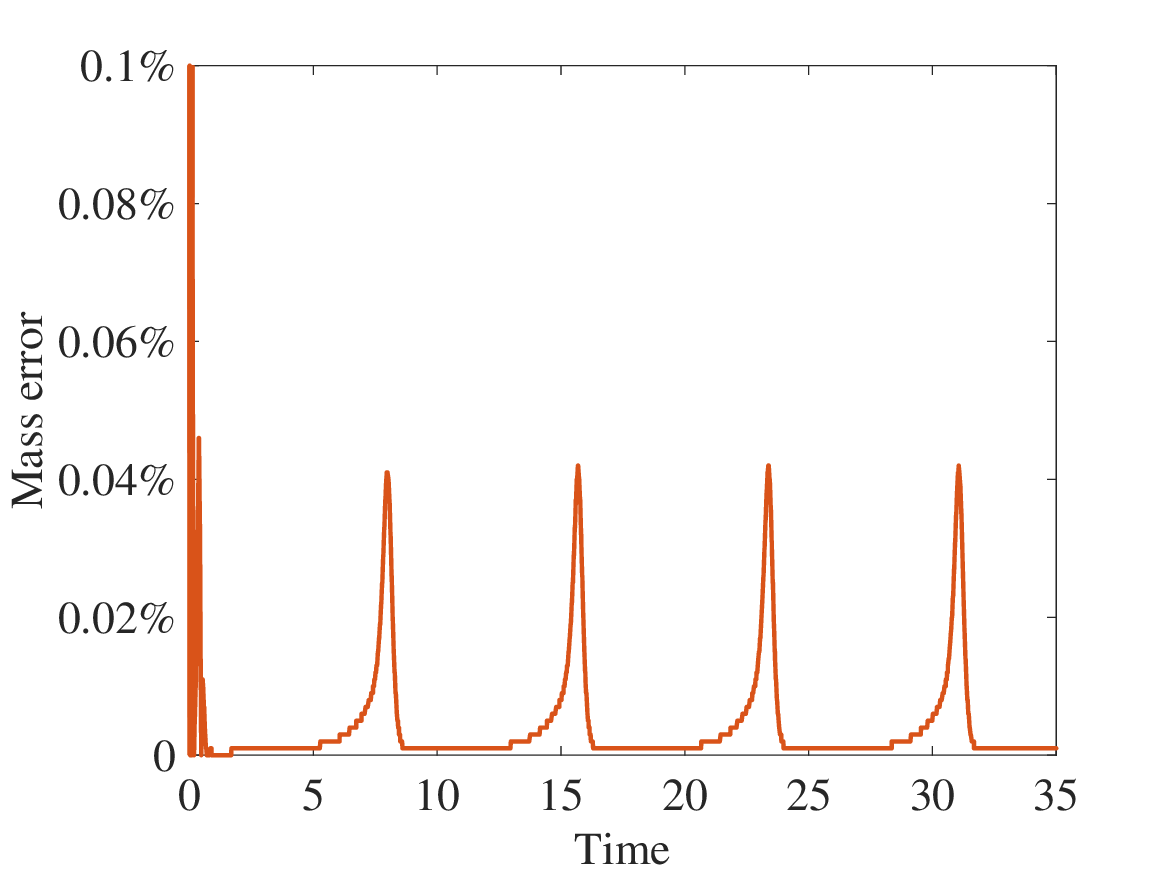}
		\caption{Tumbling motion with $\mathcal{C}_4$}
	\end{subfigure}
	\caption{Swinging and tumbling vesicles in shear flow. Evolution of the area and mass conservation error with respect to initial mass value over time for $\mathcal{C}_2$ and $\mathcal{C}_4$.}	
	\label{fig:shearFlowM}	 		
\end{figure} \par
\begin{figure}[H]
	\centering
	\includegraphics [scale=0.4]{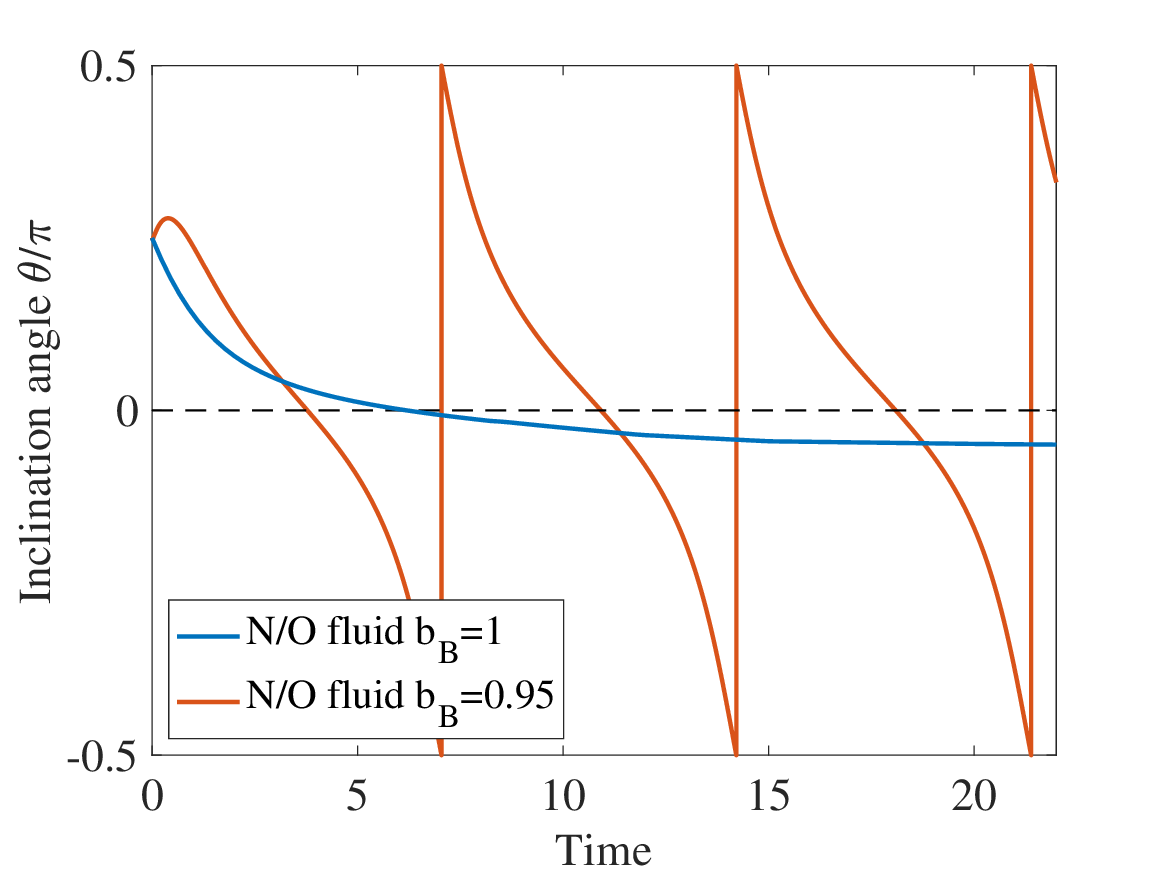}
	\caption{Comparison of homogeneous and multicomponent vesicles in N/O shear flow with $\alpha_{\mu_n} = 19$. Evolution of the inclination angle over time.}
	\label{fig:M2DN19IA}	 		
\end{figure}

We also examine the effect of bending rigidity variation on the negative inclination angle tank-treading case discussed in Section~\ref{H2DNO}. We keep the parameters the same as the unmatched homogeneous vesicle case ($\mathcal{W}i=1$), but set $b_B = 0.95$ for the soft phase of the multicomponent vesicle. The evolution of the inclination angle is shown in \fref{fig:M2DN19IA}, where the blue curve represents the homogeneous vesicle and the red curve corresponds to the multicomponent vesicle. In Section~\ref{H2DNO}, we stated that the viscoelastic effect counterbalances the high viscosity contrast $\alpha_{\mu_n}=19$, which would otherwise trigger tumbling motion, and instead results in the unique negative inclination angle tank-treading motion. However, in the multicomponent vesicle case shown in \fref{fig:M2DN19IA}, the bending rigidity variation, together with viscosity contrast, overcomes the viscoelastic effect and restores the vesicle to tumbling motion. This demonstrates the complexity of the viscoelastic effect acting on multicomponent vesicles, due to the presence of bending rigidity variation characterized by different values of $\mathcal{C}$, which result in distinct hydrodynamic behaviors as shown in this section and will be further investigated in the following section.

\subsubsection{Viscoelastic effects on the swinging motion of multicomponent vesicles} \label{SWNO}
\begin{figure}[h]
	\centering
	\includegraphics [scale=0.4]{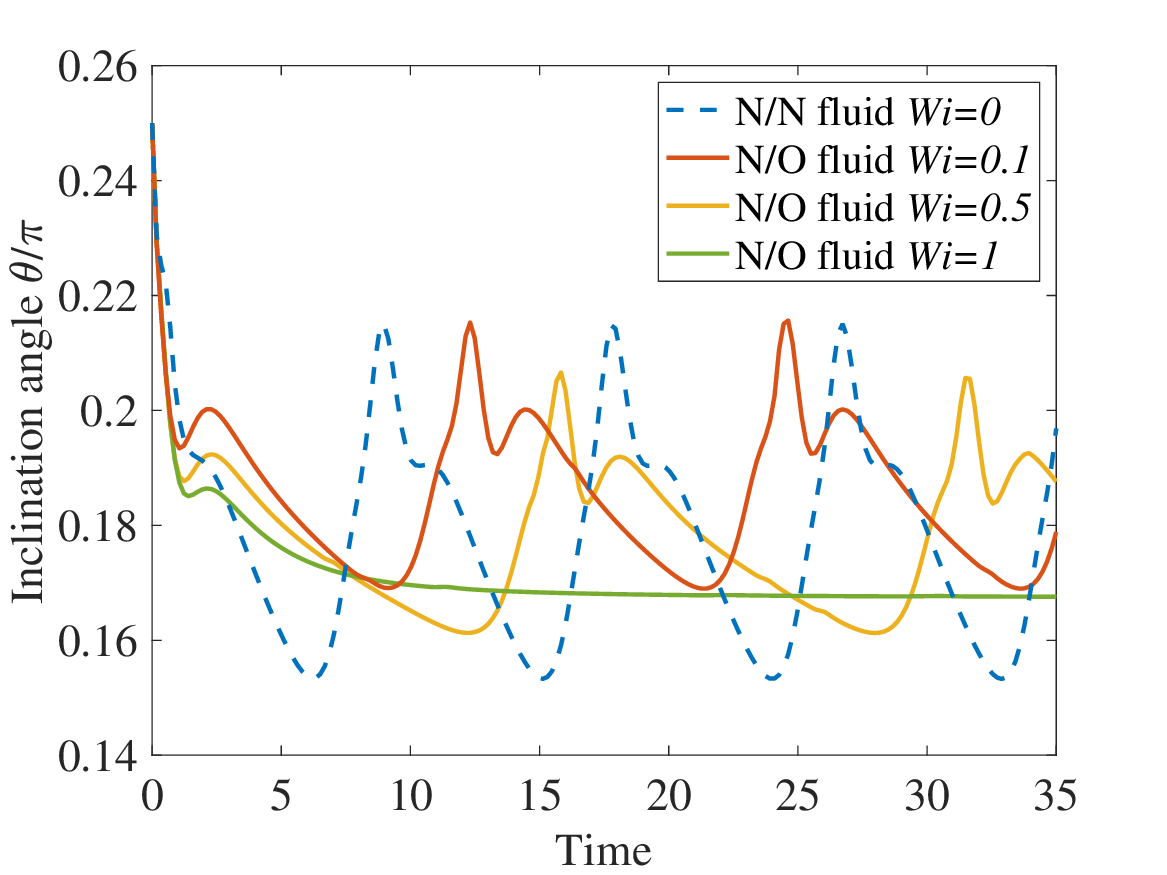}
	\caption{Multicomponent vesicles in N/O shear flow with $\mathcal{C}_2$. Evolution of the inclination angle over time.}
	\label{fig:M2DNOIA}	 		
\end{figure}

\begin{figure}[h]
	\begin{subfigure}{0.135\textwidth}
		\centering
		\includegraphics[width=\linewidth,trim={2.6cm 0.5cm 2.6cm 0.5cm},clip]{M2DSW0}
		\caption*{$t=0$}
	\end{subfigure}
	\begin{subfigure}{0.135\textwidth}
		\centering
		\includegraphics[width=\linewidth,trim={2.6cm 0.5cm 2.6cm 0.5cm},clip]{M2DSW1}
		\caption*{$t=2$}
	\end{subfigure}
	\begin{subfigure}{0.135\textwidth}
		\centering
		\includegraphics[width=\linewidth,trim={2.6cm 0.5cm 2.6cm 0.5cm},clip]{M2DSW2}
		\caption*{$t=4$}
	\end{subfigure}
	\begin{subfigure}{0.135\textwidth}
		\centering
		\includegraphics[width=\linewidth,trim={2.6cm 0.5cm 2.6cm 0.5cm},clip]{M2DSW3}
		\caption*{$t=6$}
	\end{subfigure}
	\begin{subfigure}{0.135\textwidth}
		\centering
		\includegraphics[width=\linewidth,trim={2.6cm 0.5cm 2.6cm 0.5cm},clip]{M2DSW4}
		\caption*{$t=8$}
	\end{subfigure}
	\begin{subfigure}{0.135\textwidth}
		\centering
		\includegraphics[width=\linewidth,trim={2.7cm 0.5cm 2.7cm 0.5cm},clip]{M2DSW5}
		\caption*{$t=10$}
	\end{subfigure}
	\begin{subfigure}{0.135\textwidth}
		\centering
		\includegraphics[width=\linewidth,trim={2.6cm 0.5cm 2.6cm 0.5cm},clip]{M2DSW6}
		\caption*{$t=12$}
	\end{subfigure}
	\begin{subfigure}{0.135\textwidth}
		\centering
		\includegraphics[width=\linewidth,trim={2.6cm 0.5cm 2.6cm 0.5cm},clip]{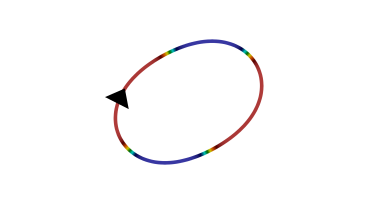}
		\caption*{$t=14$}
	\end{subfigure}
	\begin{subfigure}{0.135\textwidth}
		\centering
		\includegraphics[width=\linewidth,trim={2.6cm 0.5cm 2.6cm 0.5cm},clip]{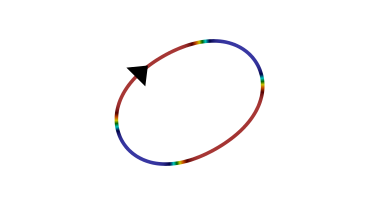}
		\caption*{$t=16$}
	\end{subfigure}
	\begin{subfigure}{0.135\textwidth}
		\centering
		\includegraphics[width=\linewidth,trim={2.6cm 0.5cm 2.6cm 0.5cm},clip]{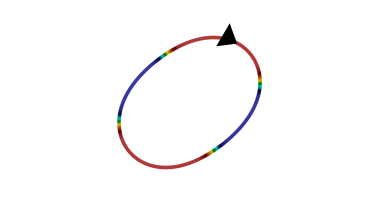}
		\caption*{$t=18$}
	\end{subfigure}
	\begin{subfigure}{0.135\textwidth}
		\centering
		\includegraphics[width=\linewidth,trim={2.6cm 0.5cm 2.6cm 0.5cm},clip]{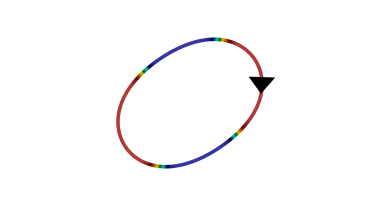}
		\caption*{$t=20$}
	\end{subfigure}
	\begin{subfigure}{0.135\textwidth}
		\centering
		\includegraphics[width=\linewidth,trim={2.6cm 0.5cm 2.6cm 0.5cm},clip]{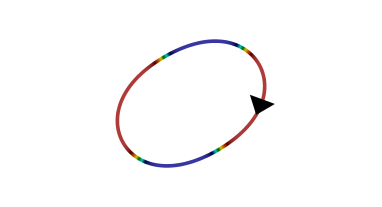}
		\caption*{$t=22$}
	\end{subfigure}
	\begin{subfigure}{0.135\textwidth}
		\centering
		\includegraphics[width=\linewidth,trim={2.6cm 0.5cm 2.6cm 0.5cm},clip]{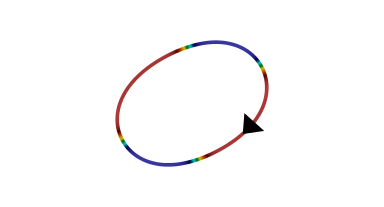}
		\caption*{$t=24$}
	\end{subfigure}
	\begin{subfigure}{0.135\textwidth}
		\centering
		\includegraphics[width=\linewidth,trim={2.6cm 0.5cm 2.6cm 0.5cm},clip]{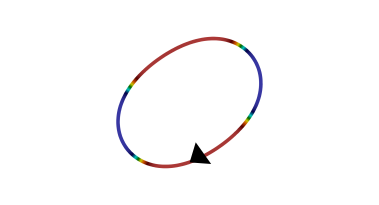}
		\caption*{$t=26$}
	\end{subfigure}
	\caption{Swinging motion of a multicomponent vesicle under N/N shear flow with $\mathcal{W}i=0$.}	
	\label{fig:M2DSWNN}	 		
\end{figure}

\begin{figure}[h]
	\begin{subfigure}{0.135\textwidth}
		\centering
		\includegraphics[width=\linewidth,trim={2.6cm 0.5cm 2.6cm 0.5cm},clip]{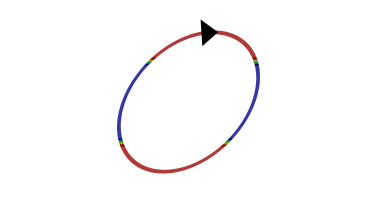}
		\caption*{$t=0$}
	\end{subfigure}
	\begin{subfigure}{0.135\textwidth}
		\centering
		\includegraphics[width=\linewidth,trim={2.6cm 0.5cm 2.6cm 0.5cm},clip]{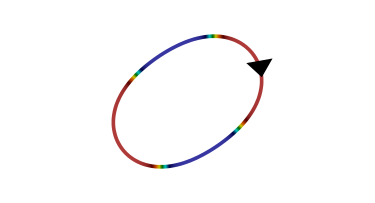}
		\caption*{$t=2$}
	\end{subfigure}
	\begin{subfigure}{0.135\textwidth}
		\centering
		\includegraphics[width=\linewidth,trim={2.6cm 0.5cm 2.6cm 0.5cm},clip]{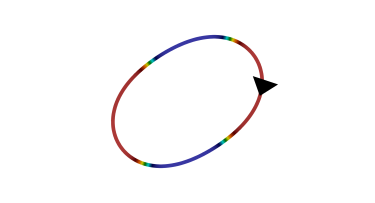}
		\caption*{$t=4$}
	\end{subfigure}
	\begin{subfigure}{0.135\textwidth}
		\centering
		\includegraphics[width=\linewidth,trim={2.6cm 0.5cm 2.6cm 0.5cm},clip]{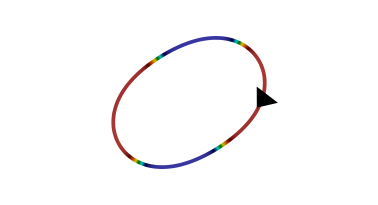}
		\caption*{$t=6$}
	\end{subfigure}
	\begin{subfigure}{0.135\textwidth}
		\centering
		\includegraphics[width=\linewidth,trim={2.6cm 0.5cm 2.6cm 0.5cm},clip]{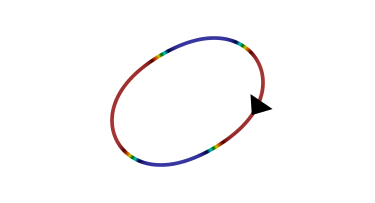}
		\caption*{$t=8$}
	\end{subfigure}
	\begin{subfigure}{0.135\textwidth}
		\centering
		\includegraphics[width=\linewidth,trim={2.6cm 0.5cm 2.6cm 0.5cm},clip]{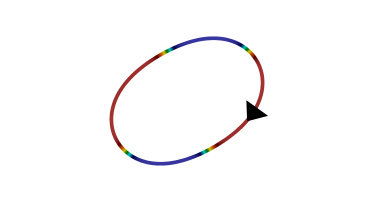}
		\caption*{$t=10$}
	\end{subfigure}
	\begin{subfigure}{0.135\textwidth}
		\centering
		\includegraphics[width=\linewidth,trim={2.6cm 0.5cm 2.6cm 0.5cm},clip]{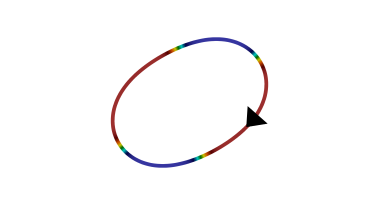}
		\caption*{$t=12$}
	\end{subfigure}
	\begin{subfigure}{0.135\textwidth}
		\centering
		\includegraphics[width=\linewidth,trim={2.6cm 0.5cm 2.6cm 0.5cm},clip]{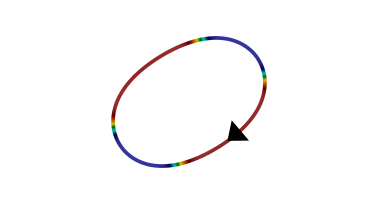}
		\caption*{$t=14$}
	\end{subfigure}
	\begin{subfigure}{0.135\textwidth}
		\centering
		\includegraphics[width=\linewidth,trim={2.6cm 0.5cm 2.6cm 0.5cm},clip]{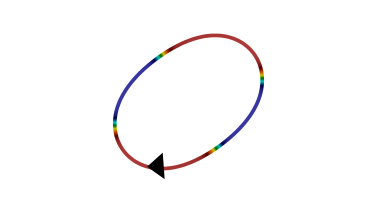}
		\caption*{$t=16$}
	\end{subfigure}
	\begin{subfigure}{0.135\textwidth}
		\centering
		\includegraphics[width=\linewidth,trim={2.6cm 0.5cm 2.6cm 0.5cm},clip]{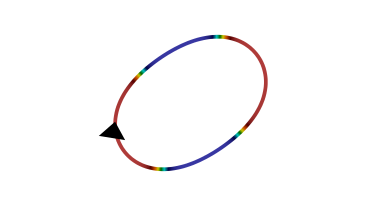}
		\caption*{$t=18$}
	\end{subfigure}
	\begin{subfigure}{0.135\textwidth}
		\centering
		\includegraphics[width=\linewidth,trim={2.6cm 0.5cm 2.6cm 0.5cm},clip]{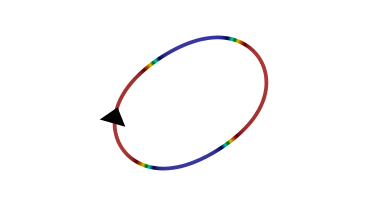}
		\caption*{$t=20$}
	\end{subfigure}
	\begin{subfigure}{0.135\textwidth}
		\centering
		\includegraphics[width=\linewidth,trim={2.6cm 0.5cm 2.6cm 0.5cm},clip]{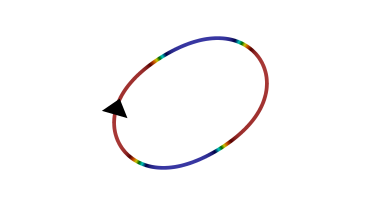}
		\caption*{$t=22$}
	\end{subfigure}
	\begin{subfigure}{0.135\textwidth}
		\centering
		\includegraphics[width=\linewidth,trim={2.6cm 0.5cm 2.6cm 0.5cm},clip]{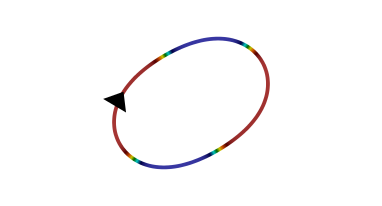}
		\caption*{$t=24$}
	\end{subfigure}
	\begin{subfigure}{0.135\textwidth}
		\centering
		\includegraphics[width=\linewidth,trim={2.6cm 0.5cm 2.6cm 0.5cm},clip]{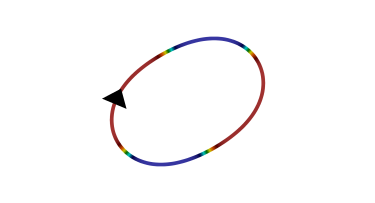}
		\caption*{$t=26$}
	\end{subfigure}
	\caption{Swinging motion of a multicomponent vesicle under N/O shear flow with $\mathcal{W}i=0.5$.}	
	\label{fig:M2DSWNOWi0.5}	 		
\end{figure}

\begin{figure}[h]
	\begin{subfigure}{0.135\textwidth}
		\centering
		\includegraphics[width=\linewidth,trim={2.6cm 0.5cm 2.6cm 0.5cm},clip]{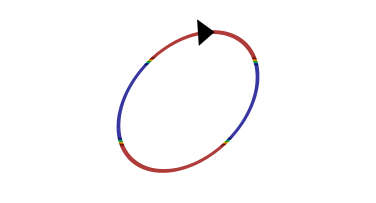}
		\caption*{$t=0$}
	\end{subfigure}
	\begin{subfigure}{0.135\textwidth}
		\centering
		\includegraphics[width=\linewidth,trim={2.6cm 0.5cm 2.6cm 0.5cm},clip]{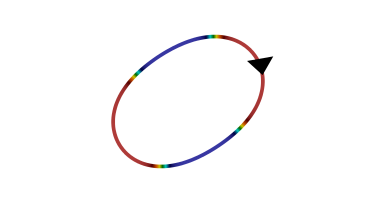}
		\caption*{$t=2$}
	\end{subfigure}
	\begin{subfigure}{0.135\textwidth}
		\centering
		\includegraphics[width=\linewidth,trim={2.6cm 0.5cm 2.6cm 0.5cm},clip]{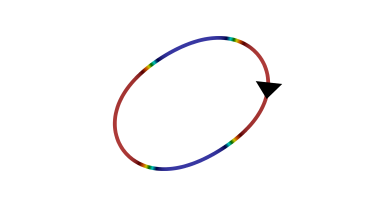}
		\caption*{$t=4$}
	\end{subfigure}
	\begin{subfigure}{0.135\textwidth}
		\centering
		\includegraphics[width=\linewidth,trim={2.6cm 0.5cm 2.6cm 0.5cm},clip]{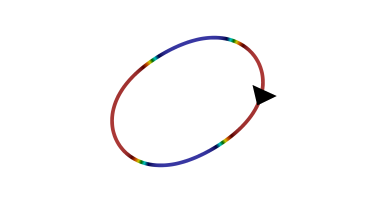}
		\caption*{$t=6$}
	\end{subfigure}
	\begin{subfigure}{0.135\textwidth}
		\centering
		\includegraphics[width=\linewidth,trim={2.6cm 0.5cm 2.6cm 0.5cm},clip]{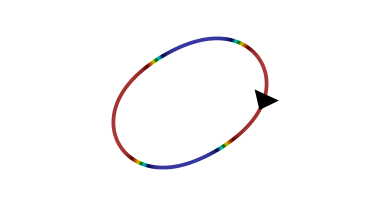}
		\caption*{$t=8$}
	\end{subfigure}
	\begin{subfigure}{0.135\textwidth}
		\centering
		\includegraphics[width=\linewidth,trim={2.6cm 0.5cm 2.6cm 0.5cm},clip]{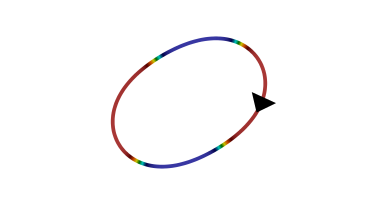}
		\caption*{$t=10$}
	\end{subfigure}
	\begin{subfigure}{0.135\textwidth}
		\centering
		\includegraphics[width=\linewidth,trim={2.6cm 0.5cm 2.6cm 0.5cm},clip]{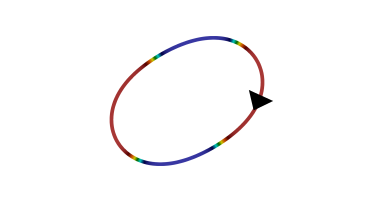}
		\caption*{$t=12$}
	\end{subfigure}
	\begin{subfigure}{0.135\textwidth}
		\centering
		\includegraphics[width=\linewidth,trim={2.6cm 0.5cm 2.6cm 0.5cm},clip]{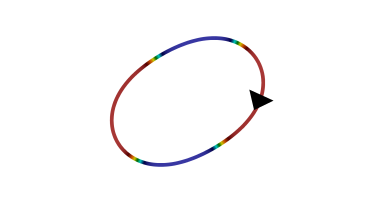}
		\caption*{$t=14$}
	\end{subfigure}
	\begin{subfigure}{0.135\textwidth}
		\centering
		\includegraphics[width=\linewidth,trim={2.6cm 0.5cm 2.6cm 0.5cm},clip]{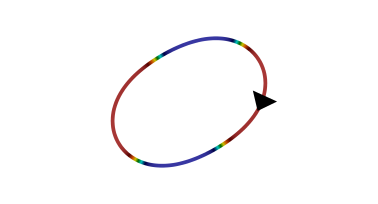}
		\caption*{$t=16$}
	\end{subfigure}
	\begin{subfigure}{0.135\textwidth}
		\centering
		\includegraphics[width=\linewidth,trim={2.6cm 0.5cm 2.6cm 0.5cm},clip]{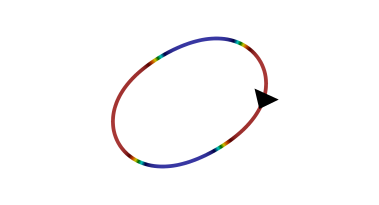}
		\caption*{$t=18$}
	\end{subfigure}
	\begin{subfigure}{0.135\textwidth}
		\centering
		\includegraphics[width=\linewidth,trim={2.6cm 0.5cm 2.6cm 0.5cm},clip]{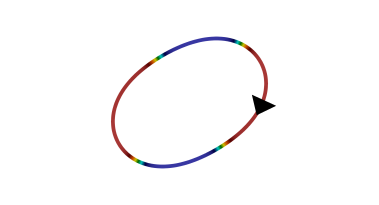}
		\caption*{$t=20$}
	\end{subfigure}
	\begin{subfigure}{0.135\textwidth}
		\centering
		\includegraphics[width=\linewidth,trim={2.6cm 0.5cm 2.6cm 0.5cm},clip]{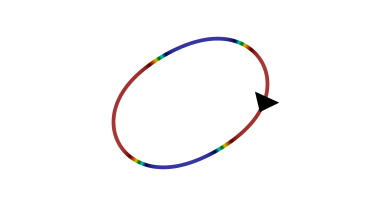}
		\caption*{$t=22$}
	\end{subfigure}
	\begin{subfigure}{0.135\textwidth}
		\centering
		\includegraphics[width=\linewidth,trim={2.6cm 0.5cm 2.6cm 0.5cm},clip]{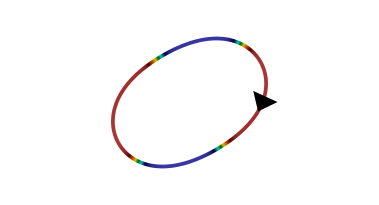}
		\caption*{$t=24$}
	\end{subfigure}
	\begin{subfigure}{0.135\textwidth}
		\centering
		\includegraphics[width=\linewidth,trim={2.6cm 0.5cm 2.6cm 0.5cm},clip]{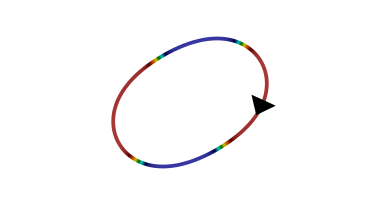}
		\caption*{$t=26$}
	\end{subfigure}
	\caption{Tank-treading motion of a multicomponent vesicle under N/O shear flow with $\mathcal{W}i=1$.}	
	\label{fig:M2DSWNOWi1}	 		
\end{figure}

In this section, we investigate the effect of viscoelasticity in N/O fluid on a multicomponent vesicle that initially exhibits swinging behavior in an N/N fluid with $\mathcal{W}i = 0$. The simulation parameters are consistent with the reference swinging case $\mathcal{C}_2$, as described in Section~\ref{M2DNN}, except that we now consider nonzero Weissenberg numbers. To assess the viscoelastic effect, we vary $\mathcal{W}i$ with values set to $0.1$, $0.5$, and $1$. The evolution of the inclination angle for these three cases, along with the reference N/N swinging case, is shown in \fref{fig:M2DNOIA}. The blue dashed line in \fref{fig:M2DNOIA} represents the reference N/N case, which exhibits periodic swinging motion due to bending rigidity variation. When viscoelastic effects are introduced, with $\mathcal{W}i = 0.1$ and $0.5$ (shown as red and yellow solid lines, respectively), the vesicle still undergoes swinging motion; however, the period of oscillation increases as the viscoelastic effect strengthens. As we impose a stronger viscoelastic effect on the multicomponent vesicle with $\mathcal{W}i = 1$, the vesicle transitions from swinging motion to tank-treading, as shown by the green solid line. We also present in \fref{fig:M2DSWNN}, \fref{fig:M2DSWNOWi0.5}, and \fref{fig:M2DSWNOWi1} the hydrodynamics and phase evolution on the membrane of multicomponent vesicles with $\mathcal{W}i = 0$, $0.5$, and $1$, respectively, corresponding to swinging, swinging with a longer period, and tank-treading motions. We find that as the vesicle transitions from swinging to longer-period swinging and eventually to tank-treading motion, the phases on the membrane surface exhibit correspondingly prolonged treading periods and ultimately become stationary.\par
These observations can be interpreted as follows: the viscoelastic behavior of the outer fluid can absorb and store a portion of the energy, thereby suppressing the energy fluctuations induced by bending rigidity variations as the phases tend to migrate along the membrane surface. As the viscoelastic effect becomes stronger and absorbs more energy, the vesicle membrane no longer needs to undergo significant deformation to maintain the overall energy balance. Consequently, the membrane stabilizes into a stationary shape with fixed curvature. Under this condition, the soft phase tends to accumulate in high-curvature regions, while the stiff phase settles in low-curvature regions and resists further treading. The overall shear stress drives the phase treading along the membrane surface, while the elastic component of the stress, together with bending effects, significantly alters the behavior of phase migration. Due to the stationary phases in the tank-treading motion shown in \fref{fig:M2DSWNOWi1}, no fluctuations in the inclination angle are observed in \fref{fig:M2DNOIA}. In contrast, minor fluctuations are present in the multicomponent vesicle case shown in \fref{fig:M2Dbc1TTIA}, resulting from line tension at the interfaces between different phases as they tread along the membrane surface. The polymeric nature of the Oldroyd-B fluid moderates the multicomponent vesicle hydrodynamics by dampening fluctuations. As a result, it alters the hydrodynamics of the multicomponent vesicle, modifying both the oscillation period and, ultimately, the type of motion exhibited.

\subsubsection{Viscoelastic effects on the tumbling motion of multicomponent vesicles}
\begin{figure}[h]
	\centering
	\includegraphics [scale=0.4]{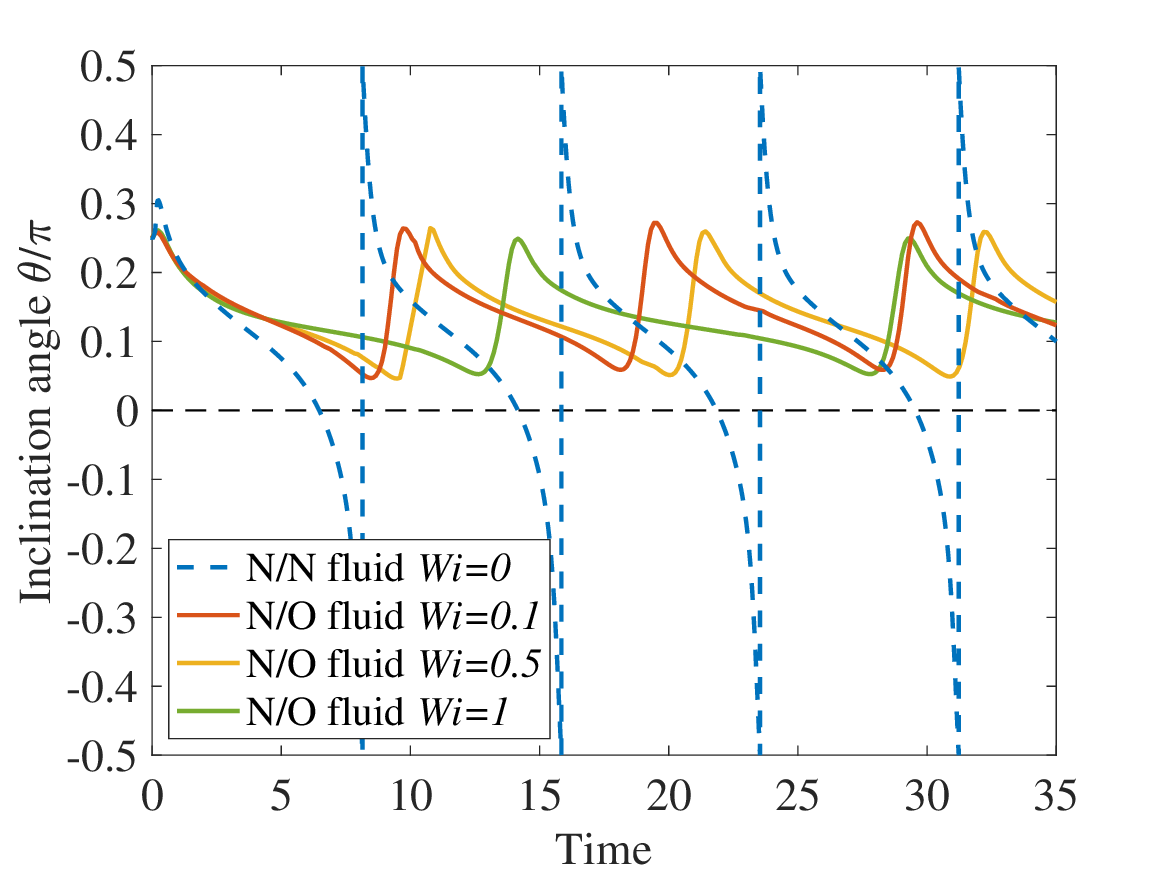}
	\caption{Multicomponent vesicles in N/O shear flow with $\mathcal{C}_4$. Evolution of the inclination angle over time.}
	\label{fig:M2DTMNOIA}	 		
\end{figure}

\begin{figure}[h]
	\begin{subfigure}{0.135\textwidth}
		\centering
		\includegraphics[width=\linewidth,trim={2.6cm 0.5cm 2.6cm 0.5cm},clip]{M2DTM0}
		\caption*{$t=0$}
	\end{subfigure}
	\begin{subfigure}{0.135\textwidth}
		\centering
		\includegraphics[width=\linewidth,trim={2.6cm 0.5cm 2.6cm 0.5cm},clip]{M2DTM1}
		\caption*{$t=2$}
	\end{subfigure}
	\begin{subfigure}{0.135\textwidth}
		\centering
		\includegraphics[width=\linewidth,trim={2.6cm 0.5cm 2.6cm 0.5cm},clip]{M2DTM2}
		\caption*{$t=4$}
	\end{subfigure}
	\begin{subfigure}{0.135\textwidth}
		\centering
		\includegraphics[width=\linewidth,trim={2.6cm 0.5cm 2.6cm 0.5cm},clip]{M2DTM3}
		\caption*{$t=6$}
	\end{subfigure}
	\begin{subfigure}{0.135\textwidth}
		\centering
		\includegraphics[width=\linewidth,trim={2.6cm 0.3cm 2.6cm 0.3cm},clip]{M2DTM4}
		\caption*{$t=8$}
	\end{subfigure}
	\begin{subfigure}{0.135\textwidth}
		\centering
		\includegraphics[width=\linewidth,trim={2.6cm 0.5cm 2.6cm 0.5cm},clip]{M2DTM5}
		\caption*{$t=10$}
	\end{subfigure}
	\begin{subfigure}{0.135\textwidth}
		\centering
		\includegraphics[width=\linewidth,trim={2.6cm 0.5cm 2.6cm 0.5cm},clip]{M2DTM6}
		\caption*{$t=12$}
	\end{subfigure}
	\begin{subfigure}{0.135\textwidth}
		\centering
		\includegraphics[width=\linewidth,trim={2.6cm 0.5cm 2.6cm 0.5cm},clip]{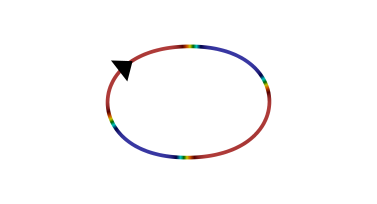}
		\caption*{$t=14$}
	\end{subfigure}
	\begin{subfigure}{0.135\textwidth}
		\centering
		\includegraphics[width=\linewidth,trim={2.6cm 0.5cm 2.6cm 0.5cm},clip]{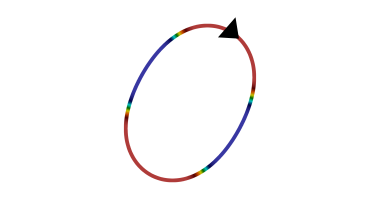}
		\caption*{$t=16$}
	\end{subfigure}
	\begin{subfigure}{0.135\textwidth}
		\centering
		\includegraphics[width=\linewidth,trim={2.6cm 0.5cm 2.6cm 0.5cm},clip]{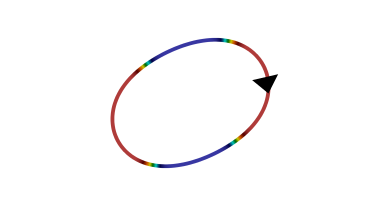}
		\caption*{$t=18$}
	\end{subfigure}
	\begin{subfigure}{0.135\textwidth}
		\centering
		\includegraphics[width=\linewidth,trim={2.6cm 0.5cm 2.6cm 0.5cm},clip]{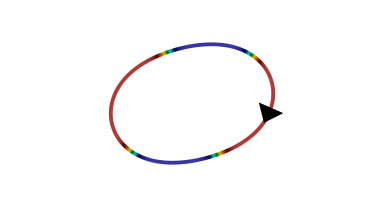}
		\caption*{$t=20$}
	\end{subfigure}
	\begin{subfigure}{0.135\textwidth}
		\centering
		\includegraphics[width=\linewidth,trim={2.6cm 0.5cm 2.6cm 0.5cm},clip]{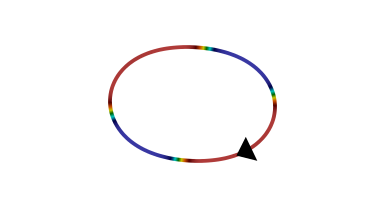}
		\caption*{$t=22$}
	\end{subfigure}
	\begin{subfigure}{0.135\textwidth}
		\centering
		\includegraphics[width=\linewidth,trim={2.6cm 0.5cm 2.6cm 0.5cm},clip]{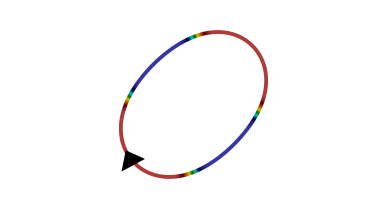}
		\caption*{$t=24$}
	\end{subfigure}
	\begin{subfigure}{0.135\textwidth}
		\centering
		\includegraphics[width=\linewidth,trim={2.6cm 0.5cm 2.6cm 0.5cm},clip]{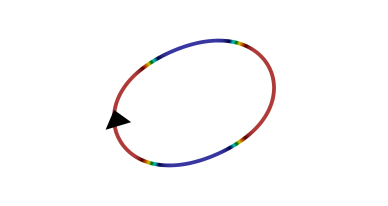}
		\caption*{$t=26$}
	\end{subfigure}
	\caption{Tumbling motion of a multicomponent vesicle under N/N shear flow with $\mathcal{W}i=0$.}	
	\label{fig:M2DTMNN}	 		
\end{figure}

\begin{figure}[h]
	\begin{subfigure}{0.135\textwidth}
		\centering
		\includegraphics[width=\linewidth,trim={2.6cm 0.5cm 2.6cm 0.5cm},clip]{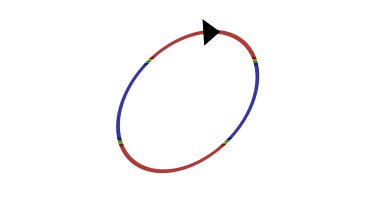}
		\caption*{$t=0$}
	\end{subfigure}
	\begin{subfigure}{0.135\textwidth}
		\centering
		\includegraphics[width=\linewidth,trim={2.6cm 0.5cm 2.6cm 0.5cm},clip]{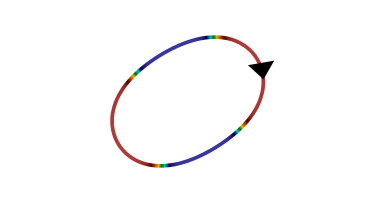}
		\caption*{$t=2$}
	\end{subfigure}
	\begin{subfigure}{0.135\textwidth}
		\centering
		\includegraphics[width=\linewidth,trim={2.6cm 0.5cm 2.6cm 0.5cm},clip]{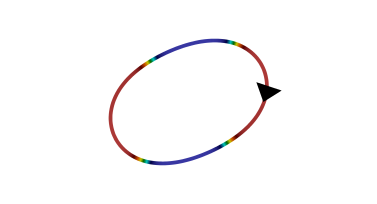}
		\caption*{$t=4$}
	\end{subfigure}
	\begin{subfigure}{0.135\textwidth}
		\centering
		\includegraphics[width=\linewidth,trim={2.6cm 0.5cm 2.6cm 0.5cm},clip]{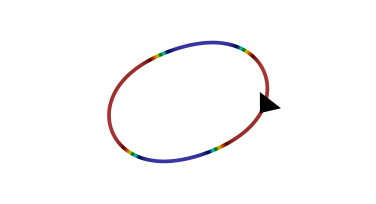}
		\caption*{$t=6$}
	\end{subfigure}
	\begin{subfigure}{0.135\textwidth}
		\centering
		\includegraphics[width=\linewidth,trim={2.6cm 0.5cm 2.6cm 0.5cm},clip]{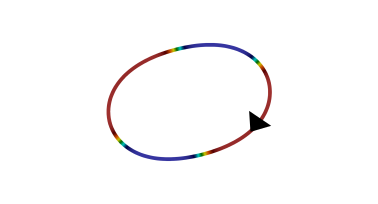}
		\caption*{$t=8$}
	\end{subfigure}
	\begin{subfigure}{0.135\textwidth}
		\centering
		\includegraphics[width=\linewidth,trim={2.6cm 0.5cm 2.6cm 0.5cm},clip]{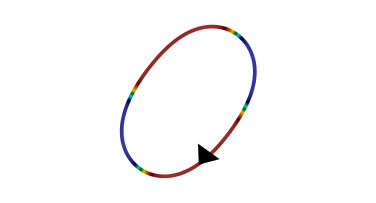}
		\caption*{$t=10$}
	\end{subfigure}
	\begin{subfigure}{0.135\textwidth}
		\centering
		\includegraphics[width=\linewidth,trim={2.6cm 0.5cm 2.6cm 0.5cm},clip]{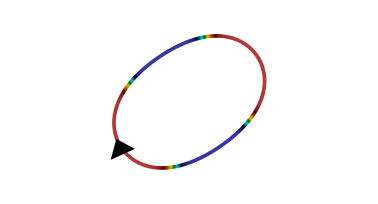}
		\caption*{$t=12$}
	\end{subfigure}
	\begin{subfigure}{0.135\textwidth}
		\centering
		\includegraphics[width=\linewidth,trim={2.6cm 0.5cm 2.6cm 0.5cm},clip]{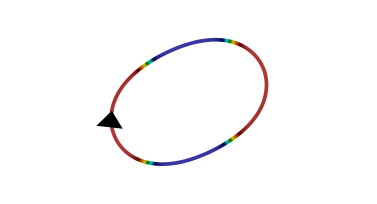}
		\caption*{$t=14$}
	\end{subfigure}
	\begin{subfigure}{0.135\textwidth}
		\centering
		\includegraphics[width=\linewidth,trim={2.6cm 0.5cm 2.6cm 0.5cm},clip]{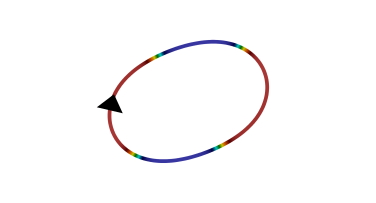}
		\caption*{$t=16$}
	\end{subfigure}
	\begin{subfigure}{0.135\textwidth}
		\centering
		\includegraphics[width=\linewidth,trim={2.6cm 0.5cm 2.6cm 0.5cm},clip]{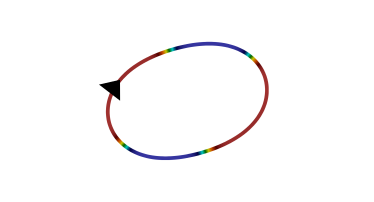}
		\caption*{$t=18$}
	\end{subfigure}
	\begin{subfigure}{0.135\textwidth}
		\centering
		\includegraphics[width=\linewidth,trim={2.6cm 0.5cm 2.6cm 0.5cm},clip]{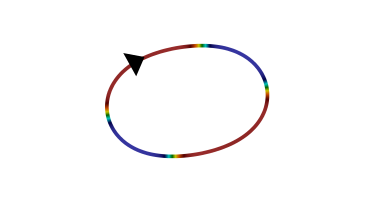}
		\caption*{$t=20$}
	\end{subfigure}
	\begin{subfigure}{0.135\textwidth}
		\centering
		\includegraphics[width=\linewidth,trim={2.6cm 0.5cm 2.6cm 0.5cm},clip]{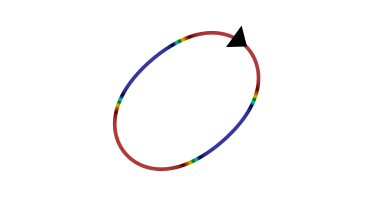}
		\caption*{$t=22$}
	\end{subfigure}
	\begin{subfigure}{0.135\textwidth}
		\centering
		\includegraphics[width=\linewidth,trim={2.6cm 0.5cm 2.6cm 0.5cm},clip]{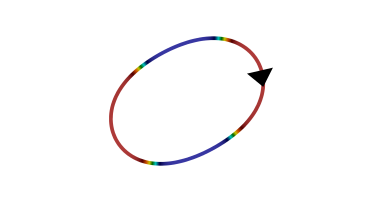}
		\caption*{$t=24$}
	\end{subfigure}
	\begin{subfigure}{0.135\textwidth}
		\centering
		\includegraphics[width=\linewidth,trim={2.6cm 0.5cm 2.6cm 0.5cm},clip]{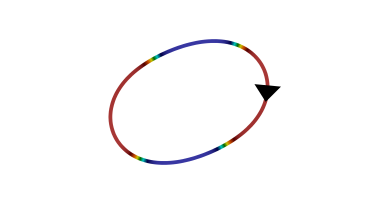}
		\caption*{$t=26$}
	\end{subfigure}
	\caption{Swinging motion of a multicomponent vesicle under N/O shear flow with $\mathcal{W}i=0.5$.}	
	\label{fig:M2DTMNOWi0.5}	 		
\end{figure}

\begin{figure}[H]
	\begin{subfigure}{0.135\textwidth}
		\centering
		\includegraphics[width=\linewidth,trim={2.6cm 0.5cm 2.6cm 0.5cm},clip]{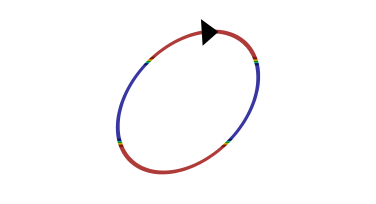}
		\caption*{$t=0$}
	\end{subfigure}
	\begin{subfigure}{0.135\textwidth}
		\centering
		\includegraphics[width=\linewidth,trim={2.6cm 0.5cm 2.6cm 0.5cm},clip]{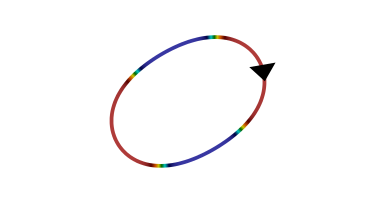}
		\caption*{$t=2$}
	\end{subfigure}
	\begin{subfigure}{0.135\textwidth}
		\centering
		\includegraphics[width=\linewidth,trim={2.6cm 0.5cm 2.6cm 0.5cm},clip]{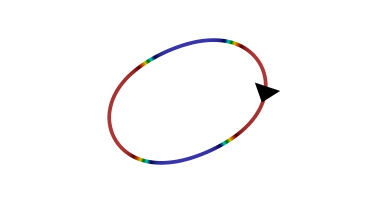}
		\caption*{$t=4$}
	\end{subfigure}
	\begin{subfigure}{0.135\textwidth}
		\centering
		\includegraphics[width=\linewidth,trim={2.6cm 0.5cm 2.6cm 0.5cm},clip]{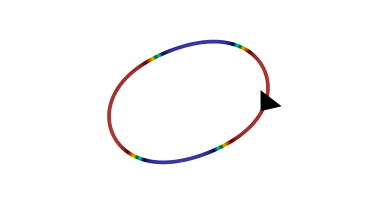}
		\caption*{$t=6$}
	\end{subfigure}
	\begin{subfigure}{0.135\textwidth}
		\centering
		\includegraphics[width=\linewidth,trim={2.6cm 0.5cm 2.6cm 0.5cm},clip]{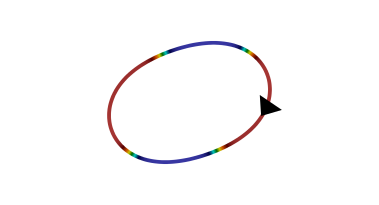}
		\caption*{$t=8$}
	\end{subfigure}
	\begin{subfigure}{0.135\textwidth}
		\centering
		\includegraphics[width=\linewidth,trim={2.6cm 0.5cm 2.6cm 0.5cm},clip]{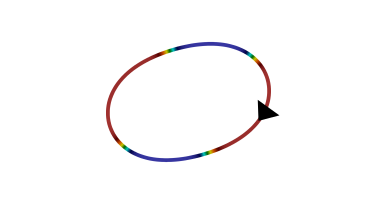}
		\caption*{$t=10$}
	\end{subfigure}
	\begin{subfigure}{0.135\textwidth}
		\centering
		\includegraphics[width=\linewidth,trim={2.6cm 0.5cm 2.6cm 0.5cm},clip]{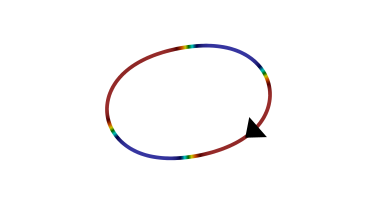}
		\caption*{$t=12$}
	\end{subfigure}
	\begin{subfigure}{0.135\textwidth}
		\centering
		\includegraphics[width=\linewidth,trim={2.6cm 0.5cm 2.6cm 0.5cm},clip]{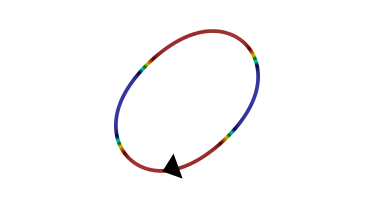}
		\caption*{$t=14$}
	\end{subfigure}
	\begin{subfigure}{0.135\textwidth}
		\centering
		\includegraphics[width=\linewidth,trim={2.6cm 0.5cm 2.6cm 0.5cm},clip]{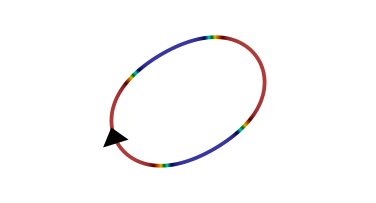}
		\caption*{$t=16$}
	\end{subfigure}
	\begin{subfigure}{0.135\textwidth}
		\centering
		\includegraphics[width=\linewidth,trim={2.6cm 0.5cm 2.6cm 0.5cm},clip]{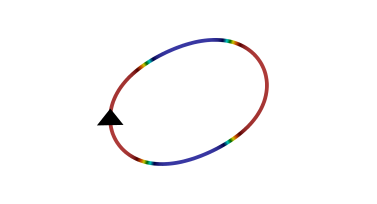}
		\caption*{$t=18$}
	\end{subfigure}
	\begin{subfigure}{0.135\textwidth}
		\centering
		\includegraphics[width=\linewidth,trim={2.6cm 0.5cm 2.6cm 0.5cm},clip]{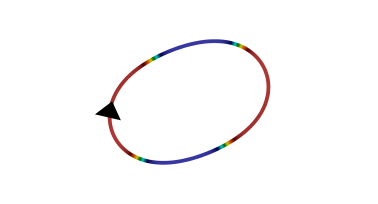}
		\caption*{$t=20$}
	\end{subfigure}
	\begin{subfigure}{0.135\textwidth}
		\centering
		\includegraphics[width=\linewidth,trim={2.6cm 0.5cm 2.6cm 0.5cm},clip]{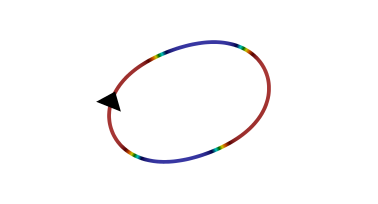}
		\caption*{$t=22$}
	\end{subfigure}
	\begin{subfigure}{0.135\textwidth}
		\centering
		\includegraphics[width=\linewidth,trim={2.6cm 0.5cm 2.6cm 0.5cm},clip]{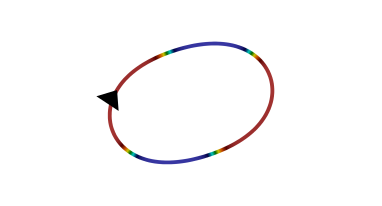}
		\caption*{$t=24$}
	\end{subfigure}
	\begin{subfigure}{0.135\textwidth}
		\centering
		\includegraphics[width=\linewidth,trim={2.6cm 0.5cm 2.6cm 0.5cm},clip]{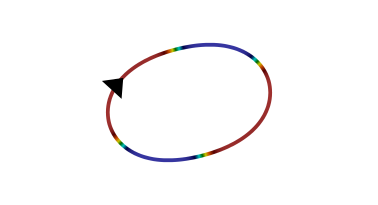}
		\caption*{$t=26$}
	\end{subfigure}
	\caption{Swinging motion of a multicomponent vesicle under N/O shear flow with $\mathcal{W}i=1$.}	
	\label{fig:M2DTMNOWi1}	 		
\end{figure}

In this section, we investigate the effect of viscoelasticity in N/O fluid on a multicomponent vesicle that initially exhibits tumbling behavior in an N/N fluid with $\mathcal{W}i = 0$. The simulation parameters are consistent with the reference swinging case $\mathcal{C}_4$, as described in Section~\ref{M2DNN}. We vary $\mathcal{W}i$ with values set to $0.1$, $0.5$, and $1$. The evolution of the inclination angle for these three cases, along with the reference N/N tumbling case, is shown in \fref{fig:M2DTMNOIA}. The blue dashed line represents the reference N/N case, which exhibits periodic tumbling motion driven by bending rigidity variation. When viscoelasticity is introduced, corresponding to $\mathcal{W}i = 0.1$, $0.5$, and $1$ (shown as red, yellow, and green solid lines, respectively), all the vesicles transition to periodic swinging motion. From \fref{fig:M2DTMNOIA}, we find that the tumbling motion of the multicomponent vesicle is highly sensitive to viscoelastic effects, as even a minor viscoelasticity ($\mathcal{W}i = 0.1$) is sufficient to reduce the tumbling motion to swinging motion. As the viscoelastic effect increases further, the swinging motion exhibits a longer period, while the amplitude remains relatively constant. In addition, the phase treading periods are extended, consistent with the longer swinging periods observed as the viscoelastic effect is strengthened. These results further support our statement that the viscoelastic effect stabilizes the hydrodynamics of multicomponent vesicles by dampening fluctuations. Consequently, it alters the vesicle’s hydrodynamic behavior, modifying both the oscillation period and, ultimately, the type of motion exhibited. We present in \fref{fig:M2DTMNN}, \fref{fig:M2DTMNOWi0.5}, and \fref{fig:M2DTMNOWi1} the hydrodynamics and phase evolution of multicomponent vesicles with $\mathcal{W}i = 0$, $0.5$, and $1$, respectively.

\subsection{Multicomponent vesicle dynamics in Poiseuille flow} \label{PF}
\begin{figure}[h]
	\centering
	\includegraphics[scale=0.5]{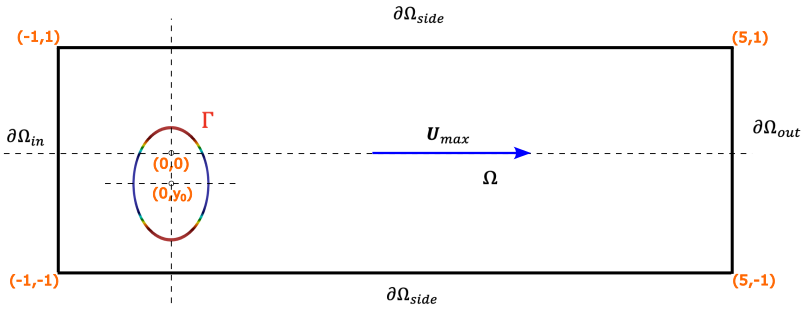}
	\caption{Problem setup for multicomponent vesicle in Poiseuille flow.}	
	\label{fig:M2DPFinitial}	 		
\end{figure}

We simulate the dynamics of a multicomponent vesicle in plane Poiseuille flow and investigate the effects of viscoelasticity on its behavior. The problem setup is illustrated in \fref{fig:M2DPFinitial}. The computational domain is $(-1,5) \times (-1,1)$ with uniform spatial discretization of $288 \times 96$. The vesicle is initialized at $(0,y_0)$ and the boundary conditions are given as follows. On the left boundary, we consider an inlet and impose:
\begin{align*}
	\bm{U}_{bc} = (U_{max}(1-(\frac{y}{W})^2),0)
\end{align*}
where $U_{max}=4$ is the maximum velocity located at the center line $y=0$. $W=1$ is the half width of the channel. On the right boundary, we consider an outlet with the $y$-component of the velocity to be zero, $u_y=0$. On the upper and lower boundaries, we consider a no-slip wall condition and have $\bm{U}_{bc}=(0,0)$. We also consider Dirichlet boundary conditions $f=0$, $\phi=-1$, and $\lambda=0$ on all the boundaries. We assume a reduced area of $R_v = 0.9412$, which results in $b_E = \frac{1}{2}$ and $a_E = \frac{1}{3}$, giving $\mathcal{A}_0 = 0.5236$. The phase initialization is presegregated, with the interface located at $y = \pm\frac{1}{3} + y_0$, where phase B constitutes approximately $\frac{1}{3}$ of the total membrane mass. The bending rigidities of phase A and phase B are set to $b_A = 1$ and $b_B = 0.1$, respectively. The Reynolds number is taken as $Re = 1 \times 10^{-2}$. A large Capillary number $Ca = 10^4$ is used, indicating that stress dominates the vesicle dynamics, consistent with the literature \cite{PhysRevLett.103.188101,Gannon2021,VALIZADEH2022114191,WEN2024117390}. We set $\alpha = 1$ so that the bending energy and line energy contribute comparably to the distribution of phases A and B. The Peclet number is fixed at $Pe = 1$, and the Cahn number is set as $Cn^2 = 0.01$ for all Poiseuille flow simulations.
 \par
\begin{figure}[H]
	\centering
	\begin{subfigure}{0.7\textwidth}
		\centering
		\includegraphics[width=\linewidth,trim={0cm 8cm 0cm 0cm},clip]{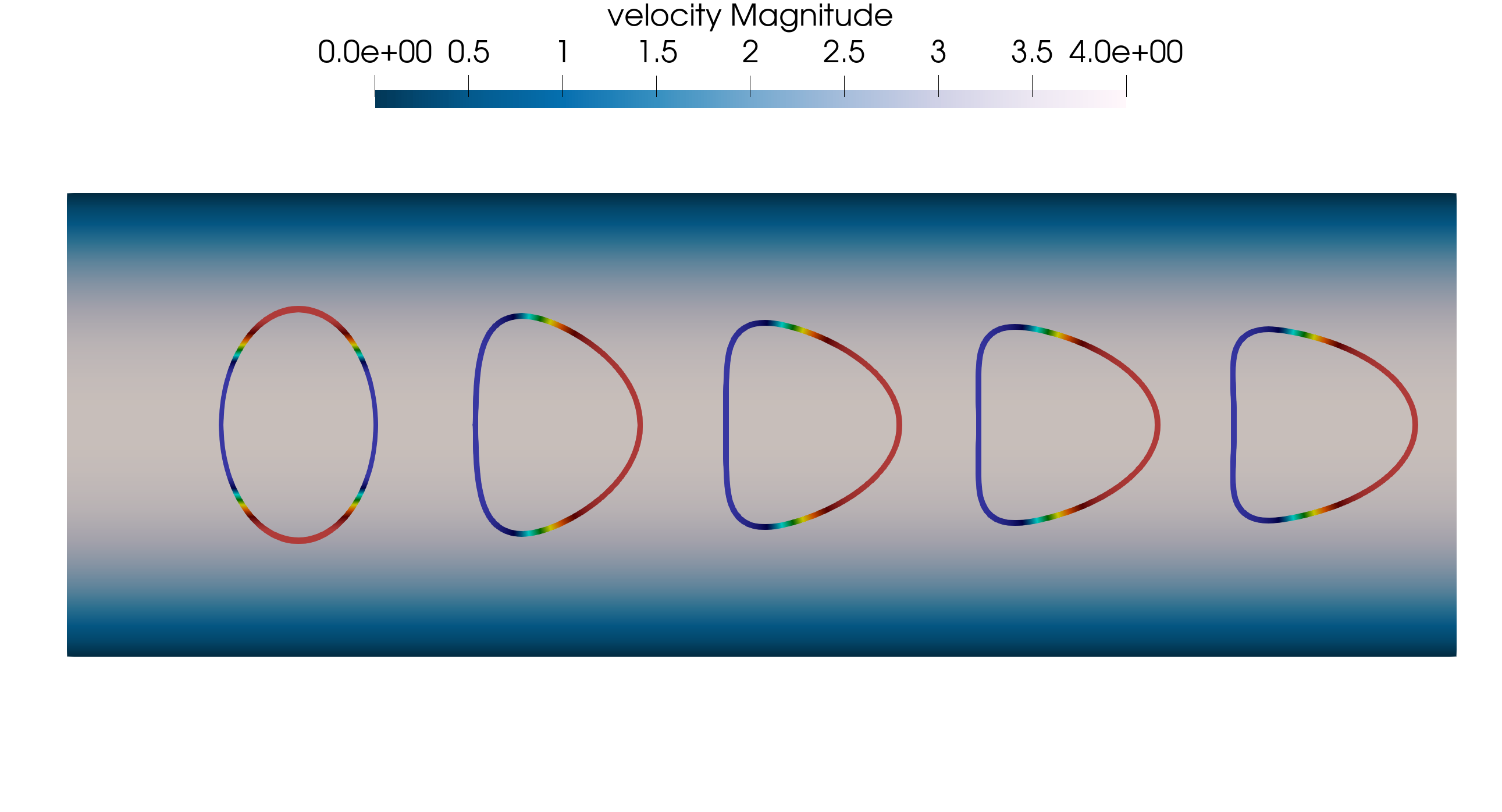}
		\caption{N/N fluid flow, $\mathcal{W}i=0$}
	\end{subfigure}
	\begin{subfigure}{0.7\textwidth}
		\centering
		\includegraphics[width=\linewidth,trim={0cm 9cm 0cm 10cm},clip]{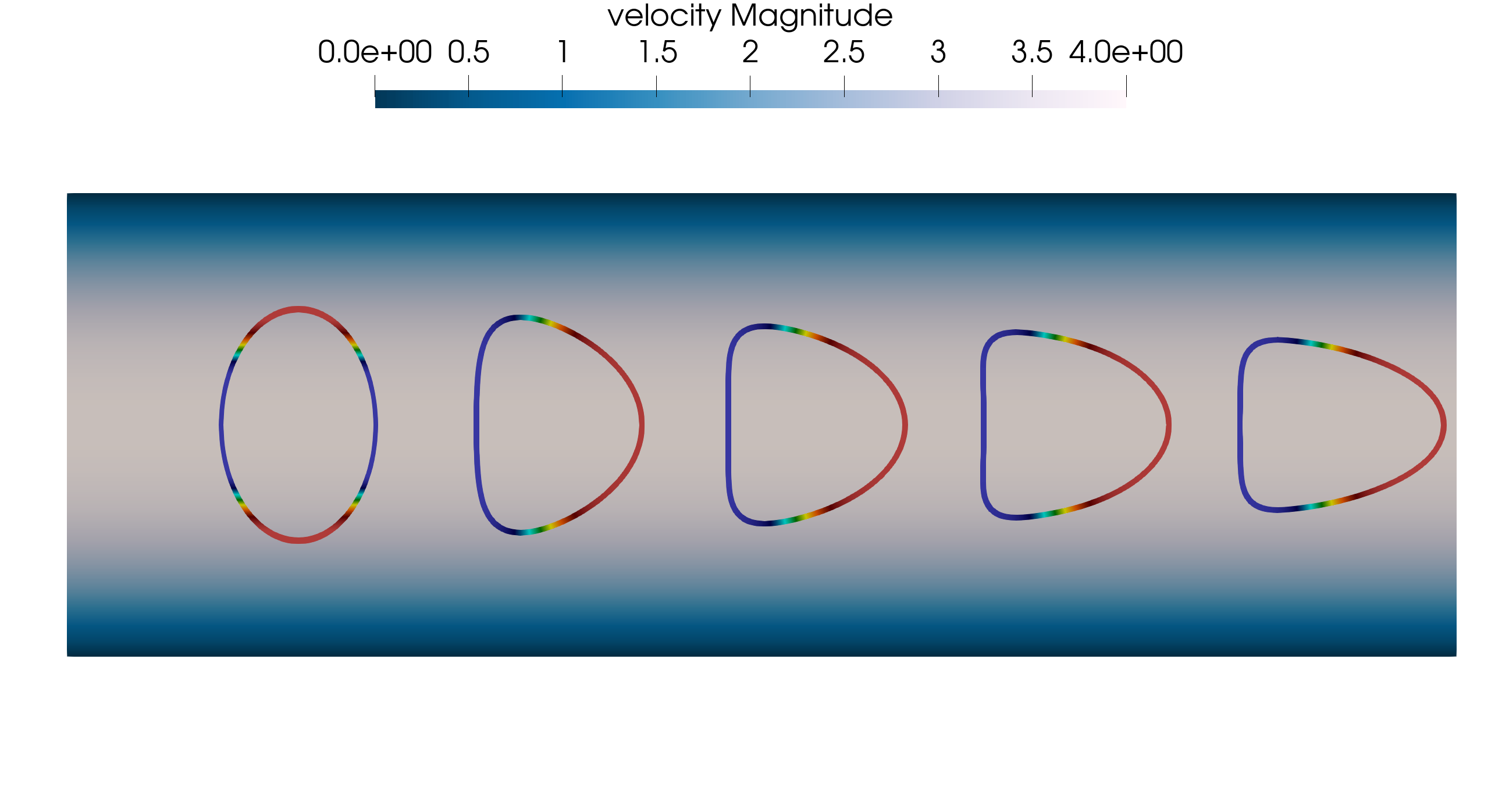}
		\caption{N/O fluid flow, $\mathcal{W}i=0.5$}
	\end{subfigure}
	\begin{subfigure}{0.7\textwidth}
		\centering
		\includegraphics[width=\linewidth,trim={0cm 9cm 0cm 10cm},clip]{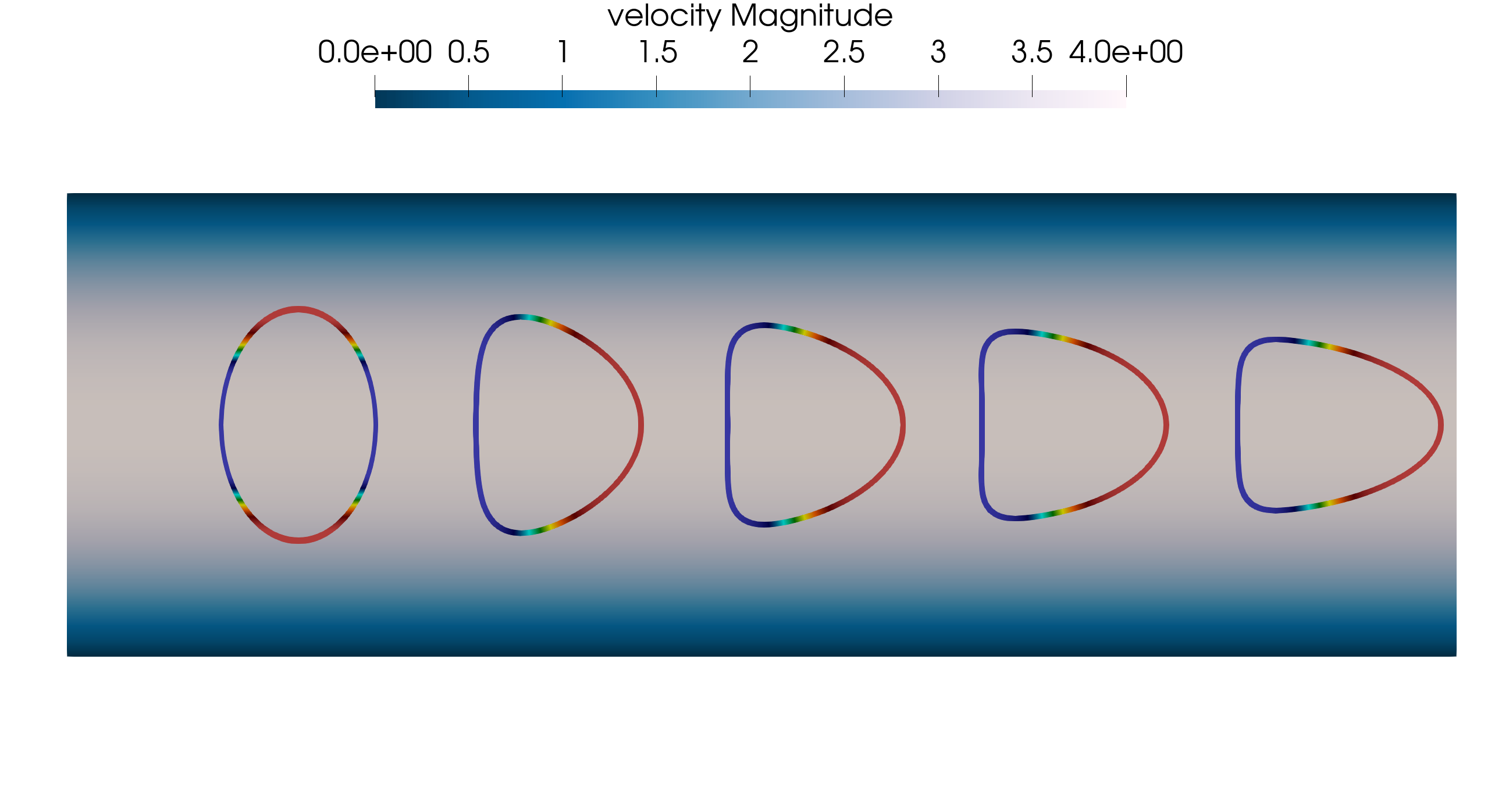}
		\caption{N/O fluid flow, $\mathcal{W}i=1$}
	\end{subfigure}       
	\caption{Time evolution of a multicomponent vesicle in Poiseuille flow with $y_0 = 0$. The configurations are shown at $t = 0$, $0.3$, $0.6$, $0.9$, and $1.2$. (a) corresponds to the N/N fluid case with $\mathcal{W}i = 0$. (b) shows the N/O fluid case with $\mathcal{W}i = 0.5$. (c) presents the N/O fluid case with $\mathcal{W}i = 1$.}	
	\label{fig:PFceter}	 		
\end{figure}

First, we investigate cases where the vesicle is initially centered, i.e., $y_0 = 0$. We study the dynamics of a multicomponent vesicle in N/N Poiseuille flow and then evaluate the viscoelastic effect by examining the corresponding N/O Poiseuille flow. For the N/N case, the evolution of membrane deformation and phases on membrane surface is shown in \fref{fig:PFceter}.(a). Upon reaching a stable state, the vesicle adopts a ``bullet'' shape, consistent with findings in the literature \cite{WEN2024117390,VALIZADEH2022114191,Gannon2021,Bartezzaghi2017,ONG2020109277,PhysRevE.84.011902,PhysRevE.89.042709} and previous experimental observations \cite{Skalak717,GUIDO2009751}. Phase A and phase B merge from four presegregated domains into two larger domains due to the coarsening effect driven by line tension at the interface between the two phases on the membrane surface. Eventually, phase B (soft) migrates toward the front where the curvature is highest, while phase A (stiff) occupies the regions with relatively lower curvature. \par
\begin{figure}[h]
	\centering
	\begin{subfigure}{0.2\textwidth}
		\centering
		\includegraphics[width=\linewidth,trim={18cm 5cm 18cm 5cm},clip]{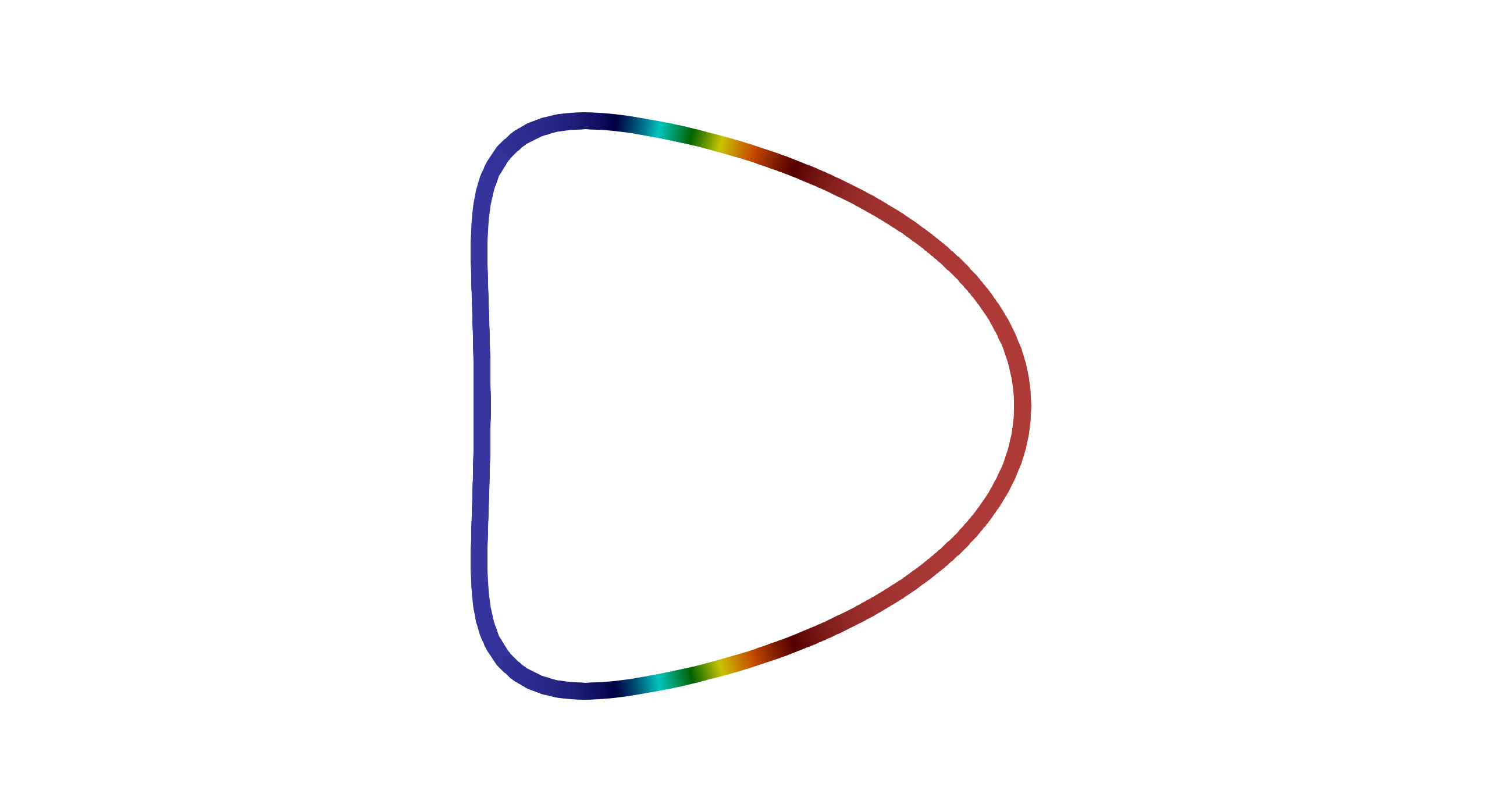}
		\caption{N/N fluid flow, $\mathcal{W}i=0$}
	\end{subfigure}\quad
	\begin{subfigure}{0.2\textwidth}
		\centering
		\includegraphics[width=\linewidth,trim={18cm 5cm 18cm 5cm},clip]{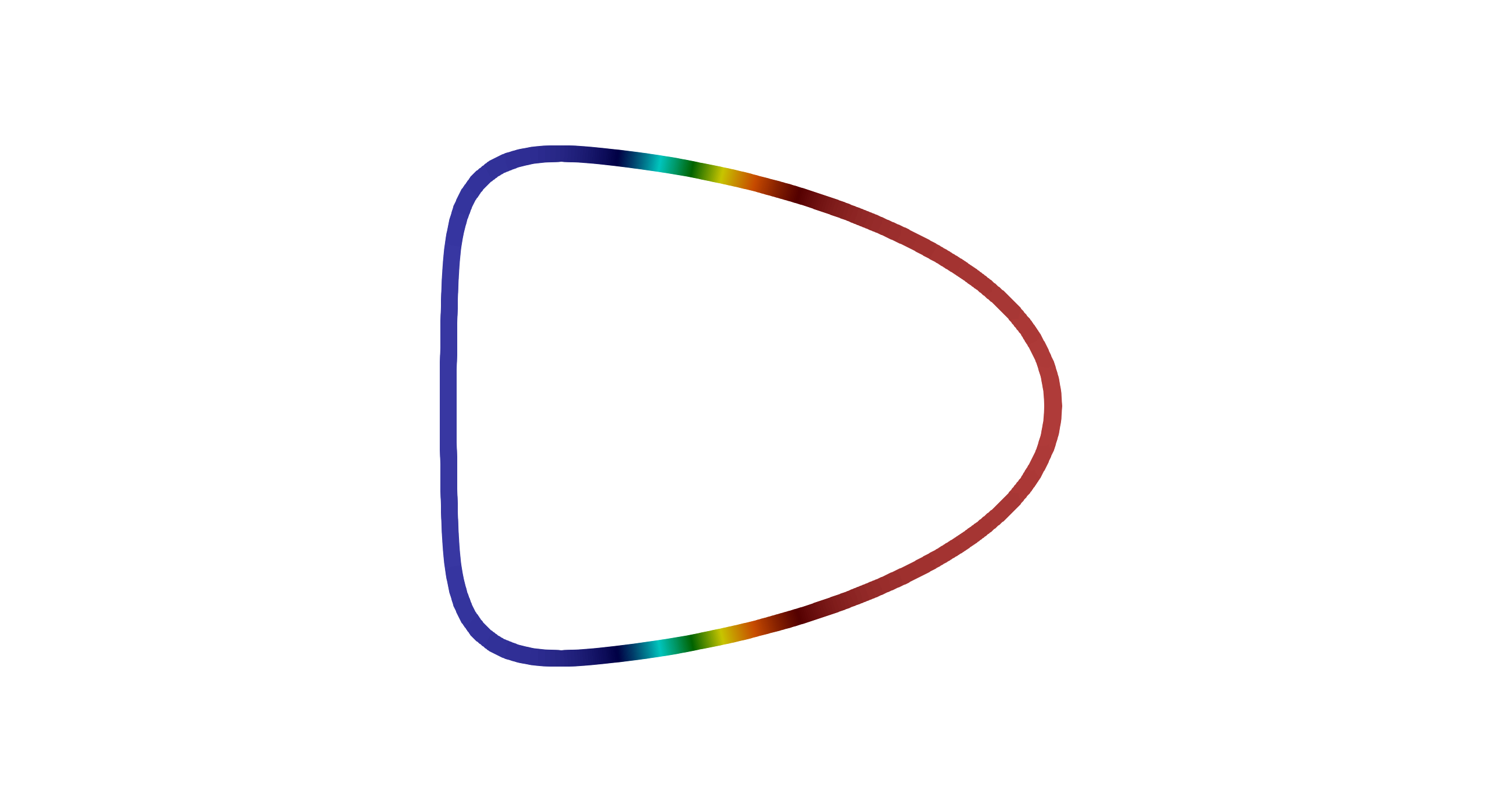}
		\caption{N/O fluid flow, $\mathcal{W}i=0.5$}
	\end{subfigure}\quad
	\begin{subfigure}{0.2\textwidth}
		\centering
		\includegraphics[width=\linewidth,trim={18cm 5cm 18cm 5cm},clip]{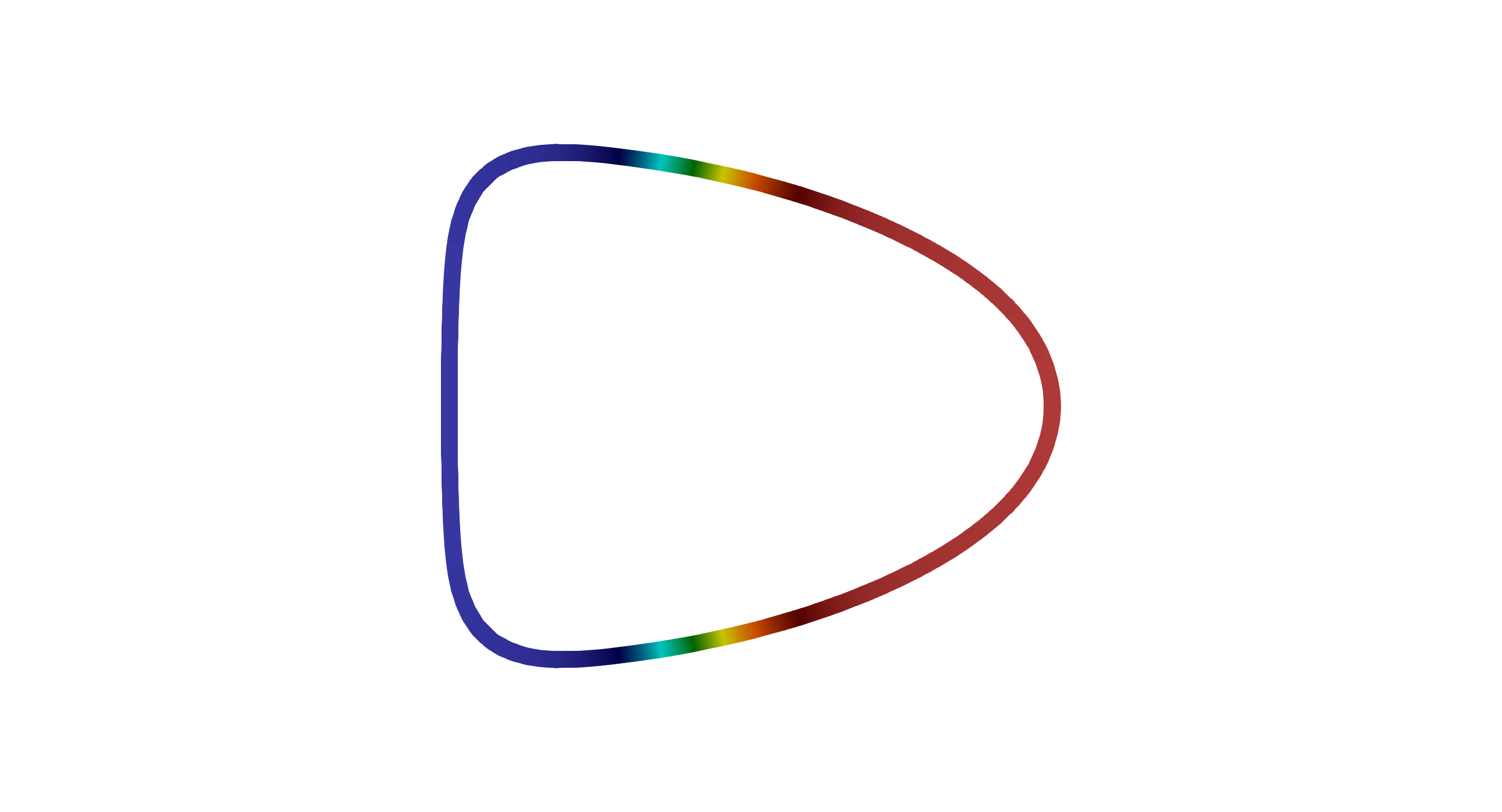}
		\caption{N/O fluid flow, $\mathcal{W}i=1$}
	\end{subfigure}
	\caption{Equivalent status ($t = 1.2$) of multicomponent vesicle in Poiseuille flow under different intensities of viscoelastic effect with symmetric initialization.}	
	\label{fig:PFy0F}	 		
\end{figure}
\begin{figure}[h]
	\centering
	\begin{subfigure}{0.45\textwidth}
		\centering
		\includegraphics[width=\linewidth,trim={0cm 0cm 0cm 0cm},clip]{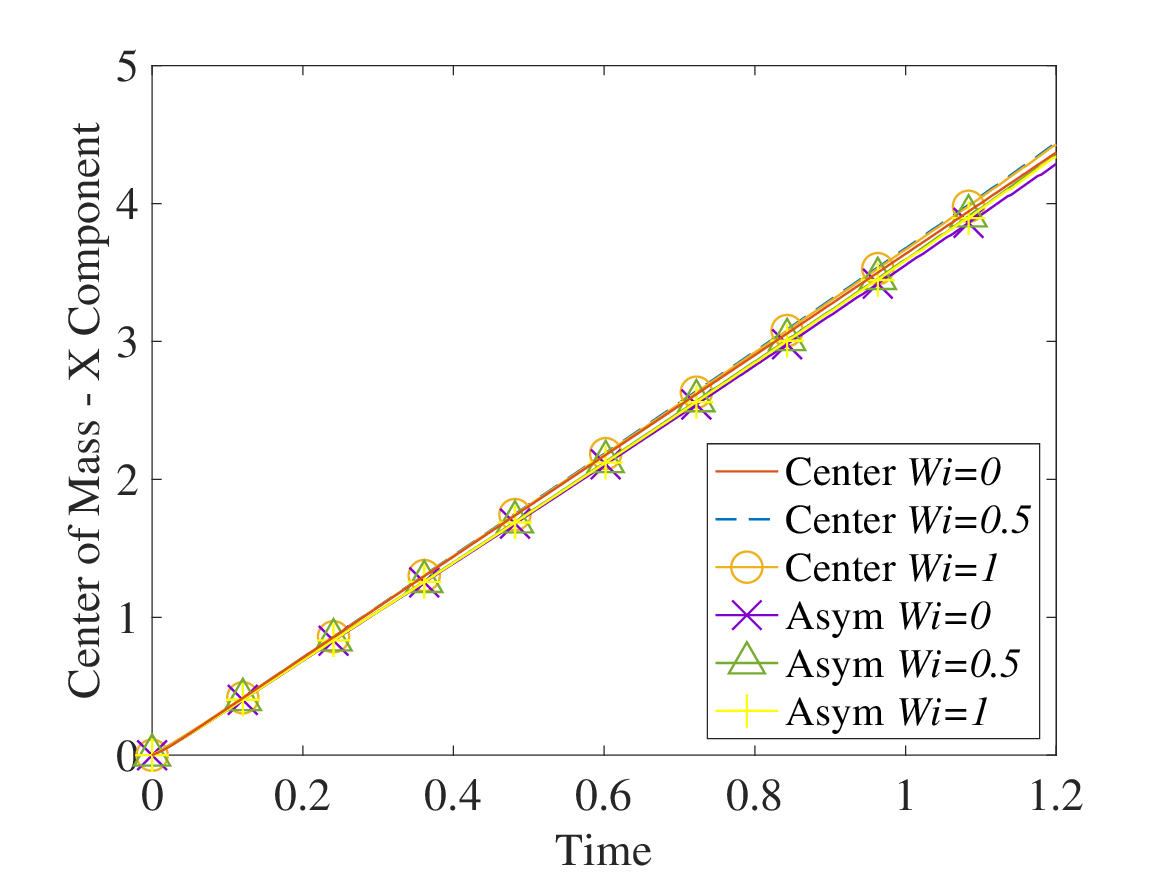}
		\caption{Evolution of vesicle mass center in $x$-direction}
	\end{subfigure}\quad
	\begin{subfigure}{0.45\textwidth}
		\centering
		\includegraphics[width=\linewidth,trim={0cm 0cm 0cm 0cm},clip]{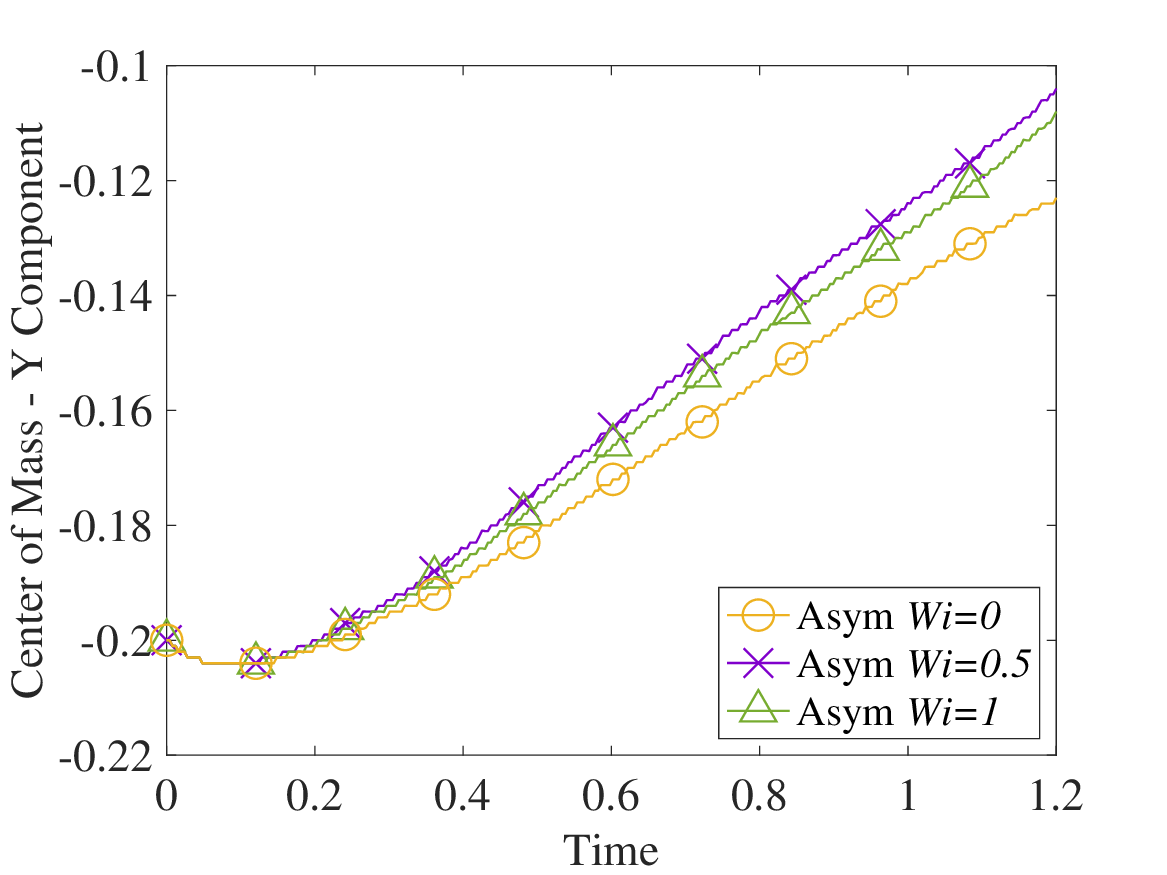}
		\caption{Evolution of vesicle mass center in $y$-direction}
	\end{subfigure}
	\caption{Multicomponent vesicles in Poiseuille flow. Evolution of the vesicle mass center over time for the cases: centered initialization ($y_0=0$) and asymmetric initialization ($y_0=-0.2$).}	
	\label{fig:PFUxy}	 		
\end{figure}

Our interest is to study the viscoelastic effect on the hydrodynamics and phase evolution of the multicomponent vesicle in Poiseuille flow; therefore, we consider two cases with $\mathcal{W}i = 0.5$ and $1$. The corresponding results are shown in \fref{fig:PFceter}.(b) and \fref{fig:PFceter}.(c), respectively. We also provide the final stable states at $t = 1.2$ for the three cases in \fref{fig:PFy0F}, allowing for a detailed comparison. From \frefs{fig:PFceter}{fig:PFy0F}, we observe that the vesicle becomes more stretched in the $x$-direction, aligned with the flow direction, when the viscoelastic effect is introduced. However, as we further increase the viscoelastic effect from $\mathcal{W}i = 0.5$ to $\mathcal{W}i = 1$, no significant difference is observed. In addition, the phase evolution patterns under viscoelastic effects remain similar compared to the case without viscoelasticity. This result is consistent with our previous statement: since there is no phase treading induced by shear stress in Poiseuille flow, the phases evolve solely in response to curvature and the coarsening effect. In the absence of phase treading, there is no variation in bending rigidity along the membrane, and thus no corresponding energy fluctuations for the Oldroyd-B fluid to damp. As a result, the viscoelastic effect does not alter the type of vesicle motion but instead exerts additional stress that causes the vesicle to deform more in alignment with the Poiseuille flow direction.
We present the evolution of the mass center of the multicomponent vesicle in the $x$-direction to illustrate the vesicle’s translational velocity along the flow, as shown in \fref{fig:PFUxy}.(a). From the plot, it is evident that the vesicle quickly adapts to the flow velocity due to the high imposed $U_{max}$. However, introducing viscoelastic effects into the outer fluid has only a negligible impact on the vesicle’s velocity in the $x$-direction.\par
\begin{figure}[h]
	\centering
	\begin{subfigure}{0.2\textwidth}
		\centering
		\includegraphics[width=\linewidth,trim={18cm 5cm 18cm 5cm},clip]{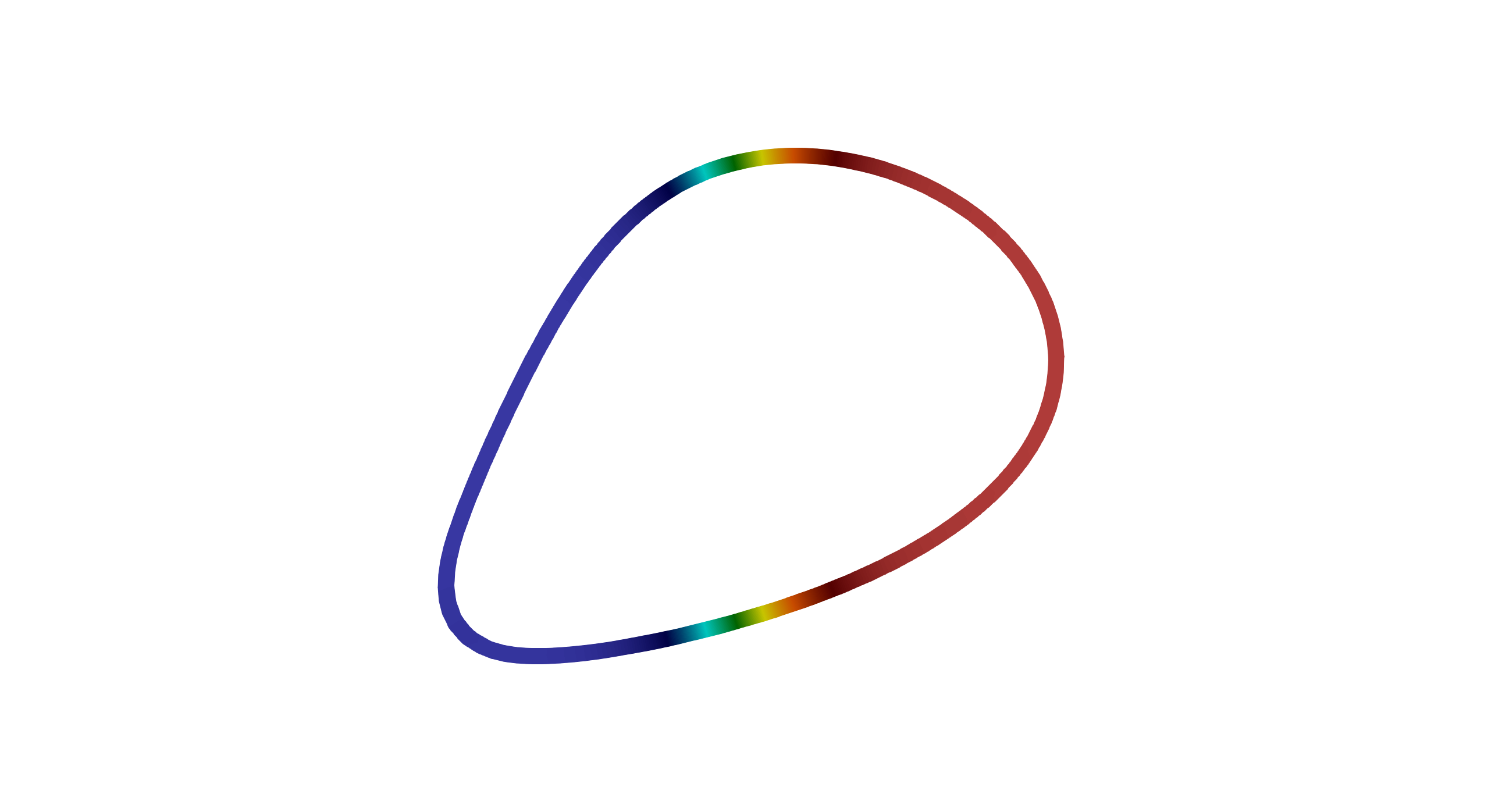}
		\caption{N/N fluid flow, $\mathcal{W}i=0$}
	\end{subfigure}\quad
	\begin{subfigure}{0.2\textwidth}
		\centering
		\includegraphics[width=\linewidth,trim={18cm 5cm 18cm 5cm},clip]{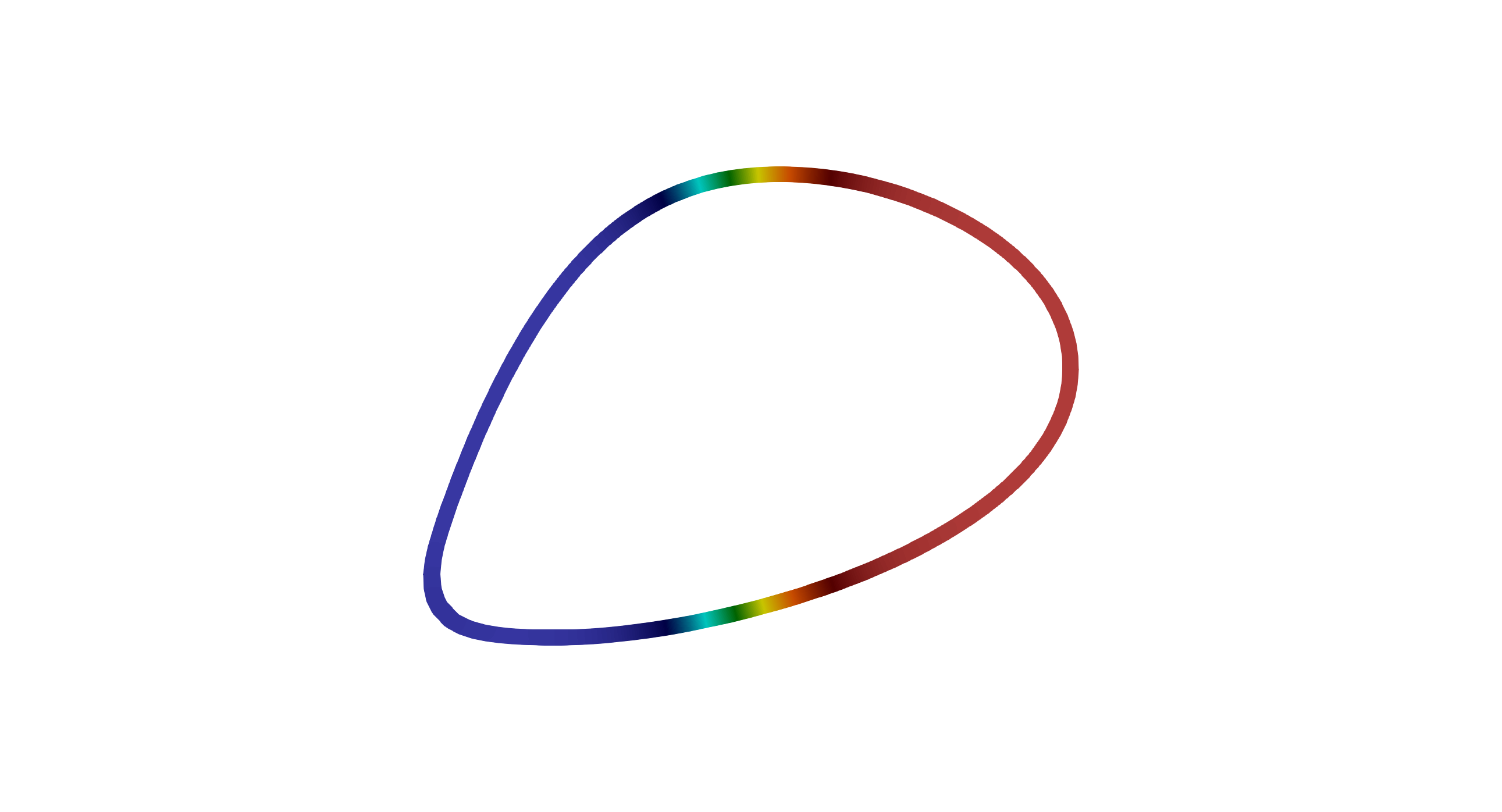}
		\caption{N/O fluid flow, $\mathcal{W}i=0.5$}
	\end{subfigure}\quad
	\begin{subfigure}{0.2\textwidth}
		\centering
		\includegraphics[width=\linewidth,trim={18cm 5cm 18cm 5cm},clip]{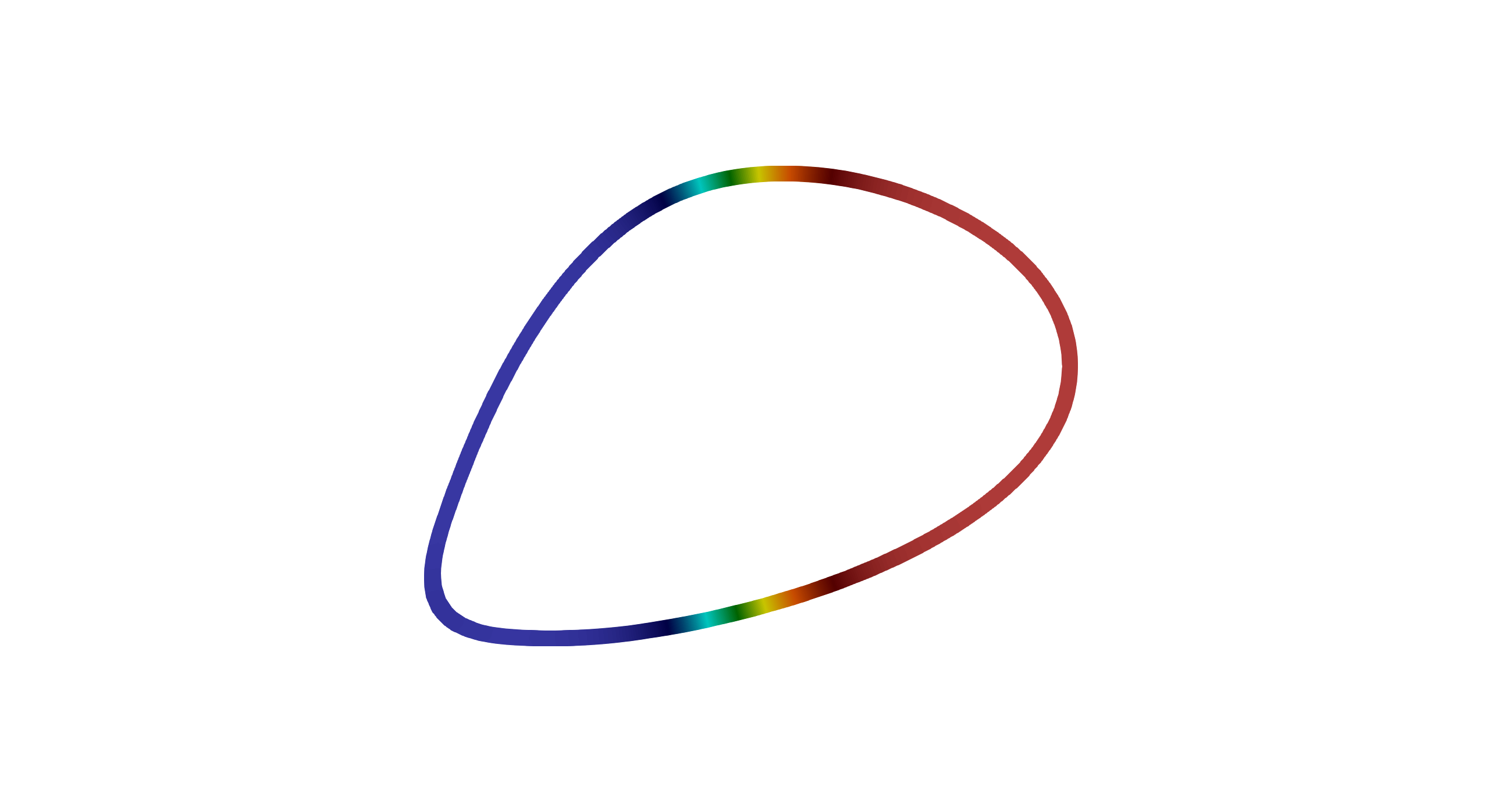}
		\caption{N/O fluid flow, $\mathcal{W}i=1$}
	\end{subfigure}
	\caption{Equivalent status ($t = 1.2$) of multicomponent vesicle in Poiseuille flow under different intensities of viscoelastic effect with asymmetric initialization.}	
	\label{fig:PFasymF}	 		
\end{figure}
\begin{figure}[H]
	\centering
	\begin{subfigure}{0.7\textwidth}
		\centering
		\includegraphics[width=\linewidth,trim={0cm 8cm 0cm 0cm},clip]{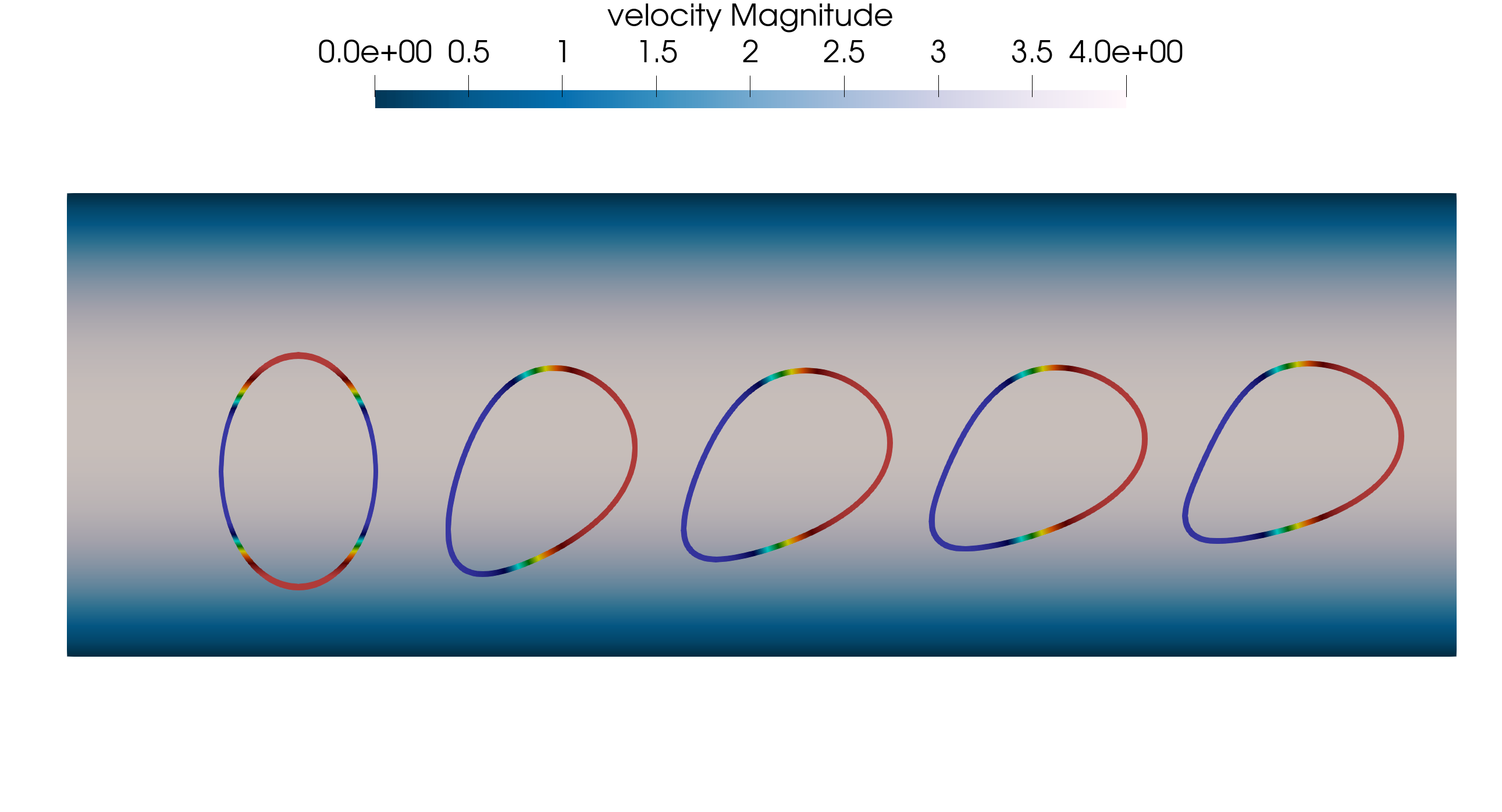}
		\caption{N/N fluid flow, $\mathcal{W}i=0$}
	\end{subfigure}
	\begin{subfigure}{0.7\textwidth}
		\centering
		\includegraphics[width=\linewidth,trim={0cm 9cm 0cm 10cm},clip]{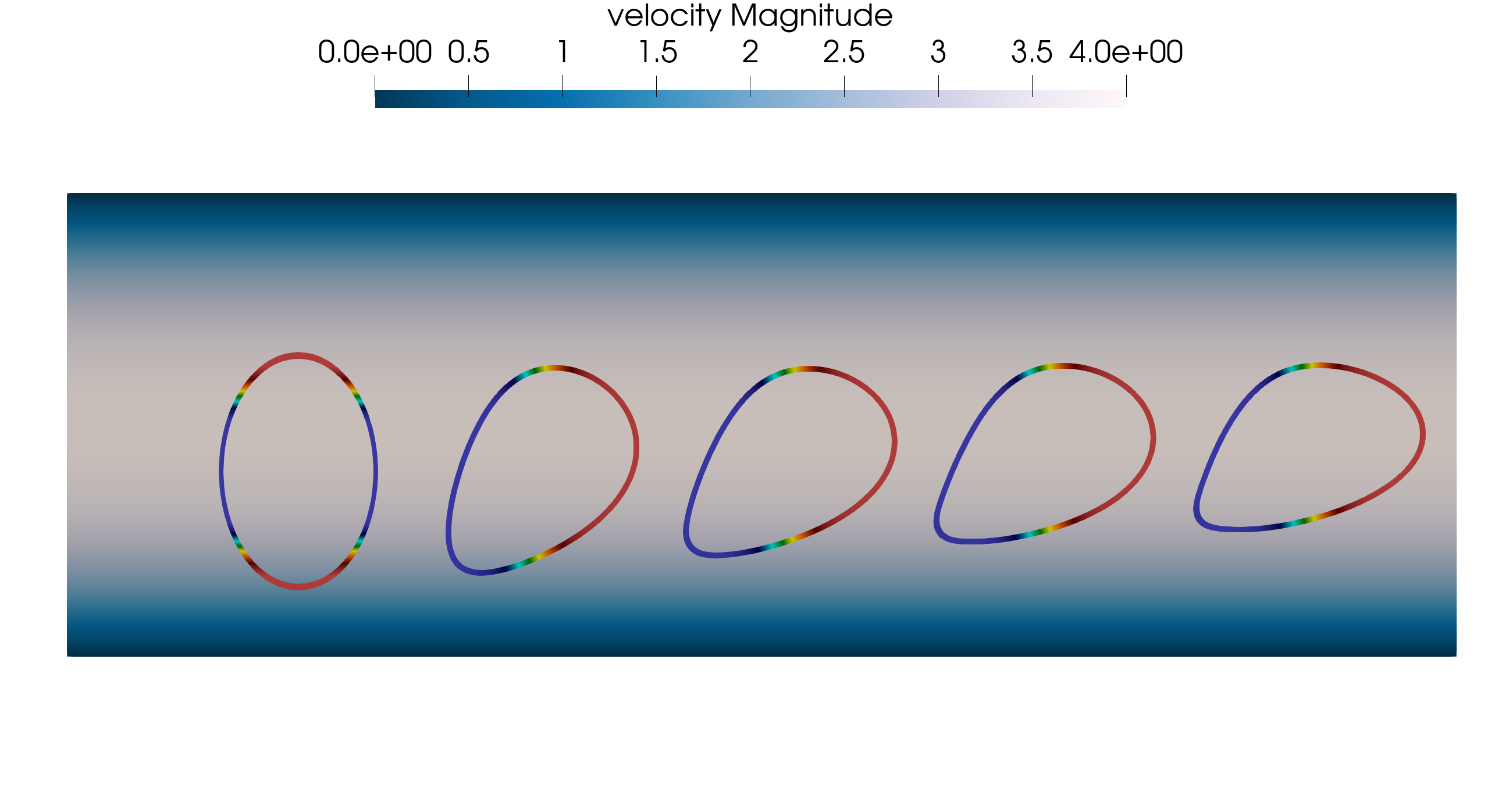}
		\caption{N/O fluid flow, $\mathcal{W}i=0.5$}
	\end{subfigure}
	\begin{subfigure}{0.7\textwidth}
		\centering
		\includegraphics[width=\linewidth,trim={0cm 9cm 0cm 10cm},clip]{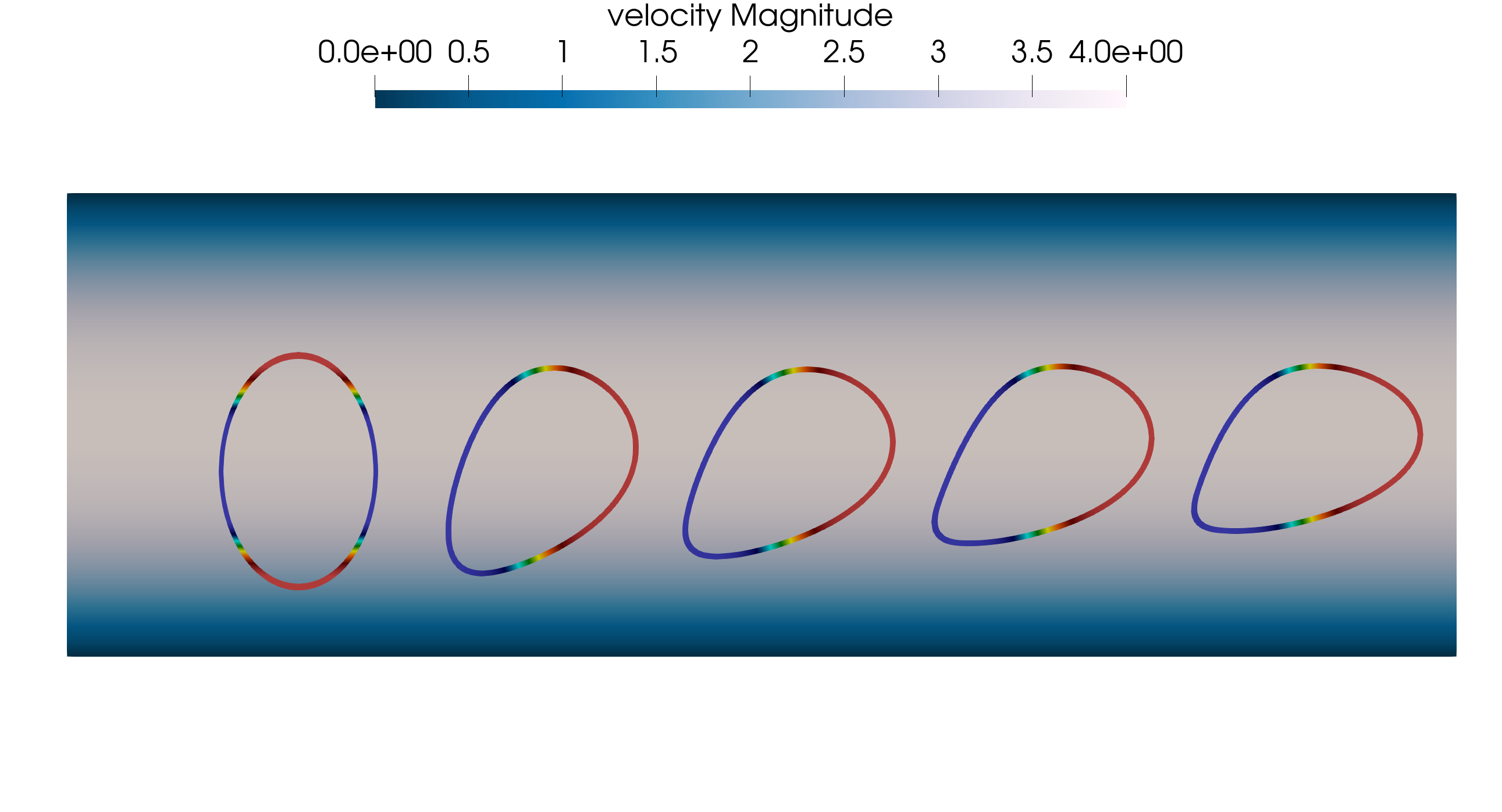}
		\caption{N/O fluid flow, $\mathcal{W}i=1$}
	\end{subfigure}       
	\caption{Time evolution of a multicomponent vesicle in Poiseuille flow with $y_0 = -0.2$. The configurations are shown at $t = 0$, $0.3$, $0.6$, $0.9$, and $1.2$. (a) corresponds to the N/N fluid case with $\mathcal{W}i = 0$. (b) shows the N/O fluid case with $\mathcal{W}i = 0.5$. (c) presents the N/O fluid case with $\mathcal{W}i = 1$.}	
	\label{fig:PFasym}	 		
\end{figure}

We now focus on the case of asymmetric initialization ($y_0 = -0.2$) of the multicomponent vesicle, both with and without viscoelastic effects. As reported in \cite{PhysRevE.84.011902,PhysRevE.89.042709}, a vesicle positioned away from the centerline tends to adopt a ``slipper'' shape, in contrast to the ``bullet'' shape observed in the centered case shown in \frefs{fig:PFceter}{fig:PFy0F}. Furthermore, as discussed in \cite{PhysRevE.89.042709}, the shifted vesicle gradually migrates back toward the centerline under Poiseuille flow. We validate this behavior using our model by initially placing a multicomponent vesicle at $y_0 = -0.2$ in N/N fluid flow with $\mathcal{W}i = 0$. We then investigate the viscoelastic effects by setting $\mathcal{W}i = 0.5$ and $\mathcal{W}i = 1$ in N/O fluid flow. The time evolution of the multicomponent vesicle is shown in \fref{fig:PFasym}, and the final stable configurations at $t = 1.2$ are shown in \fref{fig:PFasymF}. The evolution of the mass center of the multicomponent vesicle in the $x$- and $y$-directions, representing its translational velocity along the flow and migration toward the centerline, are presented in \fref{fig:PFUxy}.(a) and (b), respectively.
From \fref{fig:PFUxy}.(a) and \fref{fig:PFasym}, we observe that the vesicle initially placed at $(0, -0.2)$ migrates toward the centerline under Poiseuille flow, consistent with the findings in \cite{PhysRevE.89.042709}. In addition, we find that the viscoelastic effect slightly accelerates the migration toward the centerline. However, a stronger viscoelastic effect does not necessarily result in a higher velocity in the $y$-direction towards centerline. Our stable-state “slipper” shape, shown in \fref{fig:PFasym}.(a) and \fref{fig:PFasymF}.(a), matches with the shape presented in \cite{PhysRevE.84.011902}, Fig. 4. When the viscoelastic effect is introduced, as shown in \fref{fig:PFasym}.(b), (c) and \fref{fig:PFasymF}.(b), (c), the vesicle stretches slightly in the $x$-direction, while the distributions of phase A and phase B remain similar to the N/N case. This observation is consistent with our findings for the symmetric initialization cases: under the parameter settings we adopt, the viscoelastic effect has only a limited influence on multicomponent vesicles in Poiseuille flow.

\section{Conclusions} \label{Conclusions}
In this paper, we introduce a continuum surface force (CSF) phase-field model to investigate the effects of viscoelasticity on the hydrodynamics of inextensible multicomponent vesicles under various physical conditions, including the influence of inertial forces. The fluid behavior is governed by the full Navier–Stokes equations, augmented with the Oldroyd-B constitutive equations to capture viscoelastic effects. The phase field is modeled using a nonlinear advection–diffusion equation, while the species concentration on the vesicle membrane evolves according to an advective surface Cahn–Hilliard equation. Global area and volume conservation of the vesicle are enforced via Lagrange multipliers, and local inextensibility is maintained through surface tension. The system is coupled using a continuum surface force (CSF) approach. Our framework incorporates bending force, surface tension, and line tension, which collectively govern membrane deformation, ambient fluid flow, and the evolution of surface species. Numerically, the complex system of coupled partial differential equations (PDEs) is solved using an implicit, monolithic scheme based on the generalized-$\alpha$ time integration method. Isogeometric analysis (IGA) is employed to achieve higher spatial accuracy. To ensure numerical stability across different flow regimes and to maintain smooth viscoelastic stress fields, we incorporate the Residual-Based Variational Multiscale Method (RBVMS) and the Streamline-Upwind Petrov–Galerkin (SUPG) stabilization technique in our IGA formulation. \par
Our model is extensively examined in 2D through a wide range of numerical examples to study the viscoelastic effects. The results demonstrate that the proposed model can accurately capture the complex interactions between fluid flow and the inextensible membrane structure, while preserving global area and volume conservation. In shear flow, detailed studies have been conducted on the unique bending rigidity variations of multicomponent vesicles, which result in swinging and tumbling motions primarily governed by bending rigidity variations associated with the multicomponent membrane. More importantly, we identified a damping effect of the viscoelastic fluid, which suppresses the swinging and tumbling motions, leading to prolonged swinging periods and eventual transition to a tank-treading motion. This viscoelastic damping also consistently increases the phase treading period of the components on the multicomponent vesicle membrane. To the best of our knowledge, this is the first study to examine the influence of viscoelastic effects on the hydrodynamics of multicomponent vesicles. In Poiseuille flow, we examined the deformation of the multicomponent vesicle membrane in response to the fluid flow under both symmetric and asymmetric initial conditions, along with the corresponding phase distributions on the membrane surface. We found that viscoelasticity exerts only a minor influence in this flow, due to the limited bending rigidity variation, mainly acting to stretch the vesicle further along the flow direction. Importantly, our results show very good qualitative and quantitative agreement with previous numerical studies and experimental observations. We believe that this model holds significant potential for further exploration in related fluid–structure interaction problems. For example, our model enables extensive future numerical studies on the influence of material parameters and varying Reynolds numbers on the hydrodynamics of multicomponent vesicles under viscoelastic effects. Moreover, more topologically complex cases can be investigated by leveraging our phase-field formulation and IGA framework augmented with various stabilization techniques. Our model is designed for both two- and three-dimensional simulations; with further improvements in numerical efficiency, three-dimensional multicomponent vesicles can be systematically explored.
Investigating a wider range of fluid viscoelasticity, along with more sophisticated stabilization techniques, would also be valuable for understanding vesicle hydrodynamics under broader biophysical conditions. 

\appendix
\renewcommand{\thesection}{Appendix \Alph{section}}  
\section{Explicit formulation of the final weak form}\label{App:A}
In appendix, we provide explicit formulation of the final weak form \eref{eq:RBVMS1}:
\begin{align*} 
	B(\{\bm{w}^h,q^h,r^h,&l^h,s^h,z^h,\bm{y}^h\},\{\bm{u}^h,p^h,\phi^h,f^h,\lambda^h,c^h,\bm{\sigma}^h\})=\\
	&\int_{\Omega} \bm{w}^h \cdot \left( \partial_t \bm{u}^h + \bm{u}^h \cdot \nabla\bm{u}^h \right) ~\text{d}v + \int_{\Omega} \bm{D}(\bm{w}^h):\bm{S}(\bm{u}^h,p^h,\bm{\sigma}^h) ~\text{d}v \\
	& +\int_{\Omega} q^h \nabla \cdot \bm{u}^h ~\text{d}v \\
	&+ \sum^{n_{el}}_{e=1} \int_{\Omega^e} \tau_{M} \left(\bm{u}^h \cdot \nabla\bm{w}^h +\nabla q^h\right) \cdot \bm{r}_{M}(\bm{u}^h,p^h,\bm{\sigma}^h)~\text{d}v \\
	&+ \sum^{n_{el}}_{e=1} \int_{\Omega^e} \tau_{C} \nabla\cdot\bm{w}^h \, r_{\mathrm{C}}(\bm{u}^h)~\text{d}v\\
	& - \sum^{n_{el}}_{e=1} \int_{\Omega^e} \tau_M \bm{w}^h \cdot \left(\bm{r}_{M}(\bm{u}^h,p^h,\bm{\sigma}^h) \cdot \nabla \bm{u}^h\right)~\text{d}v\\
	&- \sum^{n_{el}}_{e=1} \int_{\Omega^e} \nabla\bm{w}^h: \Bigl(\tau_{M} \bm{r}_{M}(\bm{u}^h,p^h,\bm{\sigma}^h) \Bigr) \otimes \Bigl(\tau_{M} \bm{r}_{M}(\bm{u}^h,p^h,\bm{\sigma}^h) \Bigr)~\text{d}v\\
	&+\int_{\Omega} r^h \left(\partial_t\phi^h+\bm{u}^h\cdot\nabla\phi^h\right) ~\text{d}v +\int_{\Omega} \gamma \; r^h \left(\tilde{g}^h-\lambda_{\text{global}}f^h\right) ~\text{d}v  \\
	&+\int_{\Omega} l^h \, f^h ~\text{d}v + \int_{\Omega} \epsilon \nabla l^h \cdot \nabla \phi^h ~\text{d}v +\int_{\Omega} \frac{1}{\epsilon} l^h \left((\phi^h)^2-1\right)\phi^h ~\text{d}v \\
	&- \int_{\Omega} \xi\,\epsilon^2 \left(\nabla s^h \cdot \nabla \lambda^h\right) (\phi^h)^2~\text{d}v
	+ \int_{\Omega} s^h \,|\nabla\phi^h| \, \bm{P}^h:\nabla\bm{u}^h ~\text{d}v\\
	&+\int_{\Omega}z^h \partial_t c^h\delta_t^h~\text{d}v-\int_{\Omega} \nabla z^h\cdot\bm{u}^h c^h\delta_c^h~\text{d}v\\
	&-\int_{\Omega}\frac{1}{Pe} M_c \left(\Delta z^h\delta^h_c+\nabla z^h\cdot\nabla\phi^h(\delta^h_c)'\right)\beta~\text{d}v\\
	&+ \int_{\Omega} \alpha_P \, (\nabla z^h\cdot\bm{n}^h)(\nabla c^h \cdot\bm{n}^h)\, \delta_c^h~\text{d}v\\
	&+\int_{\Omega} \bm{y}^h : \mathcal{W}i\left(\partial_t \bm{\sigma}^h + \bm{u}^h \cdot \nabla\bm{\sigma}^h-\left(\nabla\bm{u}^h\bm{\sigma}^h+\bm{\sigma}^h\nabla^T\bm{u}^h\right)\right)~\text{d}v\\
	&-\int_{\Omega}\bm{y}^h : \alpha_{\mu_p} \bm{D}(\bm{u}^h)~\text{d}v\\
	&+ \sum^{n_{el}}_{e=1} \int_{\Omega^e} \tau_{\sigma} \mathcal{W}i\left(\bm{u}^h \cdot \nabla\bm{y}^h \right) \cdot \bm{r}_{\sigma}(\bm{u}^h,\bm{\sigma}^h)~\text{d}v,
\end{align*} 
\begin{align}\label{eq:RBVMS3}
	L(\{\bm{w}^h,r^h\})&=\int_{\Omega} \bm{w}^h \cdot \bm{F}^h ~\text{d}v
	+ \int_{\partial\Omega_{N}} \bm{w}^h \cdot \bm{h}^h_N ~\text{d}a +\int_{\Omega} \gamma \, r^h \lambda_{\text{volume}}   ~\text{d}v.
\end{align} 
Here $\Omega$ is divided into $n_{el}$ elements denoted by $\Omega^e$. The weak form \eref{eq:RBVMS1} is obtained under certain simplifying assumptions, leading to the RBVMS formulation for the Navier–Stokes equations (cf. \cite{Bazilevs2007173}) and SUPG stabilization for the Oldroyd--B equation.

\bibliographystyle{unsrt}
\bibliography{references}  

@String{arxiv = "Arxiv.org"}

@BOOK{Bird1987,
	title = {{Dynamics of Polymeric Liquids}},
	publisher = {John Wiley \& Sons, Ltd},
	year = {1987},
	author = {Bird, R. B. and Armstrong, R. C. and Hassager, O.},
}

@book{liu2011advances,
	title={Advances in planar lipid bilayers and liposomes},
	author={Liu, A Leitmannova},
	volume={5},
	year={2011},
	publisher={Elsevier}
}

@inbook{Gomez2017,
	author = {Gomez, Hector and van der Zee, Kristoffer G.},
	publisher = {John Wiley \& Sons},
	isbn = {9781119176817},
	title = {Computational Phase-Field Modeling, in Encyclopedia of Computational Mechanics Second Edition},
	booktitle = {Encyclopedia of Computational Mechanics Second Edition},
	pages = {1-35},
	doi = {https://doi.org/10.1002/9781119176817.ecm2118},
	url = {https://onlinelibrary.wiley.com/doi/abs/10.1002/9781119176817.ecm2118},
	eprint = {https://onlinelibrary.wiley.com/doi/pdf/10.1002/9781119176817.ecm2118},
	year = {2017},
}

@book{wang2005phase,
	title={Phase field models and simulations of vesicle bio-membranes},
	author={Wang, Xiaoqiang},
	year={2005},
	publisher={The Pennsylvania State University}
}

@Book{bazilevs2013computational,
	title         = {Computational Fluid-Structure Interaction: Methods and Applications},
	publisher     = {Wiley},
	year          = {2013},
	author        = {Bazilevs, Y. and Takizawa, K. and Tezduyar, T.E.},
	series        = {Wiley Series in Computational Mechanics},
	document_type = {Book},
	isbn          = {9781118483572},
	page_count    = {384},
	source        = {Scopus},
}

@article{BRACKBILL1992335,
	title = {A continuum method for modeling surface tension},
	journal = {Journal of Computational Physics},
	volume = {100},
	number = {2},
	pages = {335-354},
	year = {1992},
	issn = {0021-9991},
	doi = {https://doi.org/10.1016/0021-9991(92)90240-Y},
	url = {https://www.sciencedirect.com/science/article/pii/002199919290240Y},
	author = {J.U Brackbill and D.B Kothe and C Zemach},
}

@article{RATH2024117348,
	title = {An efficient phase-field framework for contact dynamics between deformable solids in fluid flow},
	journal = {Computer Methods in Applied Mechanics and Engineering},
	volume = {432},
	pages = {117348},
	year = {2024},
	issn = {0045-7825},
	doi = {https://doi.org/10.1016/j.cma.2024.117348},
	url = {https://www.sciencedirect.com/science/article/pii/S0045782524006030},
	author = {Biswajeet Rath and Xiaoyu Mao and Rajeev K. Jaiman}
}

@article{LOPEZ2022114564,
	title = {An isogeometric phase–field based shape and topology optimization for flexoelectric structures},
	journal = {Computer Methods in Applied Mechanics and Engineering},
	volume = {391},
	pages = {114564},
	year = {2022},
	issn = {0045-7825},
	doi = {https://doi.org/10.1016/j.cma.2021.114564},
	url = {https://www.sciencedirect.com/science/article/pii/S0045782521007386},
	author = {Jorge Lopez and Navid Valizadeh and Timon Rabczuk}
}

@article{VALIZADEH2025117618,
	title = {A monolithic finite element method for phase-field modeling of fully Eulerian fluid–structure interaction},
	journal = {Computer Methods in Applied Mechanics and Engineering},
	volume = {435},
	pages = {117618},
	year = {2025},
	issn = {0045-7825},
	doi = {https://doi.org/10.1016/j.cma.2024.117618},
	url = {https://www.sciencedirect.com/science/article/pii/S0045782524008727},
	author = {Navid Valizadeh and Xiaoying Zhuang and Timon Rabczuk}
}

@article{Herrmann2021,
	author = {Inge Katrin H., Matthew John Andrew W., Gregor F.},
	title = {Extracellular vesicles as a next-generation drug delivery platform},
	journal = {Nature Nanotechnology},
	volume = {16},
	number = {7},
	pages = {748-759},
	year = {2021},
	doi = {10.1038/s41565-021-00931-2},
	url = {https://doi.org/10.1038/s41565-021-00931-2}
}

@Article{Elani2014,
	author={Elani, Yuval
	and Law, Robert V.
	and Ces, Oscar},
	title={Vesicle-based artificial cells as chemical microreactors with spatially segregated reaction pathways},
	journal={Nat. Commun.},
	year={2014},
	month={Oct},
	day={29},
	publisher={Nature Publishing Group, a division of Macmillan Publishers Limited. All Rights Reserved.},
	volume={5},
	url={http://dx.doi.org/10.1038/ncomms6305}
}

@Article{Abkarian2008,
	doi = {10.1088/1748-6041/3/3/034011},
	url = {https://dx.doi.org/10.1088/1748-6041/3/3/034011},
	year = {2008},
	month = {sep},
	publisher = {},
	volume = {3},
	number = {3},
	pages = {034011},
	author = {Abkarian, Manouk and Faivre, Magalie and Horton, Renita and Smistrup, Kristian and Best-Popescu, Catherine A and Stone, Howard A},
	title = {Cellular-scale hydrodynamics},
	journal = {Biomedical Materials}
}

@Article{VLAHOVSKA2013451,
	title = {Flow dynamics of red blood cells and their biomimetic counterparts},
	journal = {Comptes Rendus Physique},
	volume = {14},
	number = {6},
	pages = {451-458},
	year = {2013},
	note = {Living fluids / Fluides vivants},
	issn = {1631-0705},
	doi = {https://doi.org/10.1016/j.crhy.2013.05.001},
	url = {https://www.sciencedirect.com/science/article/pii/S1631070513000765},
	author = {Petia M. Vlahovska and Dominique Barthes-Biesel and Chaouqi Misbah}
}

@article{Wan2011,
	author = {Wan, Jiandi and Forsyth, Alison M. and Stone, Howard A.},
	title = {Red blood cell dynamics: from cell deformation to ATP release},
	journal = {Integrative Biology},
	volume = {3},
	number = {10},
	pages = {972-981},
	year = {2011},
	month = {09},
	issn = {1757-9708},
	doi = {10.1039/c1ib00044f},
	url = {https://doi.org/10.1039/c1ib00044f},
	eprint = {https://academic.oup.com/ib/article-pdf/3/10/972/27325980/c1ib00044f.pdf},
}

@ARTICLE{Lipowsky1991475,
	author = {Lipowsky, Reinhard},
	title = {The conformation of membranes},
	year = {1991},
	journal = {Nature},
	volume = {349},
	number = {6309},
	pages = {475 – 481},
	doi = {10.1038/349475a0},
	url = {https://www.scopus.com/inward/record.uri?eid=2-s2.0-0025968124&doi=10.1038%2f349475a0&partnerID=40&md5=3eed8b0935f37998b2831352cae01c92}
}

@ARTICLE{Seifert199713,
	author={Seifert, U.},
	title={Configurations of fluid membranes and vesicles},
	journal={Advances in Physics},
	year={1997},
	volume={46},
	pages={13-137},
	 number={1},
}

@article{Keller_Skalak_1982,
	title = {Motion of a tank-treading ellipsoidal particle in a shear flow},
	journal = {Journal of Fluid Mechanics},
	volume = {120},
	pages = {27–47},
	year = {1982},
	issn = {0022-1120},
	doi = {10.1017/S0022112082002651},
	url = {https://www.cambridge.org/core/product/61BEBFECC4835F57A06AEC7F907E728A},
	author = {Keller, Stuart R. and Skalak, Richard},
}

@article{Noguchi2007,
	author = {Noguchi, Hiroshi and Gompper, Gerhard},
	year = {2007},
	month = {04},
	pages = {128103},
	title = {Swinging and Tumbling of Fluid Vesicles in Shear Flow},
	volume = {98},
	journal = {Physical review letters},
	doi = {10.1103/PhysRevLett.98.128103}
}

@article{Gera_2022,
	title={Swinging and tumbling of multicomponent vesicles in flow},
	volume={935},
	ISSN={1469-7645},
	url={http://dx.doi.org/10.1017/jfm.2022.40},
	DOI={10.1017/jfm.2022.40},
	journal={Journal of Fluid Mechanics},
	publisher={Cambridge University Press (CUP)},
	author={Gera, Prerna and Salac, David and Spagnolie, Saverio E.},
	year={2022},
	month=feb 
}

@article{WEN2024117390,
	title = {Hydrodynamics of multicomponent vesicles: A phase-field approach},
	journal = {Computer Methods in Applied Mechanics and Engineering},
	volume = {432},
	pages = {117390},
	year = {2024},
	issn = {0045-7825},
	doi = {https://doi.org/10.1016/j.cma.2024.117390},
	url = {https://www.sciencedirect.com/science/article/pii/S0045782524006455},
	author = {Zuowei Wen and Navid Valizadeh and Timon Rabczuk and Xiaoying Zhuang},
}

@article{ALAND201432,
	title = "Diffuse interface models of locally inextensible vesicles in a viscous fluid",
	journal = "Journal of Computational Physics",
	volume = "277",
	pages = "32 - 47",
	year = "2014",
	issn = "0021-9991",
	doi = "https://doi.org/10.1016/j.jcp.2014.08.016",
	url = "http://www.sciencedirect.com/science/article/pii/S0021999114005622",
	author = "Sebastian Aland and Sabine Egerer and John Lowengrub and Axel Voigt",
}

@article{GERA2018362,
	title = {Modeling of multicomponent three-dimensional vesicles},
	journal = {Computers \& Fluids},
	volume = {172},
	pages = {362-383},
	year = {2018},
	issn = {0045-7930},
	doi = {https://doi.org/10.1016/j.compfluid.2018.04.003},
	url = {https://www.sciencedirect.com/science/article/pii/S0045793018301919},
	author = {Prerna Gera and David Salac}
}

@article{VALIZADEH2022114191,
	title = {Isogeometric analysis of hydrodynamics of vesicles using a monolithic phase-field approach},
	journal = {Computer Methods in Applied Mechanics and Engineering},
	volume = {388},
	pages = {114191},
	year = {2022},
	issn = {0045-7825},
	doi = {https://doi.org/10.1016/j.cma.2021.114191},
	url = {https://www.sciencedirect.com/science/article/pii/S0045782521005223},
	author = {Navid Valizadeh and Timon Rabczuk}
}

@article{Gannon2021,
	author = {Quaife, Bryan and Gannon, Ashley and Young, Yuan-Nan},
	year = {2021},
	month = {07},
	pages = {},
	title = {Hydrodynamics of a semipermeable inextensible membrane under flow and confinement},
	volume = {6},
	journal = {Physical Review Fluids},
	doi = {10.1103/PhysRevFluids.6.073601}
}

@article{ESCOTT2024105262,
	title = {Rheology of a suspension of deformable spheres in a weakly viscoelastic fluid},
	journal = {Journal of Non-Newtonian Fluid Mechanics},
	volume = {330},
	pages = {105262},
	year = {2024},
	issn = {0377-0257},
	doi = {https://doi.org/10.1016/j.jnnfm.2024.105262},
	url = {https://www.sciencedirect.com/science/article/pii/S0377025724000788},
	author = {Liam J. Escott and Helen J. Wilson}
}

@article{Naseer_Izbassarov_Ahmed_Muradoglu_2024, title={Lateral migration of a deformable fluid particle in a square channel flow of viscoelastic fluid}, volume={996}, DOI={10.1017/jfm.2024.583}, journal={Journal of Fluid Mechanics}, author={Naseer, Hafiz Usman and Izbassarov, Daulet and Ahmed, Zaheer and Muradoglu, Metin}, year={2024}, pages={A31}}

@article{YUE_FENG_LIU_SHEN_2005, title={Viscoelastic effects on drop deformation in steady shear}, volume={540}, DOI={10.1017/S0022112005006166}, journal={Journal of Fluid Mechanics}, author={YUE, PENGTAO and FENG, JAMES J. and LIU, CHUN and SHEN, JIE}, year={2005}, pages={427–437}}

@article{Cahn1958,
	author = {Cahn, John W. and Hilliard, John E.},
	title = "{Free Energy of a Nonuniform System. I. Interfacial Free Energy}",
	journal = {The Journal of Chemical Physics},
	volume = {28},
	number = {2},
	pages = {258-267},
	year = {1958},
	month = {02},
	issn = {0021-9606},
	doi = {10.1063/1.1744102},
	url = {https://doi.org/10.1063/1.1744102},
	eprint = {https://pubs.aip.org/aip/jcp/article-pdf/28/2/258/18813541/258\_1\_online.pdf},
}

@Article{C6SM02452A,
	author ="Liu, Kai and Marple, Gary R. and Allard, Jun and Li, Shuwang and Veerapaneni, Shravan and Lowengrub, John",
	title  ="Dynamics of a multicomponent vesicle in shear flow",
	journal  ="Soft Matter",
	year  ="2017",
	volume  ="13",
	issue  ="19",
	pages  ="3521-3531",
	publisher  ="The Royal Society of Chemistry",
	doi  ="10.1039/C6SM02452A",
	url  ="http://dx.doi.org/10.1039/C6SM02452A",
}

@article{Laadhari2012,
	author = {Laadhari,Aymen  and Saramito,Pierre  and Misbah,Chaouqi },
	title = {Vesicle tumbling inhibited by inertia},
	journal = {Physics of Fluids},
	volume = {24},
	number = {3},
	pages = {031901},
	year = {2012},
	doi = {10.1063/1.3690862},
	URL = { 
	https://doi.org/10.1063/1.3690862
	},
	eprint = { 
	https://doi.org/10.1063/1.3690862
	}
}

@article{salac_miksis_2012, title={Reynolds number effects on lipid vesicles}, volume={711}, DOI={10.1017/jfm.2012.380}, journal={Journal of Fluid Mechanics}, publisher={Cambridge University Press}, author={Salac, David and Miksis, Michael J.}, year={2012}, pages={122–146}}

@article{LAADHARI2014328,
	title = "Computing the dynamics of biomembranes by combining conservative level set and adaptive finite element methods",
	journal = "Journal of Computational Physics",
	volume = "263",
	pages = "328 - 352",
	year = "2014",
	issn = "0021-9991",
	doi = "https://doi.org/10.1016/j.jcp.2013.12.032",
	url = "http://www.sciencedirect.com/science/article/pii/S0021999113008395",
	author = "Aymen Laadhari and Pierre Saramito and Chaouqi Misbah",
}

@article{SEOL20191009,
	title = {An immersed boundary method for simulating Newtonian vesicles in viscoelastic fluid},
	journal = {Journal of Computational Physics},
	volume = {376},
	pages = {1009-1027},
	year = {2019},
	issn = {0021-9991},
	doi = {https://doi.org/10.1016/j.jcp.2018.10.027},
	url = {https://www.sciencedirect.com/science/article/pii/S0021999118306946},
	author = {Yunchang Seol and Yu-Hau Tseng and Yongsam Kim and Ming-Chih Lai}
}

@article{SALAC20118192,
	title = {A level set projection model of lipid vesicles in general flows},
	journal = {Journal of Computational Physics},
	volume = {230},
	number = {22},
	pages = {8192-8215},
	year = {2011},
	issn = {0021-9991},
	doi = {https://doi.org/10.1016/j.jcp.2011.07.019},
	url = {https://www.sciencedirect.com/science/article/pii/S0021999111004384},
	author = {D. Salac and M. Miksis}
}

@article{DU2009923,
	title = "Energetic variational approaches in modeling vesicle and fluid interactions",
	journal = "Physica D: Nonlinear Phenomena",
	volume = "238",
	number = "9",
	pages = "923 - 930",
	year = "2009",
	issn = "0167-2789",
	doi = "https://doi.org/10.1016/j.physd.2009.02.015",
	url = "http://www.sciencedirect.com/science/article/pii/S0167278909000669",
	author = "Qiang Du and Chun Liu and Rolf Ryham and Xiaoqiang Wang"
}

@article{Du2007,
	title = {Analysis of a phase field Navier-Stokes vesicle-fluid interaction model},
	author = {Qiang Du and Manlin Li and Chun Liu},
	journal = {Discrete \& Continuous Dynamical Systems-B},
	volume = {8},
	issue = {3},
	pages = {539-556},
	year = {2007},
}

@article{SOHN2010119,
	title = {Dynamics of multicomponent vesicles in a viscous fluid},
	journal = {Journal of Computational Physics},
	volume = {229},
	number = {1},
	pages = {119-144},
	year = {2010},
	issn = {0021-9991},
	doi = {https://doi.org/10.1016/j.jcp.2009.09.017},
	url = {https://www.sciencedirect.com/science/article/pii/S0021999109005099},
	author = {Jin Sun Sohn and Yu-Hau Tseng and Shuwang Li and Axel Voigt and John S. Lowengrub},
}

@article{Bachini_Krause_Nitschke_Voigt_2023, title={Derivation and simulation of a two-phase fluid deformable surface model}, volume={977}, DOI={10.1017/jfm.2023.943}, journal={Journal of Fluid Mechanics}, author={Bachini, Elena and Krause, Veit and Nitschke, Ingo and Voigt, Axel}, year={2023}, pages={A41}}

@article{Barrett2017,
	author = {{Barrett, John W.} and {Garcke, Harald} and {N\"{u}rnberg, Robert}},
	title = {Finite element approximation  for the dynamics of fluidic two-phase biomembranes},
	DOI= "10.1051/m2an/2017037",
	url= "https://doi.org/10.1051/m2an/2017037",
	journal = {ESAIM: M2AN},
	year = 2017,
	volume = 51,
	number = 6,
	pages = "2319-2366",
}

@misc{venkatesh2024shapedynamicsnearlyspherical,
	title={Shape dynamics of nearly spherical, multicomponent vesicles under shear flow}, 
	author={Anirudh Venkatesh and Vivek Narsimhan},
	year={2024},
	eprint={2409.15102},
	archivePrefix={arXiv},
	primaryClass={physics.flu-dyn},
	url={https://arxiv.org/abs/2409.15102}, 
	note={arXiv:2409.15102 [physics.flu-dyn]},
	howpublished={\url{https://arxiv.org/abs/2409.15102}}
}

@article{PhysRevLett.94.148101,
	title = {Miscibility Phase Diagrams of Giant Vesicles Containing Sphingomyelin},
	author = {Veatch, Sarah L. and Keller, Sarah L.},
	journal = {Phys. Rev. Lett.},
	volume = {94},
	issue = {14},
	pages = {148101},
	numpages = {4},
	year = {2005},
	month = {Apr},
	publisher = {American Physical Society},
	doi = {10.1103/PhysRevLett.94.148101},
	url = {https://link.aps.org/doi/10.1103/PhysRevLett.94.148101}
}

@article{Davis2009,
	title = {Phase Equilibria in DOPC/DPPC-d<sub>62</sub>/Cholesterol Mixtures},
	author = {Davis and James, H.},
	journal = {Biophysical Journal},
	volume = {96},
	issue = {2},
	pages = {521 - 539},
	year = {2009},
	publisher = {Elsevier},
	doi = {10.1016/j.bpj.2008.09.042},
	url = {https://doi.org/10.1016/j.bpj.2008.09.042}
}

@article{Pradeep2010,
	title = {Orientation of Tie-Lines in the Phase Diagram of DOPC/DPPC/Cholesterol Model Biomembranes},
	author = {Uppamoochikkal, P. and Tristram-Nagle, S. and Nagle, J. F.},
	journal = {Langmuir},
	volume = {26},
	issue = {22},
	pages = {17363 - 17368},
	year = {2010},
	publisher = {American Chemical Society},
	doi = {10.1021/la103024f},
	url = {https://doi.org/10.1021/la103024f}
}

@article{VEATCH2005172,
	title = {Seeing spots: Complex phase behavior in simple membranes},
	journal = {Biochimica et Biophysica Acta (BBA) - Molecular Cell Research},
	volume = {1746},
	number = {3},
	pages = {172-185},
	year = {2005},
	note = {Lipid Rafts: From Model Membranes to Cells},
	issn = {0167-4889},
	doi = {https://doi.org/10.1016/j.bbamcr.2005.06.010},
	url = {https://www.sciencedirect.com/science/article/pii/S0167488905001473},
	author = {Sarah L. Veatch and Sarah L. Keller}
}

@article{CASQUERO2017646,
	title = "NURBS-based numerical proxies for red blood cells and circulating tumor cells in microscale blood flow",
	journal = "Computer Methods in Applied Mechanics and Engineering",
	volume = "316",
	pages = "646 - 667",
	year = "2017",
	note = "Special Issue on Isogeometric Analysis: Progress and Challenges",
	issn = "0045-7825",
	doi = "https://doi.org/10.1016/j.cma.2016.09.031",
	url = "http://www.sciencedirect.com/science/article/pii/S0045782516312233",
	author = "Hugo Casquero and Carles Bona-Casas and Hector Gomez",
}

@article{CASQUERO2021109872,
	title = {The divergence-conforming immersed boundary method: Application to vesicle and capsule dynamics},
	journal = {Journal of Computational Physics},
	volume = {425},
	pages = {109872},
	year = {2021},
	issn = {0021-9991},
	doi = {https://doi.org/10.1016/j.jcp.2020.109872},
	url = {https://www.sciencedirect.com/science/article/pii/S002199912030646X},
	author = {Hugo Casquero and Carles Bona-Casas and Deepesh Toshniwal and Thomas J.R. Hughes and Hector Gomez and Yongjie Jessica Zhang}
}

@Article{Veerapaneni20115610,
	author  = {Veerapaneni, S.K. and Rahimian, A. and Biros, G. and Zorin, D.},
	title   = {A fast algorithm for simulating vesicle flows in three dimensions},
	journal = {Journal of Computational Physics},
	year    = {2011},
	volume  = {230},
	number  = {14},
	pages   = {5610-5634},
	doi     = {10.1016/j.jcp.2011.03.045},
}

@article{LAADHARI2017271,
	title = {Fully implicit methodology for the dynamics of biomembranes and capillary interfaces by combining the level set and Newton methods},
	journal = {Journal of Computational Physics},
	volume = {343},
	pages = {271-299},
	year = {2017},
	issn = {0021-9991},
	doi = {https://doi.org/10.1016/j.jcp.2017.04.019},
	url = {https://www.sciencedirect.com/science/article/pii/S0021999117302905},
	author = {Aymen Laadhari and Pierre Saramito and Chaouqi Misbah and Gábor Székely},
}

@article{GONG2025111039,
	title = {A phase-field study on thermo-mechanical coupled damage evolution and failure mechanisms of sintered silver interconnections},
	journal = {Engineering Fracture Mechanics},
	volume = {320},
	pages = {111039},
	year = {2025},
	issn = {0013-7944},
	doi = {https://doi.org/10.1016/j.engfracmech.2025.111039},
	url = {https://www.sciencedirect.com/science/article/pii/S0013794425002401},
	author = {Yanpeng Gong and Yuguo Kou and Qiang Yue and Xiaoying Zhuang and Navid Valizadeh and Fei Qin and Qiao Wang and Timon Rabczuk}
}

@article{XU2020112648,
	title = {Phase-field model of vascular tumor growth: Three-dimensional geometry of the vascular network and integration with imaging data},
	journal = {Computer Methods in Applied Mechanics and Engineering},
	volume = {359},
	pages = {112648},
	year = {2020},
	issn = {0045-7825},
	doi = {https://doi.org/10.1016/j.cma.2019.112648},
	url = {https://www.sciencedirect.com/science/article/pii/S0045782519305328},
	author = {Jiangping Xu and Guillermo Vilanova and Hector Gomez},
}

@article{VALIZADEH2019599,
	title = "Isogeometric analysis for phase-field models of geometric {PDE}s and high-order {PDE}s on stationary and evolving surfaces",
	journal = "Computer Methods in Applied Mechanics and Engineering",
	volume = "351",
	pages = "599 - 642",
	year = "2019",
	issn = "0045-7825",
	doi = "https://doi.org/10.1016/j.cma.2019.03.043",
	url = "http://www.sciencedirect.com/science/article/pii/S0045782519301823",
	author = "Navid Valizadeh and Timon Rabczuk",
}

@article{Morris2001,
	title = {Cell Surface Area Regulation and Membrane Tension},
	journal = {The Journal of Membrane Biology},
	volume = {179},
	pages = {79-102},
	year = {2001},
	issn = {1432-1424},
	doi = {https://doi.org/10.1007/s002320010040},
	url = {https://doi.org/10.1007/s002320010040},
	author = {Morris, C.E. and Homann, U.},
}

@Article{C8SM01087K,
	author ="Gera, Prerna and Salac, David",
	title  ="Three-dimensional multicomponent vesicles: dynamics and influence of material properties",
	journal  ="Soft Matter",
	year  ="2018",
	volume  ="14",
	issue  ="37",
	pages  ="7690-7705",
	publisher  ="The Royal Society of Chemistry",
	doi  ="10.1039/C8SM01087K",
	url  ="http://dx.doi.org/10.1039/C8SM01087K"
}

@article{Dolow2013,
	author = {Embar, Anand and Dolbow, John and Fried, Eliot},
	year = {2013},
	month = {08},
	pages = {597-615},
	title = {Microdomain evolution on giant unilamellar vesicles},
	volume = {12},
	journal = {Biomechanics and modeling in mechanobiology},
	doi = {10.1007/s10237-012-0428-1}
}

@article{Abels2012,
	author = {Abels, HELMUT and Garcke, HARALD and Gr\"{u}n, G\"{U}NTHER},
	title = {THERMODYNAMICALLY CONSISTENT, FRAME INDIFFERENT DIFFUSE INTERFACE MODELS FOR INCOMPRESSIBLE TWO-PHASE FLOWS WITH DIFFERENT DENSITIES},
	journal = {Mathematical Models and Methods in Applied Sciences},
	volume = {22},
	number = {03},
	pages = {1150013},
	year = {2012},
	doi = {10.1142/S0218202511500138},
	URL = { 
	https://doi.org/10.1142/S0218202511500138
	},
	eprint = { 
	https://doi.org/10.1142/S0218202511500138
	}
}

@article{DU2006757,
	title = "Simulating the deformation of vesicle membranes under elastic bending energy in three dimensions",
	journal = "Journal of Computational Physics",
	volume = "212",
	number = "2",
	pages = "757 - 777",
	year = "2006",
	issn = "0021-9991",
	doi = "https://doi.org/10.1016/j.jcp.2005.07.020",
	url = "http://www.sciencedirect.com/science/article/pii/S0021999105003566",
	author = "Qiang Du and Chun Liu and Xiaoqiang Wang"
}

@article{LOWENGRUB2016112,
	title = {Numerical simulation of endocytosis: Viscous flow driven by membranes with non-uniformly distributed curvature-inducing molecules},
	journal = {Journal of Computational Physics},
	volume = {309},
	pages = {112-128},
	year = {2016},
	issn = {0021-9991},
	doi = {https://doi.org/10.1016/j.jcp.2015.12.055},
	url = {https://www.sciencedirect.com/science/article/pii/S0021999115008803},
	author = {John Lowengrub and Jun Allard and Sebastian Aland},
}

@article{GREER2006216,
	title = {Fourth order partial differential equations on general geometries},
	journal = {Journal of Computational Physics},
	volume = {216},
	number = {1},
	pages = {216-246},
	year = {2006},
	issn = {0021-9991},
	doi = {https://doi.org/10.1016/j.jcp.2005.11.031},
	url = {https://www.sciencedirect.com/science/article/pii/S0021999105005498},
	author = {John B. Greer and Andrea L. Bertozzi and Guillermo Sapiro},
}

@article{Voigt05,
	title = "P{DE}'s on surfaces -- a diffuse interface approach",
	journal = "Communications in Mathematical Sciences",
	author = "Andreas Raetz and Axel Voigt",
	year = "2006",
	volume = "4",
	number = "3",
	pages = "575-590",
	doi = "http://dx.doi.org/10.4310/CMS.2006.v4.n3.a5"
}

@article{KLOPPE2024117090,
	title = {A phase-field model of elastic and viscoelastic surfaces in fluids},
	journal = {Computer Methods in Applied Mechanics and Engineering},
	volume = {428},
	pages = {117090},
	year = {2024},
	issn = {0045-7825},
	doi = {https://doi.org/10.1016/j.cma.2024.117090},
	url = {https://www.sciencedirect.com/science/article/pii/S0045782524003463},
	author = {Maximilian Kloppe and Sebastian Aland}
}

@article{BROOKS1982199,
	title = {Streamline upwind/Petrov-Galerkin formulations for convection dominated flows with particular emphasis on the incompressible Navier-Stokes equations},
	journal = {Computer Methods in Applied Mechanics and Engineering},
	volume = {32},
	number = {1},
	pages = {199-259},
	year = {1982},
	issn = {0045-7825},
	doi = {https://doi.org/10.1016/0045-7825(82)90071-8},
	url = {https://www.sciencedirect.com/science/article/pii/0045782582900718},
	author = {Alexander N. Brooks and Thomas J.R. Hughes}
}

@ARTICLE{Bazilevs2007173,
	author={Bazilevs, Y. and Calo, V.M. and Cottrell, J.A. and Hughes, T.J.R. and Reali, A. and Scovazzi, G.},
	title={Variational multiscale residual-based turbulence modeling for large eddy simulation of incompressible flows},
	journal={Computer Methods in Applied Mechanics and Engineering},
	year={2007},
	volume={197},
	number={1-4},
	pages={173-201},
	url={http://www.scopus.com/inward/record.url?eid=2-s2.0-35348895586&partnerID=40&md5=6b50d35e2ff330bf7561a306582cac4a},
	document_type={Article},
	source={Scopus},
}

@misc{aland2023phasefieldmodelactivecontractile,
	title={A phase-field model for active contractile surfaces}, 
	author={Sebastian Aland and Claudia Wohlgemuth},
	year={2023},
	eprint={2306.16796},
	archivePrefix={arXiv},
	primaryClass={q-bio.CB},
	url={https://arxiv.org/abs/2306.16796}, 
}

@article{XU2006590,
	title = {A level-set method for interfacial flows with surfactant},
	journal = {Journal of Computational Physics},
	volume = {212},
	number = {2},
	pages = {590-616},
	year = {2006},
	issn = {0021-9991},
	doi = {https://doi.org/10.1016/j.jcp.2005.07.016},
	url = {https://www.sciencedirect.com/science/article/pii/S0021999105003475},
	author = {Jian-Jun Xu and Zhilin Li and John Lowengrub and Hongkai Zhao}
}

@ARTICLE{Chung1993371,
	author={Chung, J. and Hulbert, G.M.},
	title={A time integration algorithm for structural dynamics with improved numerical dissipation: The generalized-$\alpha$ method},
	journal={Journal of Applied Mechanics, Transactions ASME},
	year={1993},
	volume={60},
	number={2},
	pages={371-375},
	doi={10.1115/1.2900803},
	url={https://www.scopus.com/inward/record.uri?eid=2-s2.0-0027610910&doi=10.1115%2f1.2900803&partnerID=40&md5=671986739d00a55fe53faf4f1712b7e9},
	document_type={Article},
	source={Scopus},
}

@ARTICLE{Jansen2000305,
	author={Jansen, K.E. and Whiting, C.H. and Hulbert, G.M.},
	title={Generalized-$\alpha$ method for integrating the filtered {N}avier-{S}tokes equations with a stabilized finite element method},
	journal={Computer Methods in Applied Mechanics and Engineering},
	year={2000},
	volume={190},
	number={3-4},
	pages={305-319},
	url={https://www.scopus.com/inward/record.uri?eid=2-s2.0-0034287408&partnerID=40&md5=acf023e0c7e2fdeb9df4744031e999ee},
	document_type={Article},
	source={Scopus},
}

@BOOK{Hughes2009,
	author = {Cottrell, J. and Hughes, Thomas and Bazilevs, Yuri},
	year = {2009},
	month = {09},
	title = {Isogeometric Analysis: Toward integration of CAD and FEA},
	isbn = {0470748737},
	journal = {Isogeometric Analysis: Toward Integration of CAD and FEA},
	doi = {10.1002/9780470749081.ch7}
}

@article{PetIGA,
	author = "L. Dalcin and N. Collier and P. Vignal and A.M.A. Côrtes and V.M. Calo",
	title = "PetIGA: A framework for high-performance isogeometric analysis",
	journal = "Computer Methods in Applied Mechanics and Engineering",
	volume = "308",
	pages = "151--181",
	year = "2016",
	issn = "0045-7825",
	doi = "https://doi.org/10.1016/j.cma.2016.05.011",
}

@Misc{petsc-web-page,
	author = {Satish Balay and Shrirang Abhyankar and Mark~F. Adams and Jed Brown and Peter Brune
	and Kris Buschelman and Lisandro Dalcin and Victor Eijkhout and William~D. Gropp
	and Dinesh Kaushik and Matthew~G. Knepley
	and Lois Curfman McInnes and Karl Rupp and Barry~F. Smith
	and Stefano Zampini and Hong Zhang},
	title =  {{PETS}c {W}eb page},
	url =    {http://www.mcs.anl.gov/petsc},
	howpublished = {\url{http://www.mcs.anl.gov/petsc}},
	year = {2015}
}

@TechReport{petsc-user-ref,
	author = {Satish Balay and Shrirang Abhyankar and Mark~F. Adams and Jed Brown and Peter Brune
	and Kris Buschelman and Lisandro Dalcin and Victor Eijkhout and William~D. Gropp
	and Dinesh Kaushik and Matthew~G. Knepley and Dave~A. May and Lois Curfman McInnes
	and Richard Tran Mills and Todd Munson and Karl Rupp and Patrick Sanan
	and Barry~F. Smith and Stefano Zampini and Hong Zhang and Hong Zhang},
	title  = {{PETS}c Users Manual},
	institution = {Argonne National Laboratory},
	year   = 2018,
	number = {ANL-95/11 - Revision 3.9},
	url    = {http://www.mcs.anl.gov/petsc}
}

@article{PhysRevLett.103.188101,
	title = {Why Do Red Blood Cells Have Asymmetric Shapes Even in a Symmetric Flow?},
	author = {Kaoui, Badr and Biros, George and Misbah, Chaouqi},
	journal = {Phys. Rev. Lett.},
	volume = {103},
	issue = {18},
	pages = {188101},
	numpages = {4},
	year = {2009},
	month = {Oct},
	publisher = {American Physical Society},
	doi = {10.1103/PhysRevLett.103.188101},
	url = {https://link.aps.org/doi/10.1103/PhysRevLett.103.188101}
}

@phdthesis{Bartezzaghi2017,
	author  = {Bartezzaghi, Andrea},
	title   = {Isogeometric Analysis for High Order Geometric Partial Differential Equations with Applications},
	year    = {2017},
	school = {\'{E}cole Polytechnique F\'{e}d\'{e}rale de Lausanne (EPFL), Switzerland}
}

@article{ONG2020109277,
	title = "An immersed boundary projection method for simulating the inextensible vesicle dynamics",
	journal = "Journal of Computational Physics",
	volume = "408",
	pages = "109277",
	year = "2020",
	issn = "0021-9991",
	doi = "https://doi.org/10.1016/j.jcp.2020.109277",
	url = "http://www.sciencedirect.com/science/article/pii/S0021999120300516",
	author = "Kian Chuan Ong and Ming-Chih Lai"
}

@article {Skalak717,
	author = {Skalak, R. and Branemark, P I},
	title = {Deformation of Red Blood Cells in Capillaries},
	volume = {164},
	number = {3880},
	pages = {717--719},
	year = {1969},
	doi = {10.1126/science.164.3880.717},
	publisher = {American Association for the Advancement of Science},
	issn = {0036-8075},
	URL = {https://science.sciencemag.org/content/164/3880/717},
	eprint = {https://science.sciencemag.org/content/164/3880/717.full.pdf},
	journal = {Science}
}

@article{GUIDO2009751,
	title = "Microconfined flow behavior of red blood cells in vitro",
	journal = "Comptes Rendus Physique",
	volume = "10",
	number = "8",
	pages = "751 - 763",
	year = "2009",
	note = "Complex and biofluids",
	issn = "1631-0705",
	doi = "https://doi.org/10.1016/j.crhy.2009.10.002",
	url = "http://www.sciencedirect.com/science/article/pii/S1631070509001480",
	author = "Stefano Guido and Giovanna Tomaiuolo",
}

@article{PhysRevE.84.011902,
	title = {Symmetry breaking of vesicle shapes in Poiseuille flow},
	author = {Farutin, Alexander and Misbah, Chaouqi},
	journal = {Phys. Rev. E},
	volume = {84},
	issue = {1},
	pages = {011902},
	numpages = {7},
	year = {2011},
	month = {Jul},
	publisher = {American Physical Society},
	doi = {10.1103/PhysRevE.84.011902},
	url = {https://link.aps.org/doi/10.1103/PhysRevE.84.011902}
}

@article{PhysRevE.89.042709,
	title = {Symmetry breaking and cross-streamline migration of three-dimensional vesicles in an axial Poiseuille flow},
	author = {Farutin, Alexander and Misbah, Chaouqi},
	journal = {Phys. Rev. E},
	volume = {89},
	issue = {4},
	pages = {042709},
	numpages = {11},
	year = {2014},
	month = {Apr},
	publisher = {American Physical Society},
	doi = {10.1103/PhysRevE.89.042709},
	url = {https://link.aps.org/doi/10.1103/PhysRevE.89.042709}
}

@ARTICLE{Canham1970,
	author={Canham, P.B.},
	title={The minimum energy of bending as a possible explanation of the biconcave shape of the human red blood cell},
	journal={Journal of Theoretical Biology},
	year={1970},
	volume={26},
	number={1},
	pages={61-76,IN7-IN8,77-81},
	doi={10.1016/S0022-5193(70)80032-7},
}

@ARTICLE{Helfrich1973693,
	author={Helfrich, W.},
	title={Elastic properties of lipid bilayers: theory and possible experiments.},
	journal={Zeitschrift fur Naturforschung. Teil C: Biochemie, Biophysik, Biologie, Virologie},
	year={1973},
	volume={28},
	number={11},
	pages={693-703},
}






\end{document}